\documentclass{article}

\usepackage[includeheadfoot,
            bindingoffset = 0mm,
            inner = 25mm,
            top = 20mm,
            outer = 25mm,
            bottom = 10mm,
            paperwidth = 210mm,
            paperheight = 297mm,
            ]{geometry}
\usepackage[margin={2.0cm,0cm},oneside,labelfont={sf,bf},singlelinecheck=false]{caption}

\usepackage{graphicx}

\usepackage[fleqn]{amsmath}
\usepackage{accents}
\usepackage{amssymb}
\usepackage{stmaryrd}
\SetSymbolFont{stmry}{bold}{U}{stmry}{m}{n}
\usepackage{mathrsfs}

\usepackage{enumitem}
\usepackage{cite}
\usepackage[perpage]{footmisc}
\usepackage{hyphenat}
\usepackage[english]{babel}
\usepackage[section]{placeins}
\usepackage{upgreek}
\usepackage{color}

\usepackage{amsthm}
\theoremstyle{definition}

\usepackage{titlesec}
\titleformat{\section}[hang]{\Large\bfseries\raggedright\sffamily}{\thesection}{1em}{}
\titleformat{\subsection}[hang]{\large\bfseries\raggedright\sffamily}{\thesubsection}{1em}{}
\titleformat{\subsubsection}[hang]{\normalsize\bfseries\raggedright\sffamily}{\thesubsubsection}{1em}{}
\usepackage{abstract}

\usepackage{authblk}

\newcommand{\vectorn}[1]{\ensuremath{ \vec{#1} }}

\newcommand{\tensor}[1]{\ensuremath{ \boldsymbol{#1} }}
\newcommand{\tensorThree}[1]{\ensuremath{ {}^3\!\boldsymbol{#1} }}
\newcommand{\tensorFour}[1]{\ensuremath{ {}^4\!\boldsymbol{#1} }}
\newcommand{\transp}{\ensuremath{ ^\mathrm{T} }}
\newcommand{\transpRight}{\ensuremath{ ^\mathrm{RT} }}
\newcommand{\transpm}{\ensuremath{ ^\mathrm{-T} }}

\newcommand{\dif}{\ensuremath{ \operatorname{d}\! }}

\newcommand{\determ}[1]{\ensuremath{ \operatorname{det}\left(#1\right) }}

\begin{document}

\title{ \huge\bfseries\sffamily Transformation front kinetics in deformable ferromagnets }

\author{Michael Poluektov}
\affil{School of Computing and Mathematical Sciences, University of Greenwich,\\Park Row, London SE10 9LS, UK}
\affil{Corresponding author, email: m.poluektov@greenwich.ac.uk}

\date{ \large\normalfont\sffamily DRAFT: \today }

\maketitle

\setlength{\absleftindent}{2.0cm}
\setlength{\absrightindent}{2.0cm}
\setlength{\absparindent}{0em}
\begin{abstract}
Materials such as magnetic shape-memory alloys possess an intrinsic coupling between material's magnetisation and mechanical deformation. These materials also undergo structural phase transitions, with phase boundaries separating different phases. The kinetics of the phase boundaries is governed by the magnetic field and the mechanical stresses. There is a multiplicity of other materials revealing similar phenomena, e.g.\ magnetic perovskites. To model the propagation of the phase boundaries in deformable magnetic materials at the continuum scale, three ingredients are required: a set of governing equations for the bulk behaviour with coupled magnetic and mechanical degrees of freedom, a dependency of the phase boundary velocity on the governing factors, and a reliable computational method. The expression for the phase boundary velocity is usually obtained within the continuum thermodynamics setting, where the entropy production due to phase boundary propagation is derived, which gives a thermodynamic driving force for the phase boundary kinetics. For deformable ferromagnets, all three elements (bulk behaviour, interface kinetics, and computational approaches) have been explored, but under a number of limitations. The present paper focuses on the derivation of the thermodynamic driving force for transformation fronts in a general magneto-mechanical setting, adapts the cut-finite-element method for transformation fronts in magneto-mechanics, which allows for an exceptionally efficient handling of the propagating interfaces, without modifying the finite-element mesh, and applies the developments to qualitative modelling of magneto-mechanics of magnetic shape-memory alloys. magneto-structural phase transitions; phase boundary kinetics;\\ cut-finite-element method; magnetic shape-memory alloys.
\end{abstract}

\section{Introduction}
\label{sec:intro}

There is a wide range of materials that possess a coupling between magnetic and mechanical behaviour. A classical example is the magnetostriction effect in ferromagnets \cite{Aharoni1996,RenukaBalakrishna2024}: upon magnetisation of a ferromagnet, it can expand/shrink in the magnetisation direction. Materials with magneto-mechanical coupling can also undergo phase transitions. If both magnetic and mechanical properties of the phases are different, such phase transitions are called magneto-structural. They are experimentally observed for example for magnetic perovskites \cite{Kuwahara1995,Eremenko2001}.

Modelling of magneto-structural phase transitions requires a rigorous continuum theory handling electromagnetism, mechanics of deformable solids, and magnetisation dynamics in a unified way. The major challenge for development of such theory is writing the underlying system of equations in a \emph{thermodynamically consistent} way. This was largely solved in 1970s-80s by G\'{e}rard Maugin and is summarised in monograph \cite{Maugin1988} or in a more recent review \cite{Maugin2011}. The original theory did not cover magneto-structural phase transitions and, in particular, the case of phase co-existence and propagation of sharp phase boundaries separating the phases. This subject was researched over a series of publications in 1990s \cite{Fomethe1996,Fomethe1997,Maugin1997}. However, the derivations were limited to elastic material behaviour and magnetostatics. In the classical continuum thermodynamics approach (without coupling to electromagnetism), entropy production due to phase boundary propagation is derived without any assumptions on constitutive relations \cite{Abeyaratne2006}, which allows obtaining the most general expression for the configurational force that governs the kinetics of phase boundaries.

The aim of the present paper is (a) to provide a more general derivation of the entropy production due to propagation of transformation fronts in deformable ferromagnets, without reliance on assumptions regarding material's constitutive behaviour or absence of some dynamic terms, (b) to simplify the governing equations for the case of non-dissipative solids and quasistatistics and to show that solving the latter is equivalent to minimisation of an energy functional, (c) to adapt the cut-finite-element method for resolving transformation fronts' propagation problems in electromagnetic solids, (d) to apply the developments to model qualitatively the twin boundary kinetics in magnetic shape-memory alloys (MSMAs).

\begin{figure}
  \begin{center}
    \includegraphics{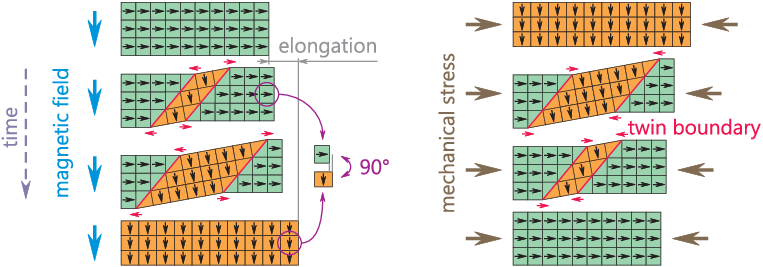}
  \end{center}
  \caption{A schematic representation of coupled magneto-mechanics of MSMAs. Black arrows represent the orientation of magnetisation coinciding with the magnetic anisotropy axis.}
  \label{fig:msmas}
\end{figure}

The major physical mechanism of MSMAs' deformation is the propagation of the boundaries between structural domains \cite{Planes2009}, so-called twin-related martensitic variants, as schematically illustrated in figure \ref{fig:msmas}. These materials are characterised by high magnetic anisotropy, along which the spin magnetic moments of atoms tend to align. Upon application of the external magnetic field, there is a reorientation of the structural domains by movement of the twin boundaries, such that the magnetic anisotropy axes of the domains align with the external magnetic field. The application of the mechanical stress, in turn, can also lead to the twin boundary motion. Therefore, the mechanical and the magnetic processes are coupled leading to the stress-affected and magnetic-field-affected twin boundary kinetics. For MSMAs, it has been modelled using various approaches: the phase-field models, e.g.\ \cite{Mennerich2011,Gebbia2018}, the sharp-interface models with the kinetics fitted to high-precision experiments, e.g.\ \cite{Faran2011,Faran2013}, and the sharp-interface models with the driving force, acting on the twin boundary, derived by minimisation of the total energy functional, e.g.\ \cite{Wang2014,Wang2016,Wang2016a}. 

Resolving kinetics of phase boundaries in materials is the classical problem of computational physics, in which partial differential equations (PDEs) with non-stationary interfaces must be solved. The class of fictitious-domain methods (also called non-conforming mesh methods or unfitted methods) became a popular approach for such problems. In this setting, the computational mesh is independent of the geometry, the interfaces can arbitrarily cut through the mesh and, in the case of non-stationary problems, can move independently of the mesh. One such method is the cut-finite-element method or CutFEM \cite{Burman2015}. It is important to mention that the method originates from \cite{Hansbo2002}, where it has been proposed to use the Nitsche's method \cite{Nitsche1971} to enforce the boundary/interface conditions, and from \cite{Burman2012}, where the numerical stabilisation has been proposed. There is a vast number of papers on CutFEM, advancing the method in numerous directions and adapting it to different PDEs; therefore, the reader is referred to a brilliant and exhaustive recent review paper \cite{Burman2025}. It is important to mention prior work on adaptation of CutFEM for linear \cite{Hansbo2017,Burman2018,Burman2019} and non-linear \cite{Poluektov2019,Poluektov2022} elasticity problems.

\section{Theory}
\label{sec:theo}

The standard continuum mechanics approach is employed, where the current and the reference configurations are considered. The position of a material point in the current configuration is denoted as $\vectorn{x} = \vectorn{x}\big(\vectorn{X},t\big)$, where $\vectorn{X}$ is the position in the reference configuration and $t$ is the time; $\vectorn{u} = \vectorn{x} - \vectorn{X}$ is the displacement field; $\vectorn{v} = \dot{\vectorn{u}}$ is the velocity field, where dot denotes the material time derivative\footnote{For arbitrary scalar $a = a\big(\vectorn{x},t\big)$, it is of course related to the spatial time derivative: $\dot{a} = \tfrac{\dif a}{\dif t} = \tfrac{\partial a}{\partial t} + \vectorn{v} \cdot \nabla a$.}. Nabla operators $\nabla$ and $\nabla_0$ are defined with respect to the current and the reference configurations, respectively.

Path-connected arbitrary domain $\varOmega$ in the reference configuration is considered. It maps onto domain $\omega$ in the current configuration. The boundaries of domains $\varOmega$ and $\omega$ are denoted as $\varGamma = \partial \varOmega$ and $\gamma = \partial \omega$, respectively.

To introduce the Maxwell's equations, path-connected arbitrary surface $\varTheta$ in the reference configuration is considered. It maps onto surface $\theta$ in the current configuration. The boundary curves of surfaces $\varTheta$ and $\theta$ are denoted as $\varPhi = \partial \varTheta$ and $\phi = \partial \theta$, respectively.

Two continua are considered --- the lattice continuum and the spin continuum. The former is the standard continuum describing the movement of material points in space. The latter describes the distribution of the magnetisation throughout the body. The continua are coupled --- each material point of the body is associated with a time-dependent magnetisation vector uniquely. Thus, the spin continuum expands and contracts with the lattice continuum and occupies the same volume. The magnetisation field per unit of mass is denoted as $\vectorn{m}$. It can be written as dependent on the position of a material point in the current configuration or in the reference configuration: $\vectorn{m} = \vectorn{m}\big(\vectorn{x},t\big) = \vectorn{m}\big(\vectorn{X},t\big)$. 

It is assumed that the magnetisation is saturated and has time-independent magnitude: $\vectorn{m}\cdot\vectorn{m} = m_\mathrm{S}^2$, where $m_\mathrm{S}$ is constant. This assumption has deep implications for further formalisation of the theory. In particular, differentiating this expression by time results in $\dot{\vectorn{m}}\cdot\vectorn{m} = 0$, which means that $\dot{\vectorn{m}}$ and $\vectorn{m}$ are perpendicular, leading to $\dot{\vectorn{m}}$ necessarily taking the following form:
\begin{equation}
  \dot{\vectorn{m}} = \vectorn{m} \times \vectorn{a} ,
  \label{eq:mdotgen}
\end{equation}
where $\vectorn{a}$ is some vector.

The magnetisation at the material point is directly related to the angular momentum at the material point via the gyromagnetic relationship:
\begin{equation}
  \vectorn{l} = \gamma_0^{-1} \vectorn{m} ,
\end{equation}
where $\gamma_0$ is the gyromagnetic ratio\footnote{It has units of ($\mathrm{C}\cdot\mathrm{kg}^{-1}$) and links the magnetic moment of a unit of mass, measured in ($\mathrm{J}\cdot\mathrm{T}^{-1}\cdot\mathrm{kg}^{-1}$), and the angular momentum of a unit of mass, measured in ($\mathrm{m}^2\cdot\mathrm{s}^{-1}$).} and $\vectorn{l}$ is the angular momentum per unit of mass. Thus, each material point of the spin-lattice coupled continuum intrinsically possesses the angular momentum.

Before proceeding, it is useful to reiterate two transport theorems commonly encountered in continuum mechanics \cite{Maugin1988}. The time derivative of an integral of scalar $g$ over material volume $\omega$ is given by
\begin{equation}
  \frac{\dif}{\dif t} \int_\omega g \dif \omega = \int_\omega \left( \frac{\partial g}{\partial t} + \nabla \cdot \left( \vectorn{v} g \right) \right) \dif \omega .
  \label{eq:transpVol}
\end{equation}
If the second term in brackets in the right-hand side is transformed via the Gauss theorem to the surface integral, then the Reynolds transport theorem is obtained. The time derivative of an integral of the normal component of vector $\vectorn{g}$ over material surface $\theta$ is given by
\begin{equation}
  \frac{\dif}{\dif t}\int_\theta \vectorn{g}\cdot\vectorn{n} \dif \theta = \int_\theta \left( \frac{\partial\vectorn{g}}{\partial t} + \nabla\times\left(\vectorn{g}\times\vectorn{v}\right) + \left(\nabla\cdot\vectorn{g}\right) \vectorn{v}\right) \cdot\vectorn{n} \dif \theta ,
  \label{eq:transpSurf}
\end{equation}
where the expression in brackets in the right-hand side is called the \emph{convective time derivative}. It is not difficult to prove relations \eqref{eq:transpVol} and \eqref{eq:transpSurf} by mapping the left-hand-side integral to the reference configuration, moving the time derivative inside, and mapping back to the current configuration.

\subsection{Time-dependent interface}

\subsubsection{Interface cutting through a volume}

An interface that sweeps through the material points of the body is considered. Domain $\varOmega$ cuts out a part of the interface that is denoted as $\varSigma = \varSigma\left(t\right)$ in the reference configuration and maps onto $\sigma = \sigma\left(t\right)$ in the current configuration. This surface splits domain $\varOmega$ into $\varOmega_+$ and $\varOmega_-$, the external boundaries of which are denoted as $\varGamma_+ = \partial\varOmega_+$ and $\varGamma_- = \partial\varOmega_-$, respectively; the intersection of these boundaries is the considered interface surface: $\varGamma_+ \cap \varGamma_- = \varSigma$. External normals to $\varGamma_+$ and $\varGamma_-$ are denoted as $\vectorn{N}_+$ and $\vectorn{N}_-$, respectively; external normal to $\varGamma$ is denoted as $\vectorn{N}_\varGamma$. At the interface, the normal is denoted as
\begin{equation}
  \vectorn{N}_* = \vectorn{N}_+ = -\vectorn{N}_- , \quad\quad \text{on } \varSigma.
\end{equation}

The balance laws below have boundary terms with a tensor/vector field multiplied by the normal. As usual, these boundary terms are transformed into the bulk terms using the Gauss theorem. It should be emphasised that the theorem requires the tensor/vector fields to be continuously differentiable. For the considered continua, this is true within domains $\varOmega_+$ and $\varOmega_-$ only --- these tensor/vector fields can be discontinuous across the interface. Therefore, the standard transformation employing the Gauss theorem within separate phases can be performed:
\begin{equation}
  \int_\varGamma \tensor{A}\cdot\vectorn{N}_\varGamma \dif \varGamma =
  \int_{\varOmega_+} \nabla_0\cdot\tensor{A}_+\transp \dif \varOmega_+ + \int_{\varOmega_-} \nabla_0\cdot\tensor{A}_-\transp \dif \varOmega_- - \int_\varSigma \left\llbracket\tensor{A}\right\rrbracket\cdot\vectorn{N}_* \dif \varSigma , 
  \label{eq:genBalG}
\end{equation}
where $\tensor{A}$ is some second-order tensor that is discontinuous across $\varSigma$. Restrictions of $\tensor{A}$ to domains $\varOmega_+$ and $\varOmega_-$ are denoted as $\tensor{A}_+$ and $\tensor{A}_-$, respectively. The double square brackets and the angular brackets denote the jump and the average of the quantity, respectively: $\left\llbracket\tensor{A}\right\rrbracket = \tensor{A}_+ - \tensor{A}_-$ and $\left\langle\tensor{A}\right\rangle = \tfrac{1}{2}\left(\tensor{A}_+ + \tensor{A}_-\right)$. 

The balance laws below also have terms with a time derivative of an integral over $\varOmega$ of a vector/scalar field. Again, these fields are continuously differentiable within the phases, but discontinuous across the interface. Therefore, the standard transformation employing the Reynolds transport theorem within separate phases can be performed:
\begin{equation}
  \frac{\dif}{\dif t} \int_\varOmega \vectorn{a} \dif \varOmega = 
  \int_{\varOmega_+} \dot{\vectorn{a}}_+ \dif \varOmega_+ + \int_{\varOmega_-} \dot{\vectorn{a}}_- \dif \varOmega_- + \int_\varSigma \left\llbracket\vectorn{a}\right\rrbracket \vectorn{W}\cdot\vectorn{N}_* \dif \varSigma ,
  \label{eq:genBalR}
\end{equation}
where $\vectorn{W}$ is the velocity of the interface in the reference configuration, $\vectorn{a}$ is some vector that is discontinuous across $\varSigma$. Restrictions of $\vectorn{a}$ to domains $\varOmega_+$ and $\varOmega_-$ are denoted as $\vectorn{a}_+$ and $\vectorn{a}_-$, respectively. To obtain relation \eqref{eq:genBalR}, the reference configuration of the body (with domains $\varOmega_\pm$ that have evolving boundaries) is considered as the current configuration of another body (with preimages of $\varOmega_\pm$ that are fixed), and transport relation \eqref{eq:transpVol} is applied together with the Gauss theorem for the second term in brackets.

As usual in phase transition problems, without loss of generality, the interface velocity is taken to be aligned with the interface normal \cite{Truesdell1960}. It is convenient to denote the normal component of the interface velocity as $W_* = \vectorn{W}\cdot\vectorn{N}_*$. Thus, $\vectorn{W} = W_* \vectorn{N}_*$.

For any scalar $g$ and any vector $\vectorn{g}$ that are continuous across the interface, the kinematic compatibility conditions are written as \cite{Truesdell1960}:
\begin{align}
  &\llbracket\nabla_0 g \rrbracket = \vectorn{N}_*\xi , \label{eq:genGradForm} \\
  &\llbracket\dot{\vectorn{g}}\rrbracket = -W_* \vectorn{N}_* \cdot \llbracket\nabla_0\vectorn{g}\rrbracket ,
  \label{eq:genCompat}
\end{align}
where $\xi$ can be called an amplitude of the jump of the gradient of $g$. The first condition shows that the jump of the gradient of a continuous quantity is aligned with the interface normal.

Finally, it is useful to note that for arbitrary $\vectorn{a}$ and $\tensor{A}$ that are discontinuous across the interface, the following equality can be trivially proven:
\begin{equation}
  \llbracket\vectorn{a}\cdot\tensor{A}\rrbracket = \langle\vectorn{a}\rangle\cdot\llbracket\tensor{A}\rrbracket + \llbracket\vectorn{a}\rrbracket\cdot\langle\tensor{A}\rangle .
  \label{eq:intIdent}
\end{equation}

\subsubsection{Interface cutting through a surface}

Surface $\varTheta$ with normal $\vectorn{N}_\varTheta$ intersects the interface along a curve that is denoted as $\varPsi = \varPsi\left(t\right)$ in the reference configuration. This interface curve splits surface $\varTheta$ into $\varTheta_+$ and $\varTheta_-$, the boundary curves of which are denoted as $\varPhi_+ = \partial\varTheta_+$ and $\varPhi_- = \partial\varTheta_-$, respectively; the intersection of these boundaries is the considered interface curve: $\varPhi_+ \cap \varPhi_- = \varPsi$. Tangent vectors along $\varPhi_+$ and $\varPhi_-$ are denoted as $\vectorn{T}_+$ and $\vectorn{T}_-$, respectively; tangent vector along $\varPhi$ is denoted as $\vectorn{T}_\varPhi$. The orientation of the tangent vectors is consistent with the orientation of normal $\vectorn{N}_\varTheta$ according to the right-hand rule. At the interface curve, the tangent is denoted as
\begin{equation}
  \vectorn{T}_* = \vectorn{T}_+ = -\vectorn{T}_- , \quad\quad \text{on } \varPsi.
\end{equation}

As above, for a discontinuous quantity, the standard transformation employing the Stokes theorem within separate phases can be performed:
\begin{equation}
  \int_\varPhi \vectorn{a}\cdot\vectorn{T}_\varPhi \dif \varPhi =
  \int_{\varTheta_+} \big(\nabla_0\times\vectorn{a}_+\big)\cdot\vectorn{N}_\varTheta \dif \varTheta_+ + \int_{\varTheta_-}  \big( \nabla_0 \times\vectorn{a}_-\big)\cdot\vectorn{N}_\varTheta \dif \varTheta_- - \int_\varPsi \llbracket\vectorn{a}\rrbracket\cdot\vectorn{T}_* \dif \varPsi , 
  \label{eq:genBalS}
\end{equation}
where $\vectorn{a}$ is some vector that is discontinuous across $\varPsi$. 

In the case when a surface integral with a discontinuity curve is differentiated by time, it can be shown that
\begin{equation}
  \frac{\dif}{\dif t} \int_\varTheta \vectorn{a}\cdot\vectorn{N}_\varTheta \dif \varTheta =
  \int_{\varTheta_+} \dot{\vectorn{a}}_+ \cdot\vectorn{N}_\varTheta \dif \varTheta_+ + \int_{\varTheta_-} \dot{\vectorn{a}}_- \cdot\vectorn{N}_\varTheta \dif \varTheta_- + \int_\varPsi \big(\llbracket\vectorn{a}\rrbracket\times\vectorn{W}\big)\cdot\vectorn{T}_* \dif \varPsi , 
  \label{eq:genBalX}
\end{equation}
where $\vectorn{a}$ is some vector that is discontinuous across $\varPsi$ and is also divergence-free, $\nabla_0\cdot\vectorn{a}_\pm = 0$. Similarly to the way how equation \eqref{eq:genBalR} is obtained, to get equation \eqref{eq:genBalX}, the reference configuration of the body (with surfaces $\varTheta_\pm$ that have evolving boundary curves) is considered as the current configuration of another body (with preimages of $\varTheta_\pm$ that are fixed), and transport relation \eqref{eq:transpSurf} is applied together with the Stokes theorem for the second term in brackets. The third term in brackets vanishes due to the divergence-free requirement on $\vectorn{a}$. It can be remarked that without the latter, the mean curvature of $\varTheta$ emerges in the equations \cite{Maugin1988}, but consideration of such case in not needed for the purposes of the present paper.

\subsection{Balance laws in the current configuration}

The balance laws outlined below follow \cite{Maugin1988}. The change of the total linear momentum of the solid confined to domain $\omega$ is due to the tractions at the boundary, the volumetric forces, and the surface forces acting at the interface:
\begin{equation}
  \frac{\dif}{\dif t} \int_\omega \rho \vectorn{v} \dif \omega = \int_\gamma \vectorn{t} \dif \gamma + \int_\omega \rho \vectorn{b} \dif \omega + \int_{\omega\backslash\sigma} \rho \vectorn{b}_\mathrm{E} \dif \omega + \int_\sigma \vectorn{t}_\mathrm{E} \dif \sigma,
  \label{eq:momentLInt}
\end{equation}
where $\rho$ is the mass density of the solid per unit of volume in the current configuration, $\vectorn{t}$ is the mechanical traction, $\vectorn{b}$ is the mechanical body force per unit of mass (e.g.\ gravity), $\vectorn{b}_\mathrm{E}$ is the electromagnetic body force per unit of mass, and $\vectorn{t}_\mathrm{E}$ is the surface force concentrated at the interface. The volume integral of $\rho\vectorn{b}_\mathrm{E}$ explicitly excludes surface $\sigma$, which is discussed below. 

The change of the total angular momentum of the lattice continuum is due to the tractions, the volumetric forces, the interaction with the spin continuum, and the forces at the interface:
\begin{align}
  &\frac{\dif}{\dif t} \int_\omega \vectorn{x} \times \rho \vectorn{v} \dif \omega = \int_\gamma \vectorn{x} \times \vectorn{t} \dif \gamma + 
  \int_\omega \big( \vectorn{x} \times \rho \vectorn{b} + \rho \vectorn{c}_\mathrm{SL} \big) \dif \omega + 
  \vphantom{a} \nonumber \\ &\vphantom{a}\hspace{5cm} +
  \int_{\omega\backslash\sigma} \vectorn{x} \times \rho \vectorn{b}_\mathrm{E} \dif \omega +
  \int_\sigma \vectorn{x} \times \vectorn{t}_\mathrm{E} \dif \sigma,
  \label{eq:momentALInt}
\end{align}
where $\vectorn{c}_\mathrm{SL}$ is the couple per unit of mass due to the spin continuum acting on the lattice continuum.

The change of the total angular momentum of the spin continuum is due to the contact forces at the boundary, the volumetric couples, and the interaction with the lattice continuum:
\begin{equation}
  \frac{\dif}{\dif t} \int_\omega \rho \gamma_0^{-1} \vectorn{m} \dif \omega = \int_\gamma \vectorn{m} \times \vectorn{q} \dif \gamma + \int_\omega \big( \rho \vectorn{c}_\mathrm{E} + \rho \vectorn{c}_\mathrm{LS} \big) \dif \omega ,
  \label{eq:momentASInt}
\end{equation}
where $\vectorn{q}$ is the surface exchange contact force\footnote{Vector $\vectorn{q}$ should be understood as some local quantity analogous to a magnetic induction field. It acts on magnetic moment $\vectorn{m}$ of a unit of mass. Due to this analogy, it produces local torque $\vectorn{\tau} = \vectorn{m}\times\vectorn{q}$ (per unit of surface).}, $\vectorn{c}_\mathrm{E}$ is the electromagnetic couple per unit of mass, $\vectorn{c}_\mathrm{LS}$ is the couple per unit of mass due to the lattice continuum acting on the spin continuum. Without loss of generality\footnote{As will be seen later from the transformation to the local form of the balance laws, this can be done because the time-evolution of $\vectorn{m}$ must have the form given by equation \eqref{eq:mdotgen}.}, it can be assumed that the couples that link both continua result from the existence of local magnetic induction field $\vectorn{B}_\mathrm{L}$:
\begin{equation}
  \vectorn{c}_\mathrm{LS} = -\vectorn{c}_\mathrm{SL} = \vectorn{m} \times \vectorn{B}_\mathrm{L} .
\end{equation}

The change of the total energy of the spin-lattice coupled continuum is due to the work of the forces\footnote{Due the analogy between $\vectorn{q}$ and a magnetic induction field, the power generated by the surface exchange contact force is written as the power of a magnetic induction field acting on a magnetic moment.}, the heat outflow and generation, the electromagnetic work on the coupled continuum in the bulk and at the interface:
\begin{align}
  &\frac{\dif}{\dif t} \int_\omega \big( \rho \epsilon + \tfrac{1}{2} \rho \vectorn{v}\cdot\vectorn{v} \big) \dif \omega = \int_\gamma \big( \vectorn{v} \cdot \vectorn{t} + \dot{\vectorn{m}} \cdot \vectorn{q} - g \big) \dif \gamma + \int_\omega \big( \rho \vectorn{v} \cdot \vectorn{b} + \rho r \big) \dif \omega + 
  \vphantom{a} \nonumber \\ &\vphantom{a}\hspace{8cm} +
  \int_{\omega\backslash\sigma} \rho p_\mathrm{E} \dif \omega + \int_\sigma f_\mathrm{E} \dif \sigma,
  \label{eq:energInt}
\end{align}
where $\epsilon$ is the internal energy per unit of mass, $g$ is the heat outflow through the boundary, $r$ is the internal heat source per unit of mass, $p_\mathrm{E}$ is the electromagnetic power per unit of mass, and $f_\mathrm{E}$ is the power generated at the interface. Inertia of the magnetisation change does not enter the energy balance due to the gyroscopic nature of equation \eqref{eq:mdotgen}.  

The entropy production should be non-negative, which results in the following standard inequality:
\begin{equation}
  \frac{\dif}{\dif t} \int_\omega \rho s \dif \omega \geq - \int_\gamma \frac{g}{T} \dif \gamma + \int_\omega \frac{\rho r}{T} \dif \omega ,
  \label{eq:entrInt}
\end{equation} 
where $s$ is the entropy per unit of mass, $T$ is the temperature. 

The electromagnetic quantities are governed by the Maxwell's equations. Because the present paper targets the behaviour of the solids, it is reasonable to consider the quasistatic form of the equations (due to the difference between the time scales of the electromagnetic dynamics and the dynamics of solids); however, this cannot be done right from the start, since important transformations of the volumetric electromagnetic quantities require the complete equations. Nevertheless, it can be assumed that the electric polarisation, the electric charge density, and the current density are absent right from the start, which should still cover a large class of deformable ferromagnets. Thus, the Maxwell's equations take the following form:
\begin{align}
  &\int_\gamma \vectorn{B}\cdot\vectorn{n}_\gamma \dif \gamma = 0 , \quad\quad\quad
  \int_\phi \big( \vectorn{H} - \vectorn{v}\times\vectorn{D} \big) \cdot\vectorn{t}_\phi \dif \phi = \frac{\dif}{\dif t}\int_\theta \vectorn{D}\cdot\vectorn{n}_\theta \dif \theta ,
  \label{eq:MaxwBcur} \\
  &\int_\gamma \vectorn{D}\cdot\vectorn{n}_\gamma \dif \gamma = 0 , \quad\quad\quad
  \int_\phi \big( \vectorn{E} + \vectorn{v}\times\vectorn{B} \big) \cdot\vectorn{t}_\phi \dif \phi = -\frac{\dif}{\dif t}\int_\theta \vectorn{B}\cdot\vectorn{n}_\theta \dif \theta , \label{eq:MaxwHcur}
\end{align}
where vectors $\vectorn{n}_\gamma$ and $\vectorn{n}_\theta$ are the external normals to surfaces $\gamma$ and $\theta$, respectively, vector $\vectorn{t}_\phi$ is the tangent vector to boundary curve $\phi$, vectors $\vectorn{B}$, $\vectorn{H}$, $\vectorn{D}$, $\vectorn{E}$ are the magnetic flux density\footnote{Vector $\vectorn{B}$ is also commonly named as ``magnetic induction'' or ``magnetic induction field''.}, the magnetic field, the electric flux density, and the electric field, respectively. By definition, fields $\vectorn{H}$ and $\vectorn{D}$ are (in the absence of the polarisation)
\begin{equation}
  \vectorn{H} = \frac{1}{\mu_0} \vectorn{B} - \rho \vectorn{m} , \quad\quad\quad
  \vectorn{D} = \varepsilon_0 \vectorn{E} ,
  \label{eq:BHlink}
\end{equation}
where $\mu_0$ is the vacuum magnetic permeability and $\varepsilon_0$ is the vacuum permittivity.

In equations \eqref{eq:MaxwBcur} and \eqref{eq:MaxwHcur}, the balance laws for $\vectorn{H}$ and $\vectorn{E}$ may look non-conventional due to the presence of the terms with the cross product with $\vectorn{v}$. However, here, \emph{material} surfaces and curves are considered, as traditionally done in continuum mechanics. These surfaces and curves are `glued' to the material points of the deforming continuum and move together with these points. In classical electromagnetism, the balance laws are written using \emph{spatial} surfaces and curves, i.e.\ that are not moving. It can be easily shown that equations \eqref{eq:MaxwBcur} and \eqref{eq:MaxwHcur} are equivalent to the classical formulation (the standard local form of the Maxwell's equations) by applying transport relation \eqref{eq:transpSurf} to the right-hand sides of the equations for $\vectorn{H}$ and $\vectorn{E}$, together with the Stokes theorem, and applying the Gauss theorem to the equations for $\vectorn{B}$ and $\vectorn{D}$.

The volumetric electromagnetic quantities in the balance laws above ($\vectorn{b}_\mathrm{E}$, $\vectorn{c}_\mathrm{E}$, $p_\mathrm{E}$) are derived from the Lorentz force acting on a collection of charges, neglecting quadrupole (and higher-order) moments and with subsequent averaging \cite{Maugin1988}. In absence of the electric polarisation, the electric charge density, and the current density, these quantities reduce to the following:
\begin{equation}
  \vectorn{b}_\mathrm{E} = \big( \nabla \vectorn{B} \big) \cdot \vectorn{m} , \quad\quad\quad\quad
  \vectorn{c}_\mathrm{E} = \vectorn{m} \times \vectorn{B} , \quad\quad\quad\quad
  p_\mathrm{E} = \vectorn{v}\cdot\big( \nabla \vectorn{B} \big) \cdot \vectorn{m} -\vectorn{m} \cdot \dot{\vectorn{B}} .
  \label{eq:EMterms}
\end{equation}
The reader can find the full expressions and derivations in \cite{Eringen1990} and references therein.

The full set of the Maxwell's equations allows rewriting the expressions for the electromagnetic body force per unit of volume and the electromagnetic power per unit of volume. The former leads to the emergence of the well-known Maxwell stress tensor, denoted here as $\tensor{\tau}_\mathrm{E}$. This step is repeated many times in literature, e.g.\ in \cite{Eringen1990}; therefore, only the final result is provided here (as above, omitting the electric polarisation, the electric charge density, and the current density):
\begin{align}
  &\rho\vectorn{b}_\mathrm{E} = \nabla \cdot \tensor{\tau}_\mathrm{E} - \rho \frac{\dif}{\dif t} \big( \rho^{-1} \varepsilon_0 \vectorn{E}\times\vectorn{B} \big) , \label{eq:volForceVMaxwell} \\
  &\tensor{\tau}_\mathrm{E} = \frac{1}{\mu_0} \vectorn{B}\vectorn{B} + \rho\vectorn{m}\cdot\vectorn{B} \tensor{I} - \vectorn{B}\rho\vectorn{m} + \varepsilon_0 \vectorn{E}\vectorn{E} - \zeta_\mathrm{E} \tensor{I} + \varepsilon_0 \vectorn{v} \vectorn{E}\times\vectorn{B} , \label{eq:volForceTau} \\
  &\rho p_\mathrm{E} = \nabla \cdot \big( \vectorn{H}\times\vectorn{E} + \zeta_\mathrm{E} \vectorn{v} \big) - \rho \frac{\dif}{\dif t} \big( \rho^{-1} \zeta_\mathrm{E} \big) , \label{eq:volEnergVMaxwell} \\
  &\zeta_\mathrm{E} = \frac{1}{2\mu_0}\vectorn{B}\cdot\vectorn{B} + \frac{\varepsilon_0}{2} \vectorn{E}\cdot\vectorn{E} ,
\end{align}
where $\tensor{I}$ is the second-order identity tensor. It must also be emphasised that expressions \eqref{eq:volForceVMaxwell} and \eqref{eq:volEnergVMaxwell} are valid only within the individual phases ($\omega_+$ and $\omega_-$ in the current configuration) because their derivation involves application of the Gauss and the Stokes theorems to the Maxwell's equations in the integral form, requiring fields to be continuously differentiable\footnote{This can always be done because domain $\varOmega$ is arbitrary and can be chosen to exclude the interface.}. Both forms of $\vectorn{b}_\mathrm{E}$ and $p_\mathrm{E}$ will be used below.

A counter-intuitive element of the magneto-mechanical coupled theory is the emergence of the surface force of an electromagnetic nature at the interface, denoted above as $\vectorn{t}_\mathrm{E}$. It is the consequence of the structure of the electromagnetic body force in the bulk as given by equation \eqref{eq:EMterms}, which is the primary quantity derived from the Lorentz force acting on a collection of charges\footnote{Not to be confused with the secondary form given by equation \eqref{eq:volForceVMaxwell} that is derived from \eqref{eq:EMterms}.}. Body force $\vectorn{b}_\mathrm{E}$ contains field $\vectorn{B}$, which can be discontinuous across $\sigma$. At first glance, one may say that $\vectorn{b}_\mathrm{E}$ just appears in the individual phases and the presence of $\sigma$ does not change anything, which will be incorrect. If one imagines that the interface is the limit case of a thin finite-width slice of the volume with $\vectorn{B}$ continuously differentiable everywhere, then the gradient of $\vectorn{B}$ is `large' within the slice and increases to infinity with the thickness of the slice going to zero. Hence, the zero-thickness limit of the volumetric integral of $\vectorn{b}_\mathrm{E}$ over the slice can be finite. 

An even more counter-intuitive situation is with the presence of the time derivative of $\vectorn{B}$ in the expression for the electromagnetic power, denoted above as $p_\mathrm{E}$. When $\vectorn{B}$ is discontinuous across $\sigma$, the movement of $\sigma$ means that $\vectorn{B}$ jumps `instantaneously' at points belonging to some surface, which implies that the time derivative is infinite. When integrated over the volume, this case becomes reminiscent of an integral of a delta-function. Hence, the presence of the time derivative of $\vectorn{B}$ in $p_\mathrm{E}$ leads to concentrated power $f_\mathrm{E}$ on $\sigma$.

Due to discontinuity of $\vectorn{B}$ across $\sigma$, formally, $\vectorn{b}_\mathrm{E}$ and $p_\mathrm{E}$ are undefined on $\sigma$; therefore, $\sigma$ is explicitly excluded from the volume integration of these terms, while surface integrals of concentrated force $\vectorn{t}_\mathrm{E}$ and concentrated power $f_\mathrm{E}$ are added to the balance laws. These quantities are obtained from equations \eqref{eq:volForceVMaxwell} and \eqref{eq:volEnergVMaxwell}, respectively, by considering a limit of a thin finite-width slice of the volume with $\vectorn{B}$ continuously differentiable everywhere and are written directly in the reference configuration below.

\subsection{Balance laws in the reference configuration}

To write the balance laws in the reference configuration, the standard transformations are employed:
\begin{align}
  &\rho \dif \omega = \rho_0 \dif \varOmega , \\
  &\vectorn{t} \dif \gamma = \vectorn{T} \dif \varGamma = \tensor{P}\cdot\vectorn{N}_\varGamma \dif \varGamma , \\
  &g \dif \gamma = G \dif \varGamma = \vectorn{G}\cdot\vectorn{N}_\varGamma \dif \varGamma ,
\end{align}
where $\rho_0$ is the mass density of the solid per unit of volume in the reference configuration, $\vectorn{T}$ is the mechanical traction per unit of surface in the reference configuration, $\tensor{P}$ the first Piola-Kirchhoff stress tensor, $G$ is the heat outflow through the boundary in the reference configuration, $\vectorn{G}$ is the heat flux vector. Analogously to the mechanical tractions, the surface exchange contact force is also transformed into the reference configuration and represented by some second-order tensor multiplied by the normal:
\begin{equation}
  \vectorn{q} \dif \gamma = \vectorn{Q} \dif \varGamma = \tensor{B}\cdot\vectorn{N}_\varGamma \dif \varGamma , 
\end{equation}
where $\tensor{B}$ can be called the exchange-force tensor. Substituting all these expressions gives the following form of the balance laws:
\begin{align}
  &\frac{\dif}{\dif t} \int_\varOmega \rho_0 \vectorn{v} \dif \varOmega = \int_\varGamma \tensor{P}\cdot\vectorn{N}_\varGamma \dif \varGamma + \int_\varOmega \rho_0 \vectorn{b} \dif \varOmega + \int_{\varOmega\backslash\varSigma} \rho_0 \vectorn{b}_\mathrm{E} \dif \varOmega + \int_\varSigma \vectorn{t}_{\mathrm{E}0} \dif \varSigma , \label{eq:momentLRef} \\
  &\frac{\dif}{\dif t} \int_\varOmega \vectorn{x} \times \rho_0 \vectorn{v} \dif \varOmega = \int_\varGamma \vectorn{x} \times \tensor{P}\cdot\vectorn{N}_\varGamma \dif \varGamma + \int_\varOmega \big( \vectorn{x} \times \rho_0 \vectorn{b} + \rho_0 \vectorn{c}_\mathrm{SL} \big) \dif \varOmega + \vphantom{a} \nonumber \\
  &\vphantom{a}\hspace{4.5cm} + \int_{\varOmega\backslash\varSigma} \big( \vectorn{x} \times \rho_0 \vectorn{b}_\mathrm{E} \big) \dif \varOmega + \int_\varSigma \vectorn{x} \times \vectorn{t}_{\mathrm{E}0} \dif \varSigma, \\
  &\frac{\dif}{\dif t} \int_\varOmega \rho_0 \gamma_0^{-1} \vectorn{m} \dif \varOmega = \int_\varGamma \vectorn{m} \times \tensor{B}\cdot\vectorn{N}_\varGamma \dif \varGamma + \int_\varOmega \big( \rho_0 \vectorn{c}_\mathrm{E} + \rho_0 \vectorn{c}_\mathrm{LS} \big) \dif \varOmega , \label{eq:momentASRef} \\
  &\frac{\dif}{\dif t} \int_\varOmega \big( \rho_0 \epsilon + \tfrac{1}{2} \rho_0 \vectorn{v}\cdot\vectorn{v} \big) \dif \varOmega = \int_\varGamma \big( \vectorn{v} \cdot \tensor{P} + \dot{\vectorn{m}} \cdot \tensor{B} - \vectorn{G} \big)\cdot\vectorn{N}_\varGamma \dif \varGamma + \vphantom{a} \nonumber \\
  &\vphantom{a}\hspace{3cm} + \int_\varOmega \big( \rho_0 \vectorn{v} \cdot \vectorn{b} + \rho_0 r \big) \dif \varOmega + \int_{\varOmega\backslash\varSigma} \rho_0 p_\mathrm{E} \dif \varOmega + \int_\varSigma f_{\mathrm{E}0} \dif \varSigma, \label{eq:energRef} \\
  &\frac{\dif}{\dif t} \int_\varOmega \rho_0 s \dif \varOmega \geq - \int_\varGamma \frac{1}{T}\vectorn{G}\cdot\vectorn{N}_\varGamma \dif \varGamma + \int_\varOmega \frac{\rho_0 r}{T} \dif \varOmega , \label{eq:entrRef} 
\end{align}
where $\vectorn{t}_{\mathrm{E}0}$ and $f_{\mathrm{E}0}$ are the concentrated force and the concentrated power in the reference configuration, respectively.

To transform the Maxwell's equations to the reference configuration, it is useful to remember the Nanson's formula:
\begin{equation}
  \vectorn{n}_\gamma \dif \gamma = J \tensor{F}\transpm \cdot \vectorn{N}_\varGamma \dif \varGamma ,
  \label{eq:Nanson}
\end{equation}
where $\tensor{F} = \left(\nabla_0\vectorn{x}\right)\transp$ is the deformation gradient and $J = \determ{\tensor{F}}$ is the volume change, $\dif \omega = J \dif \varOmega$. When surface $\theta$ is mapped onto the reference configuration, tangent vector $\vectorn{t}_\phi$ of boundary curve $\phi$ transforms trivially as 
\begin{equation}
  \vectorn{t}_\phi \dif \phi = \tensor{F} \cdot \vectorn{T}_\varPhi \dif \varPhi .
\end{equation}
This gives 
\begin{align}
  &\int_\varGamma \vectorn{B}_0 \cdot\vectorn{N}_\varGamma \dif \varGamma = 0 , \quad\quad\quad
  \int_\varPhi \vectorn{H}_0 \cdot\vectorn{T}_\varPhi \dif \varPhi = \frac{\dif}{\dif t} \int_\varTheta \vectorn{D}_0 \cdot\vectorn{N}_\varTheta \dif \varTheta ,
  \label{eq:MaxwBRef} \\
  &\int_\varGamma \vectorn{D}_0 \cdot\vectorn{N}_\varGamma \dif \varGamma = 0 , \quad\quad\quad
  \int_\varPhi \vectorn{E}_0 \cdot\vectorn{T}_\varPhi \dif \varPhi = -\frac{\dif}{\dif t}\int_\varTheta \vectorn{B}_0 \cdot\vectorn{N}_\varTheta \dif \varTheta ,
  \label{eq:MaxwHRef}
\end{align}
where the electromagnetic quantities in the reference configuration are introduced as
\begin{align}
  &\vectorn{B}_0 = J \tensor{F}^{-1} \cdot \vectorn{B} , \quad\quad\quad  
  \vectorn{H}_0 = \tensor{F}\transp \cdot \big( \vectorn{H} - \vectorn{v}\times\vectorn{D}\big) , \label{eq:Bref} \\
  &\vectorn{D}_0 = J \tensor{F}^{-1} \cdot \vectorn{D} , \quad\quad\quad  
  \vectorn{E}_0 = \tensor{F}\transp \cdot \big( \vectorn{E} + \vectorn{v}\times\vectorn{B}\big). \label{eq:Href} 
\end{align}

Volumetric body force $\vectorn{b}_\mathrm{E}$ and power $p_\mathrm{E}$ transform as
\begin{align}
  &\rho_0 \vectorn{b}_\mathrm{E} = \nabla_0 \cdot \tensor{\tau}_{\mathrm{E}0} - \rho_0 \frac{\dif}{\dif t} \big( \rho^{-1} \varepsilon_0 \vectorn{E}\times\vectorn{B} \big) ,  \label{eq:volForceVMaxwellRef} \\
  &\rho_0 p_\mathrm{E} = \nabla_0 \cdot \big( J \tensor{F}^{-1} \cdot \big( \vectorn{H}\times\vectorn{E} + \zeta_\mathrm{E} \vectorn{v} \big)\big) - \rho_0 \frac{\dif}{\dif t} \big( \rho^{-1} \zeta_\mathrm{E} \big) , \label{eq:volEnergVMaxwellRef}
\end{align}
where $\tensor{\tau}_{\mathrm{E}0} = J \tensor{F}^{-1} \cdot \tensor{\tau}_\mathrm{E}$ is the Maxwell stress tensor in the reference configuration. These relations are obtained from equations \eqref{eq:volForceVMaxwell} and \eqref{eq:volEnergVMaxwell}, the relation between the Nabla operators $\nabla = \tensor{F}\transpm \cdot \nabla_0$, and identity $\nabla_0 \cdot ( J \tensor{F}^{-1} ) = \vectorn{0}$. As before, they are valid exclusively within the individual phases.

To obtain expressions for $\vectorn{t}_{\mathrm{E}0}$ and $f_{\mathrm{E}0}$, the volumetric integrals of $\vectorn{b}_\mathrm{E}$ and $p_\mathrm{E}$ should be considered within a finite-width slice of a volume with continuously differentiable fields, followed by taking the zero-thickness limit. In the considered assumed case, there is no interface $\varSigma$, while $\vectorn{b}_\mathrm{E}$ and $p_\mathrm{E}$ are continuously differentiable in the entire $\varOmega$ and change steeply within $\varLambda \subset \varOmega$. Then, integral over $\varOmega$ can be written as integral over $\varOmega/\varLambda$ and integral over $\varLambda$. Taking the zero-thickness limit gives integrals over $\varOmega/\varSigma$ and over $\varSigma$.

For arbitrary fields $\tensor{A}$ and $\vectorn{a}$ and for `thin' time-dependent domain $\varLambda$ bounded by `wide' surfaces $\varSigma_+$ and $\varSigma_-$ and `narrow' surface $\varUpsilon$, the following can be obtained via the Gauss and the Reynolds transport theorems:
\begin{align*}
  &\int_\varLambda \nabla_0 \cdot \tensor{A}\transp \dif \varLambda = 
  \int_{\varSigma_+} \tensor{A} \cdot \vectorn{N}_{\varSigma_+} \dif \varSigma_+ + \int_{\varSigma_-} \tensor{A} \cdot \vectorn{N}_{\varSigma_-} \dif \varSigma_- + \int_\varUpsilon \left(\ldots\right) \dif \varUpsilon , \\
  &\int_\varLambda \dot{\vectorn{a}} \dif \varLambda = \frac{\dif}{\dif t} \int_\varLambda \vectorn{a} \dif \varLambda - \int_{\varSigma_+} \vectorn{a} \vectorn{W} \cdot \vectorn{N}_{\varSigma_+} \dif \varSigma_+ - \int_{\varSigma_-} \vectorn{a} \vectorn{W} \cdot \vectorn{N}_{\varSigma_-} \dif \varSigma_- - \int_\varUpsilon \left(\ldots\right) \dif \varUpsilon , 
\end{align*}
where $\vectorn{N}_{\varSigma_+}$ and $\vectorn{N}_{\varSigma_-}$ are the external normals to $\varSigma_+$ and $\varSigma_-$, respectively, and $\vectorn{W}$ is the boundary velocity. It is now assumed that there is time-dependent surface $\varSigma$ between surfaces $\varSigma_+$ and $\varSigma_-$, such that $\varSigma_+$ and $\varSigma_-$ lie within previously-defined $\varOmega_+$ and $\varOmega_-$, respectively. Next, the limit is taken as $\varSigma_+$ and $\varSigma_-$ approach $\varSigma$, while keeping fields $\tensor{A}$ and $\vectorn{a}$ fixed on $\partial\varLambda$. Attention must be paid to the normals; as $\vectorn{N}_{\varSigma_+}$ and $\vectorn{N}_{\varSigma_-}$ are oriented towards the inside of $\varOmega_+$ and $\varOmega_-$, respectively, in the limit, they become $-\vectorn{N}_*$ and $\vectorn{N}_*$, respectively. This results in 
\begin{align*}
  &\lim_{\varSigma_\pm\to\varSigma} \int_\varLambda \nabla_0 \cdot \tensor{A}\transp \dif \varLambda = -\int_{\varSigma} \llbracket\tensor{A}\rrbracket \cdot \vectorn{N}_* \dif \varSigma , \\
  &\lim_{\varSigma_\pm\to\varSigma} \int_\varLambda \dot{\vectorn{a}} \dif \varLambda = \int_{\varSigma} \llbracket\vectorn{a}\rrbracket W_* \dif \varSigma ,
\end{align*}
where the integrals over `narrow' surface $\varUpsilon$ and the integral over volume $\varLambda$ vanish because the integrands are finite, while the `narrow' surface contracts to a curve and the volume contracts to a surface. Thus, from expressions \eqref{eq:volForceVMaxwellRef} and \eqref{eq:volEnergVMaxwellRef}, it can be seen that 
\begin{align}
  &\vectorn{t}_{\mathrm{E}0} = -\vectorn{N}_* \cdot \llbracket\tensor{\tau}_{\mathrm{E}0}\rrbracket - \rho_0 \llbracket \rho_0^{-1} J \varepsilon_0 \vectorn{E}\times\vectorn{B} \rrbracket W_* , \\
  &f_{\mathrm{E}0} = -\vectorn{N}_* \cdot \llbracket J \tensor{F}^{-1} \cdot \big( \vectorn{H}\times\vectorn{E} + \zeta_\mathrm{E} \vectorn{v} \big) \rrbracket - \rho_0 \llbracket \rho_0^{-1} J \zeta_\mathrm{E} \rrbracket W_* .
\end{align}

\subsection{Time-dependent interface --- deformable ferromagnets}

Because domain $\varOmega$ is chosen arbitrarily, the integrands of the corresponding terms of the balance laws should be equal. In particular, applying transformations \eqref{eq:genBalG} and \eqref{eq:genBalR} to equations \eqref{eq:momentLRef}, \eqref{eq:momentASRef}, \eqref{eq:energRef}, \eqref{eq:entrRef} and considering terms related to $\varSigma$ results in
\begin{align}
  &\llbracket\tensor{P}\rrbracket\cdot\vectorn{N}_* = -\llbracket\rho_0\vectorn{v}\rrbracket W_* - \vectorn{N}_* \cdot \llbracket \tensor{\tau}_{\mathrm{E}0}\rrbracket - \rho_0 \llbracket \rho_0^{-1} J \varepsilon_0 \vectorn{E}\times\vectorn{B} \rrbracket W_* , \label{eq:intJumpL} \\
  &\llbracket\vectorn{m}\times\tensor{B}\rrbracket\cdot\vectorn{N}_* = -\gamma_0^{-1}\llbracket\rho_0\vectorn{m}\rrbracket W_* , \label{eq:intJumpS} \\
  &\llbracket\vectorn{v}\cdot\tensor{P}\rrbracket\cdot\vectorn{N}_* + \llbracket\dot{\vectorn{m}}\cdot\tensor{B}\rrbracket\cdot\vectorn{N}_* - \llbracket\vectorn{G}\rrbracket\cdot\vectorn{N}_* = -\big(\llbracket\rho_0\epsilon\rrbracket + \tfrac{1}{2}\llbracket\rho_0\vectorn{v}\cdot\vectorn{v}\rrbracket\big) W_* - \vphantom{a} \nonumber \\
  &\vphantom{a}\hspace{4.5cm} - \vectorn{N}_* \cdot \llbracket J \tensor{F}^{-1} \cdot \big( \vectorn{H}\times\vectorn{E} + \zeta_\mathrm{E} \vectorn{v} \big) \rrbracket - \rho_0 \llbracket \rho_0^{-1} J \zeta_\mathrm{E} \rrbracket W_* , \label{eq:intJumpEnerg} \\
  &\llbracket \rho_0 s \rrbracket W_* - \llbracket T^{-1}\vectorn{G}\rrbracket\cdot\vectorn{N}_* \geq 0 . \label{eq:intJumpEntro} 
\end{align}
Equation \eqref{eq:intJumpL} gives one of the two interface conditions for the lattice continuum; the second condition is the continuity of the displacements: $\llbracket\vectorn{u}\rrbracket = \vectorn{0}$. Next, it is assumed that $\rho_0$, $\vectorn{m}$, $T$ are also continuous across $\varSigma$. Writing condition \eqref{eq:genCompat} for $\vectorn{u}$ and $\vectorn{m}$ then gives
\begin{align}
  &\llbracket\vectorn{v}\rrbracket = -W_* \llbracket\tensor{F}\rrbracket \cdot \vectorn{N}_* , \label{eq:compatL} \\
  &\llbracket\dot{\vectorn{m}}\rrbracket = -W_* \vectorn{N}_* \cdot \llbracket\nabla_0\vectorn{m}\rrbracket . \label{eq:compatS}
\end{align}
Furthermore, continuity of $\vectorn{m}$ and $\rho_0$ across $\varSigma$ means that equation \eqref{eq:intJumpS} transforms into
\begin{equation}
  \vectorn{m}\times\llbracket\tensor{B}\rrbracket\cdot\vectorn{N}_* = \vectorn{0} \hspace{1.3cm} \Rightarrow \hspace{1.3cm}
  \llbracket\tensor{B}\rrbracket\cdot\vectorn{N}_* = \lambda \vectorn{m}, \quad \lambda \in \mathbb{R} . \label{eq:intJumpSref}
\end{equation}
This equation together with the continuity of the magnetisation, $\llbracket\vectorn{m}\rrbracket = \vectorn{0}$, form two interface conditions for the spin continuum.

Applying transformations \eqref{eq:genBalG}, \eqref{eq:genBalS}, \eqref{eq:genBalX} to equations \eqref{eq:MaxwBRef}, \eqref{eq:MaxwHRef}, considering terms related to $\varSigma$, $\varPsi$, and using the fact that surface $\varTheta$ is arbitrary results in the interface conditions for the electromagnetic fields written in the reference configuration\footnote{Taking for example $\vectorn{H}_0$, from transformation \eqref{eq:genBalX}, one obtains $-\llbracket\vectorn{H}_0\rrbracket\cdot\vectorn{T}_* = W_*\big(\llbracket\vectorn{D}_0\rrbracket\times\vectorn{N}_*\big)\cdot\vectorn{T}_*$. This can be rewritten as $\vectorn{T}_*\cdot\left(\ldots\right) = 0$. Because $\varTheta$ is arbitrary, the latter is equivalent to $\vectorn{N}_*\times\left(\ldots\right) = \vectorn{0}$.} (transformation \eqref{eq:genBalX} can be applied because fields $\vectorn{B}_0$ and $\vectorn{D}_0$ emerge to be divergence-free):
\begin{align}
  &\llbracket\vectorn{B}_0\rrbracket\cdot\vectorn{N}_* = 0 , \quad\quad
  \llbracket\vectorn{H}_0\rrbracket\times\vectorn{N}_* = W_*\llbracket\vectorn{D}_0\rrbracket , \label{eq:MaxwBCref1} \\
  &\llbracket\vectorn{D}_0\rrbracket\cdot\vectorn{N}_* = 0 , \quad\quad
  \llbracket\vectorn{E}_0\rrbracket\times\vectorn{N}_* = -W_* \llbracket\vectorn{B}_0\rrbracket . \label{eq:MaxwBCref2} 
\end{align}

Using equations \eqref{eq:intIdent}, \eqref{eq:intJumpL}, \eqref{eq:compatL}, the following transformations can be made:
\begin{align*}
  &\llbracket\vectorn{v}\cdot\tensor{P}\rrbracket\cdot\vectorn{N}_* = -\rho_0\langle\vectorn{v}\rangle\cdot\llbracket\vectorn{v}\rrbracket W_* - \vectorn{N}_* \cdot \llbracket \tensor{\tau}_{\mathrm{E}0}\rrbracket \cdot \langle\vectorn{v}\rangle - \varepsilon_0 \langle\vectorn{v}\rangle\cdot\llbracket J \vectorn{E}\times\vectorn{B} \rrbracket W_* - \vphantom{a} \\
  &\vphantom{a}\hspace{6cm} - \vectorn{N}_* \cdot \llbracket\tensor{F}\rrbracket\transp\cdot\langle\tensor{P}\rangle\cdot\vectorn{N}_* W_* , \\
  &-\tfrac{1}{2}\rho_0\llbracket\vectorn{v}\cdot\vectorn{v}\rrbracket W_* = -\rho_0\langle\vectorn{v}\rangle\cdot\llbracket\vectorn{v}\rrbracket W_* .
\end{align*}
Using equations \eqref{eq:intIdent}, \eqref{eq:compatS}, \eqref{eq:intJumpSref}, a similar transformation for the magnetic term is made:
\begin{equation*}
  \llbracket\dot{\vectorn{m}}\cdot\tensor{B}\rrbracket\cdot\vectorn{N}_* = -\vectorn{N}_* \cdot \llbracket\nabla_0\vectorn{m}\rrbracket\cdot\langle\tensor{B}\rangle\cdot\vectorn{N}_* W_* ,
\end{equation*}
where $\langle\dot{\vectorn{m}}\rangle\cdot\vectorn{m} = 0$ is used due to continuity of $\vectorn{m}$ and equation \eqref{eq:mdotgen}. Analogously\footnote{To transform the second term, $\vectorn{N}_* \cdot J_+ \tensor{F}_+^{-1} = \vectorn{N}_* \cdot J_- \tensor{F}_-^{-1}$ is used, which is due to equation \eqref{eq:Nanson}.}, 
\begin{align*}
  &\vectorn{N}_* \cdot \llbracket J \tensor{F}^{-1} \cdot \zeta_\mathrm{E} \vectorn{v} \rrbracket = \vectorn{N}_* \cdot \llbracket J \tensor{F}^{-1} \zeta_\mathrm{E} \rrbracket \cdot \langle\vectorn{v}\rangle - \vectorn{N}_* \cdot \langle J \tensor{F}^{-1}  \zeta_\mathrm{E} \rangle \cdot \llbracket\tensor{F}\rrbracket \cdot \vectorn{N}_* W_* = \vphantom{a} \\
  &\vphantom{a} \hspace{3cm} = \vectorn{N}_* \cdot \llbracket J \tensor{F}^{-1} \zeta_\mathrm{E} \rrbracket \cdot \langle\vectorn{v}\rangle - \llbracket J \rrbracket \langle \zeta_\mathrm{E} \rangle W_* .
\end{align*}
Due to condition \eqref{eq:genGradForm}, it is easy to see that\footnote{Symbol `$:$' denotes the double inner product of two second-order tensors; in coordinate notation $\tensor{A}:\tensor{G}$ is $A_{ij} G_{ji}$.}
\begin{align*}
  &\vectorn{N}_* \cdot \llbracket\tensor{F}\rrbracket\transp\cdot\langle\tensor{P}\rangle\cdot\vectorn{N}_* = \llbracket\tensor{F}\rrbracket\transp : \langle\tensor{P}\rangle , \\
  &\vectorn{N}_* \cdot \llbracket\nabla_0\vectorn{m}\rrbracket\cdot\langle\tensor{B}\rangle\cdot\vectorn{N}_* = \llbracket\nabla_0\vectorn{m}\rrbracket : \langle\tensor{B}\rangle .
\end{align*}
Finally, through derivations outlined in appendix \ref{sec:addDeriv}, it is possible to show that 
\begin{align}
  &\vectorn{N}_* \cdot \llbracket \tensor{\tau}_{\mathrm{E}0} - J \tensor{F}^{-1} \zeta_\mathrm{E}\rrbracket \cdot \langle\vectorn{v}\rangle - \vectorn{N}_* \cdot \llbracket J \tensor{F}^{-1} \cdot \big( \vectorn{H}\times\vectorn{E} \big) \rrbracket + \varepsilon_0 \langle\vectorn{v}\rangle\cdot\llbracket J \vectorn{E}\times\vectorn{B} \rrbracket W_* = \vphantom{a} \nonumber \\
  &\vphantom{a} \hspace{1cm} = \langle J \rangle \Big( \langle \vectorn{E} \rangle \cdot \llbracket \vectorn{D} \rrbracket + \langle \vectorn{H} \rangle \cdot \llbracket \vectorn{B} \rrbracket \Big) W_* + \mathscr{D} , \label{eq:bigTermsS} \\
  &\mathscr{D} = \vectorn{N}_* \cdot \langle J \tensor{F}^{-1} \rangle \cdot \Big( \big( \langle \vectorn{v} \rangle\langle \vectorn{B} \rangle - \langle \vectorn{B} \rangle\langle \vectorn{v} \rangle\big)\cdot\big(\llbracket \vectorn{D} \rrbracket\times\langle \tensor{F} \rangle\cdot\vectorn{N}_*\big) + \vphantom{a} \nonumber \\
  &\vphantom{a} \hspace{1cm} +
  \big( \langle \vectorn{D} \rangle\langle \vectorn{v} \rangle - \langle \vectorn{v} \rangle\langle \vectorn{D} \rangle\big)\cdot\big(\llbracket \vectorn{B} \rrbracket\times\langle \tensor{F} \rangle\cdot\vectorn{N}_*\big) \Big) W_* + \langle J \rangle\llbracket \vectorn{D}\times\vectorn{B}\rrbracket\cdot\langle \vectorn{v} \rangle W_* , \label{eq:bigTermsD}
\end{align}
where $\mathscr{D}$ collects the `dynamic' terms, i.e.\ ones containing multiplication of $\langle \vectorn{v}\rangle$ and $W_*$, such that they are second-order with respect to the velocity of the points/interface. Additionally, it is easy to see that
\begin{equation*}
  \llbracket \zeta_\mathrm{E} \rrbracket - \langle \vectorn{E} \rangle \cdot \llbracket \vectorn{D} \rrbracket - \langle \vectorn{H} \rangle \cdot \llbracket \vectorn{B} \rrbracket = \langle J^{-1} \rangle \rho_0 \vectorn{m} \cdot \llbracket \vectorn{B} \rrbracket .
\end{equation*}
Substituting these relations into equation \eqref{eq:intJumpEnerg} results in
\begin{equation}
  \left(\rho_0\llbracket\epsilon\rrbracket + \langle J \rangle \langle J^{-1} \rangle \rho_0 \vectorn{m} \cdot \llbracket \vectorn{B} \rrbracket - \llbracket\tensor{F}\rrbracket\transp : \langle\tensor{P}\rangle - \llbracket\nabla_0\vectorn{m}\rrbracket : \langle\tensor{B}\rangle \right) W_* = \llbracket\vectorn{G}\rrbracket\cdot\vectorn{N}_* + \mathscr{D} . \label{eq:intVelTherm}
\end{equation}
The Helmholtz free energy per unit of mass is introduced as $\psi = \epsilon - T s + \vectorn{m}\cdot\vectorn{B}$. Finally, substituting equation \eqref{eq:intVelTherm} into \eqref{eq:intJumpEntro}  results in
\begin{equation}
  \left(-\rho_0\llbracket\psi\rrbracket + \big(1 - \langle J \rangle \langle J^{-1} \rangle\big) \rho_0 \vectorn{m}\cdot\llbracket\vectorn{B}\rrbracket + \llbracket\tensor{F}\rrbracket\transp : \langle\tensor{P}\rangle + \llbracket\nabla_0\vectorn{m}\rrbracket : \langle\tensor{B}\rangle \right) W_* + \mathscr{D} \geq 0 .  \label{eq:intJumpMain}
\end{equation}
In the expression above, one can also simplify $\big(1 - \langle J \rangle \langle J^{-1} \rangle\big)=\tfrac{1}{4}\llbracket J \rrbracket \llbracket J^{-1} \rrbracket$. 

\subsubsection{Discussion of the obtained entropy production and relation to prior work}

The obtained relation imposes constraints onto constitutive modelling of the time-dependent interfaces in deformable ferromagnets. In particular, in thermodynamics of irreversible processes, such relation is typically viewed as the product of the thermodynamic driving force and the corresponding thermodynamic flux \cite{Glansdorff1971}. The relation above, however, contains dynamic part $\mathscr{D}$, which is quadratic with respect to the velocities and is therefore later neglected in quasistatics. This is the standard practice in such problems and is done for e.g.\ chemo-mechanics \cite{Freidin2014}.

The obtained structure allows assuming specific constitutive relations, where the flux (velocity $W_*$) is a function of the driving force (expression in brackets), such that the dissipation inequality is always fulfilled \cite{Glansdorff1971}. The difference between equation \eqref{eq:intJumpMain} and the classical expression for the entropy production due to the interface propagation in deformable solids (see e.g.\ \cite{Abeyaratne2006}) is the presence of two terms containing the magnetisation and its gradient, as well as the dynamic terms containing the electromagnetic fields.

As mentioned in the introduction, the entropy production due to the phase boundary propagation in deformable ferromagnets was also considered in \cite{Fomethe1996,Fomethe1997,Maugin1997}, but in a more restrictive setting --- non-linear elasticity, magnetostatics, and absence of electric fields were assumed right from the start. The derivations above do not make these assumptions (they are made later in the present paper to obtain a system of equations for computational examples). Furthermore, derivations in \cite{Fomethe1996,Fomethe1997,Maugin1997} took a different path. Nevertheless, the resulting driving force transpires to be same, albeit with one remark. The published expressions for the driving force are equation (4.23) in \cite{Fomethe1997} or equation (4.21) in \cite{Maugin1997}, where it is denoted as $\mathcal{H}^\mathrm{fer}$. Taken with the minus sign, it coincides with the expression in brackets in equation \eqref{eq:intJumpMain} --- it can be seen that the equations contain exactly the same terms, including one with $\langle J \rangle \langle J^{-1} \rangle$, which is equivalent to $\langle \rho \rangle \langle \rho^{-1} \rangle$. There is a difference however in the Helmholtz free energy --- in \cite{Fomethe1997,Maugin1997}, it does not contain term $\vectorn{m}\cdot\vectorn{B}$, which is evident from equation (4.25) in \cite{Fomethe1997} or  equation (2.26) in \cite{Maugin1997}. This particular quantity, denoted as $W$ in the papers, is used to obtain the first Piola-Kirchhoff stress tensor and the exchange-force tensor via the derivatives by the deformation gradient and the magnetisation gradient, respectively, which is seen from equation (2.31) in \cite{Maugin1997} or equations (3.40) and (3.42) in \cite{Fomethe1996}. The reason for this is the absence of $\vectorn{m}\cdot\vectorn{B}$ from the entropy and energy balances right from the start --- from equation (3.16) in \cite{Fomethe1996}, which contradicts equation (6.2.57) in \cite{Maugin1988}. In the latter, $\hat{\psi}$ explicitly contains $\hat{e} = e + \vectorn{\mu}\cdot\vectorn{B}$ (in the notation of the book), and specifically $\hat{\psi}$ is used to obtain the constitutive relations, outlined in section 6.4 of \cite{Maugin1988}. If the initial balance laws are followed precisely, as will be seen in the present paper below, the Helmholtz free energy should contain $\vectorn{m}\cdot\vectorn{B}$, and the constitutive laws should be obtained from such expression. The author of the present paper does not understand the reason for the discrepancy, given that papers \cite{Fomethe1996,Fomethe1997,Maugin1997} follow the theory from \cite{Maugin1988}, and the probable reason is that internal energy $e$ in \cite{Fomethe1996} is already implied to contain $\vectorn{m}\cdot\vectorn{B}$.

Another derivation of the driving force for such problem, also for the non-linear elasticity case and for the magnetostatics, was done in \cite{James2002}. The derivation did not assume constant magnetisation length (i.e.\ no saturation condition was imposed) and was based on the theory of W.\ Brown \cite{Brown1966}; however, the exchange energy was neglected, which can be seen from the form of the free energy density, equation (29) in \cite{James2002}, which depends only on the deformation gradient and on the magnetisation, but not on the magnetisation gradient. The author of \cite{James2002} specifically draws attention to non-constant magnetisation length, claiming that some MSMAs exhibit the existence of phase boundaries between phases with drastically different $m_\mathrm{S}$. However, such case can be handled by the theory of G.\ Maugin, which the present paper follows: the saturation is still assumed, such that the magnetisation dynamics follows equation \eqref{eq:mdotgen}, but $m_\mathrm{S}$ is different in different phases (and spatially constant within each phase), leading to parameters $m_{\mathrm{S}+}$ and $m_{\mathrm{S}-}$. The magnetisation is then scaled in both phases, resulting in $\vectorn{m}'_\pm = \vectorn{m}_\pm / m_{\mathrm{S}\pm}$ that has unit length. The free energy densities of the phases are then written as functions of the scaled magnetisation. The possibility of doing this is clearly seen from an example of a constitutive law, equation \eqref{eq:constLaw} below, where the scaling would simply mean that $m_\mathrm{S}$ goes into $A_\mathrm{m}$ and $K_\mathrm{m}$ (the exchange and the anisotropy parameters, respectively). All equations will then be based on scaled magnetisation $\vectorn{m}'_\pm$, for which assuming continuity across the interface is equivalent to assuming the continuity of the direction of the magnetisation, but not the magnitude.

It is interesting to remark that the author of \cite{James2002} criticised works \cite{Fomethe1996,Fomethe1997,Maugin1997} for the absence of clarity, with which the author of the present paper is inclined to agree. However, the initial theory outlined in \cite{Maugin1988} and especially in its Russian translation \cite{Maugin1991} is clear and well-written. The prior work on this topic deserves criticism for not writing the Maxwell's equations in SI units, which are the accepted international standard nowadays.

\subsection{Towards governing equations}

To obtain the set of governing equations, similarly to the previous section, transformations \eqref{eq:genBalG} and \eqref{eq:genBalR} are applied to the balance laws, but now terms related to $\varOmega_\pm$ are considered. In this section, subscripts `$\pm$' are dropped for brevity. It is easy to see that balance laws \eqref{eq:momentLRef}-\eqref{eq:entrRef} become
\begin{align}
  &\rho_0 \dot{\vectorn{v}} = \nabla_0 \cdot \tensor{P}\transp + \rho_0 \vectorn{b} + \rho_0 \vectorn{b}_\mathrm{E} , \label{eq:momentLLoc} \\
  &\vectorn{x}\times\rho_0\dot{\vectorn{v}} = \vectorn{x} \times \big(\nabla_0 \cdot \tensor{P}\transp\big) + \tensorThree{E} : \big(\tensor{F}\cdot\tensor{P}\transp\big) + \vectorn{x}\times\rho_0 \vectorn{b} + \vectorn{x}\times\rho_0 \vectorn{b}_\mathrm{E} - \rho_0 \vectorn{m}\times\vectorn{B}_\mathrm{L} , \label{eq:momentALLoc} \\
  &\rho_0 \gamma_0^{-1} \dot{\vectorn{m}} = \vectorn{m} \times \big(\nabla_0 \cdot \tensor{B}\transp\big) + \tensorThree{E} : \big((\nabla_0\vectorn{m})\transp\cdot\tensor{B}\transp\big) + \rho_0 \vectorn{c}_\mathrm{E} + \rho_0 \vectorn{m}\times\vectorn{B}_\mathrm{L} , \label{eq:momentASLoc} \\
  &\rho_0 \dot{\epsilon} + \vectorn{v}\cdot\rho_0 \dot{\vectorn{v}} = \vectorn{v}\cdot\big(\nabla_0 \cdot \tensor{P}\transp\big) + \tensor{P}:\nabla_0\vectorn{v} + \dot{\vectorn{m}}\cdot\big(\nabla_0 \cdot \tensor{B}\transp\big) + \tensor{B}:\nabla_0\dot{\vectorn{m}} - \nabla_0\cdot\vectorn{G} + \vphantom{a} \nonumber \\
  &\vphantom{a}\hspace{9cm} + \vectorn{v}\cdot\rho_0 \vectorn{b} + \rho_0 r + \rho_0 p_\mathrm{E} , \label{eq:energLoc} \\
  &\rho_0 \dot{s} + \frac{1}{T}\nabla_0\cdot\vectorn{G} - \frac{1}{T^2}\vectorn{G}\cdot\nabla_0 T - \frac{\rho_0 r}{T} \geq 0 , \label{eq:entrLoc}
\end{align}
where $\tensorThree{E}$ is the third-order Levi-Civita tensor\footnote{For arbitrary vectors $\vectorn{a}$ and $\vectorn{g}$, multiplication by $\tensorThree{E}$ manifests as the cross product: $\tensorThree{E}:\vectorn{a}\vectorn{g} = \vectorn{a}\times\vectorn{g}$.}. Analogously, the Maxwell's equations \eqref{eq:MaxwBRef} and \eqref{eq:MaxwHRef} become
\begin{align}
  &\nabla_0 \cdot \vectorn{B}_0 = 0 , \quad\quad
  \nabla_0 \times \vectorn{H}_0 = \dot{\vectorn{D}}_0 , \label{eq:maxwBH} \\
  &\nabla_0 \cdot \vectorn{D}_0 = 0 , \quad\quad
  \nabla_0 \times \vectorn{E}_0 = -\dot{\vectorn{B}}_0 .
\end{align}

The linear momentum balance in the local form, equation \eqref{eq:momentLLoc}, is the first governing equation describing the coupled magneto-mechanics of the considered solid. It must be supplemented by the corresponding constitutive law, providing the expression for tensor $\tensor{P}$.

Substituting equation \eqref{eq:momentLLoc} is into \eqref{eq:momentALLoc} results in 
\begin{equation}
  \tensorThree{E} : \big(\tensor{F}\cdot\tensor{P}\transp\big) - \rho_0 \vectorn{m}\times\vectorn{B}_\mathrm{L} = \vectorn{0} , \label{eq:momentALCheck}
\end{equation}
which should be enforced by choosing the appropriate constitutive law. Next, to obtain the structure of the equation for the magnetisation dynamics that adheres to from \eqref{eq:mdotgen}, tensor $\tensor{B}$ should be such that 
\begin{equation}
  \tensorThree{E} : \big((\nabla_0\vectorn{m})\transp\cdot\tensor{B}\transp\big) = \vectorn{0} , \label{eq:momentASCheck}
\end{equation}
which should also be enforced via the constitutive laws. Substituting $\vectorn{c}_\mathrm{E}$ from equation \eqref{eq:EMterms} into \eqref{eq:momentASLoc} gives the angular momentum balance for the spin continuum in the local form:
\begin{equation}
  \dot{\vectorn{m}} = \gamma_0 \vectorn{m} \times \big( \rho_0^{-1} \nabla_0 \cdot \tensor{B}\transp + \vectorn{B} + \vectorn{B}_\mathrm{L} \big) ,
  \label{eq:momentASMain}
\end{equation}
which is the second governing equation for the coupled spin-lattice continuum. It requires specifying expressions for $\tensor{B}$ and $\vectorn{B}_\mathrm{L}$.

To obtain the expression for the entropy production, first\footnote{From equation \eqref{eq:momentASMain}, it follows that $\dot{\vectorn{m}} \cdot \big( \nabla_0 \cdot \tensor{B}\transp + \rho_0 \vectorn{B} + \rho_0 \vectorn{B}_\mathrm{L} \big) = 0$.}, the third term of the right-hand side of equation \eqref{eq:energLoc} is expressed with the use of equation \eqref{eq:momentASMain}. Second, equation \eqref{eq:momentLLoc} is substituted into \eqref{eq:energLoc}, as well as $\vectorn{b}_\mathrm{E}$ and $p_\mathrm{E}$ from equation \eqref{eq:EMterms}, resulting in
\begin{equation}
  \rho_0 \dot{\epsilon} = \tensor{P}:\nabla_0\vectorn{v} + \tensor{B}:\nabla_0\dot{\vectorn{m}} - \rho_0\vectorn{m}\cdot \dot{\vectorn{B}} - \rho_0\dot{\vectorn{m}}\cdot \vectorn{B} - \rho_0\dot{\vectorn{m}}\cdot \vectorn{B}_\mathrm{L} - \nabla_0\cdot\vectorn{G} + \rho_0 r , 
\end{equation}
which is then substituted into \eqref{eq:entrLoc}, giving the entropy production inequality:
\begin{equation}
  -\rho_0 \dot{\psi} -\rho_0 \dot{T} s + \tensor{P}:\dot{\tensor{F}}\transp + \tensor{B}:\nabla_0\dot{\vectorn{m}} - \rho_0\dot{\vectorn{m}}\cdot \vectorn{B}_\mathrm{L} - \frac{1}{T}\vectorn{G}\cdot\nabla_0 T \geq 0 . \label{eq:entrMain}
\end{equation}
It now can be used to choose the appropriate structure of the constitutive laws.

\subsection{Non-dissipative solid}
\label{sec:constitut}

The constitutive laws should of course be chosen in accordance with the entropy production inequality, equation \eqref{eq:entrMain}. Furthermore, equations \eqref{eq:momentALCheck} and \eqref{eq:momentASCheck} must be fulfilled. Finally, the principle of objectivity (frame indifference) must be satisfied. This section considers non-dissipative solids.

To determine the structure of the constitutive laws, first, the independent variables must be chosen. The Helmholtz free energy must be dependent on invariant tensors (i.e.\ it should not depend on the changes of the coordinate system). As usual, to describe material deformation, the right Cauchy-Green deformation tensor, $\tensor{C} = \tensor{F}\transp \cdot \tensor{F}$, can be used. For the spin continuum, vector $\vectorn{m}_0 = \tensor{F}\transp \cdot \vectorn{m}$, which is the magnetisation pulled back to the reference configuration, and tensor $\tensor{M} = \left( \nabla_0 \vectorn{m} \right) \cdot \left( \nabla_0 \vectorn{m} \right)\transp$, which characterises the magnetisation gradient, are convenient choices for the sought invariant tensors. Thus, the Helmholtz free energy per unit of mass can be assumed to have the following functional form:
\begin{equation}
  \psi = \psi \left( \tensor{C}, \tensor{M}, \vectorn{m}_0, T \right) . \label{eq:psi}
\end{equation}
It will now be shown that it describes a non-dissipative solid, and the structure of the constitutive relations for $\tensor{P}$, $\tensor{B}$, $\vectorn{B}_\mathrm{L}$ will be obtained.

It is easy to show that\footnote{From the tensor derivative rules, one obtains $\partial \tensor{C} / \partial\tensor{F} = \big( \tensorFour{I} + \tensorFour{I}\transpRight \big) : \big( \tensor{F}\transp\cdot \tensorFour{I}\transpRight \big)$, where $\tensorFour{I} = \vectorn{e}_s\vectorn{e}_k\vectorn{e}_k\vectorn{e}_s$ and \\$\tensorFour{I}\transpRight = \vectorn{e}_s\vectorn{e}_k\vectorn{e}_s\vectorn{e}_k$ are the fourth-order identity tensors. Hence, $\tensor{A} : \big( \partial \tensor{C} / \partial\tensor{F} \big) = 2 \tensor{F}\cdot\tensor{A}$ for arbitrary symmetric tensor $\tensor{A}$.}
\begin{equation*}
  \dot{\psi} = \left(2 \tensor{F}\cdot\frac{\partial\psi}{\partial\tensor{C}}\right):\dot{\tensor{F}}\transp + \left( 2 \left( \nabla_0 \vectorn{m} \right)\transp\cdot\frac{\partial\psi}{\partial\tensor{M}}\right): \nabla_0 \dot{\vectorn{m}} + \frac{\partial\psi}{\partial\vectorn{m}_0} \cdot \left( \dot{\tensor{F}}\transp \cdot \vectorn{m} + \tensor{F}\transp \cdot \dot{\vectorn{m}} \right) + \frac{\partial\psi}{\partial T} \dot{T} .
\end{equation*}
Substituting it into inequality \eqref{eq:entrMain} results in
\begin{align}
  &- \frac{1}{T}\vectorn{G}\cdot\nabla_0 T - \rho_0\left( s + \frac{\partial\psi}{\partial T} \right) \dot{T} + \left( \tensor{P}-2\rho_0 \tensor{F}\cdot\frac{\partial\psi}{\partial\tensor{C}} - \rho_0\vectorn{m} \frac{\partial\psi}{\partial\vectorn{m}_0} \right):\dot{\tensor{F}}\transp + \vphantom{a} \nonumber \\
  &\vphantom{a}\hspace{2cm} +
  \left(\tensor{B} - 2 \rho_0 \left( \nabla_0 \vectorn{m} \right)\transp\cdot\frac{\partial\psi}{\partial\tensor{M}}\right):\nabla_0\dot{\vectorn{m}} - \rho_0 \left(\vectorn{B}_\mathrm{L}+\frac{\partial\psi}{\partial\vectorn{m}_0} \cdot \tensor{F}\transp\right)\cdot\dot{\vectorn{m}}  \geq 0 . \label{eq:entrSplit}
\end{align}
The resulting dissipation inequality should be fulfilled under various different thermodynamic paths. Therefore, it is typically split into stronger relations --- the so-called Coleman-Noll procedure \cite{Coleman1963}. Within the current context, the specific choice of the constitutive relationships for $\tensor{P}$, $\tensor{B}$, $\vectorn{B}_\mathrm{L}$, $s$ automatically enforces all terms starting from the second to become zeros:
\begin{align}
  &\tensor{P} = 2 \rho_0 \tensor{F} \cdot \frac{\partial \psi}{\partial \tensor{C}} - \rho_0 \vectorn{m} \vectorn{B}_\mathrm{L} \cdot \tensor{F}\transpm , \label{eq:constP} \\
  &\tensor{B} = 2 \rho_0 \left( \nabla_0 \vectorn{m} \right)\transp \cdot \frac{\partial \psi}{\partial \tensor{M}} , \label{eq:constM} \\
  &\vectorn{B}_\mathrm{L} = - \tensor{F} \cdot \frac{\partial \psi}{\partial \vectorn{m}_0} , \label{eq:constB} \\
  &s = -\frac{\partial\psi}{\partial T} . 
\end{align}
The remaining entropy production is solely due to the heat flow: $-\vectorn{G}\cdot\nabla_0 T \geq 0$. This means that the material deformation and the magnetisation dynamics are non-dissipative processes under the above-made choices. The functional form of $\psi$, equation \eqref{eq:psi}, can be written with the use of experiments or atomistic modelling; however, there can be additional material symmetry conditions enforcing restrictions on the expression for $\psi$.

Finally, it is easy to verify that relations \eqref{eq:constP}, \eqref{eq:constM}, \eqref{eq:constB} satisfy equations \eqref{eq:momentALCheck} and \eqref{eq:momentASCheck}. This can be seen from the fact that the double-dot multiplication of a symmetric second-order tensor by $\tensorThree{E}$ gives $\vectorn{0}$.

It is useful to note an implication of the obtained structure of the constitutive laws on the interface conditions for the spin continuum, which are given by equation \eqref{eq:intJumpSref}. Due to the constant length of the magnetisation vector, $\left(\nabla_0 \vectorn{m}\right)\cdot\vectorn{m} = \vectorn{0}$ must be fulfilled, similarly to the orthogonality of $\vectorn{m}$ and $\dot{\vectorn{m}}$, as discussed just before equation \eqref{eq:mdotgen}. Therefore, $\vectorn{m}\cdot\tensor{B} = \vectorn{0}$ must be fulfilled. This implies that the interface conditions for the spin continuum become
\begin{equation}
  \llbracket\tensor{B}\rrbracket\cdot\vectorn{N}_* = \vectorn{0}.
\end{equation}

\subsection{Quasistatic problem}
\label{sec:stat}

For the considered applications --- modelling of the propagation of the transformation fronts, it is reasonable to assume that the dynamic terms can be neglected, since the time scale of the dynamic processes is much smaller than the time scale of the fronts' propagation. 

As usual, the structure of equation \eqref{eq:maxwBH} in the quasistatic form (i.e.\ zero right-hand side) indicates that the magnetic field has the form of a gradient of some scalar field: $\vectorn{H}_0 = -\nabla_0 \eta$. The first equation of \eqref{eq:maxwBH} then becomes the governing equation for field $\eta$, where $\vectorn{B}_0$ is expressed via $\vectorn{H}_0$ using relations \eqref{eq:BHlink}, \eqref{eq:Bref}, \eqref{eq:Href}. From here onwards, field $\vectorn{E}_0$ is neglected but it can be trivially retained within the system of equations if its effects are expected to be non-negligible.

This leads to the quasistatic form of the governing equations:
\begin{align}
  &\rho_0^{-1} \nabla_0 \cdot \big( \tensor{P}\transp + J \tensor{F}^{-1} \cdot \tensor{\tau}_\mathrm{E} \big) + \vectorn{b} = \vectorn{0} , 
  \quad\quad\quad\quad \vectorn{X}\in\varOmega_\pm , \label{eq:quasMechP} \\
  &\vectorn{m} \times \big( \rho_0^{-1} \nabla_0 \cdot \tensor{B}\transp + J^{-1}\tensor{F} \cdot\vectorn{B}_0 + \vectorn{B}_\mathrm{L} \big) = \vectorn{0} , 
  \quad\quad\quad\quad \left|\vectorn{m}\right| = m_\mathrm{S} ,
  \quad\quad\quad\quad \vectorn{X}\in\varOmega_\pm , \label{eq:quasMagP} \\
  &\nabla_0 \cdot \vectorn{B}_0 = 0 , \quad\quad\quad\quad \vectorn{X}\in\varOmega_\pm , \label{eq:quasFieP}
\end{align}
where $\tensor{P}$, $\tensor{B}$, $\vectorn{B}_\mathrm{L}$ are given by equations \eqref{eq:constP}-\eqref{eq:constB}, and $\tensor{\tau}_\mathrm{E}$, $\vectorn{B}_0$ are expressed as
\begin{align}
  &\tensor{\tau}_\mathrm{E} = \mu_0^{-1} J^{-2} \big(\tensor{F} \cdot \vectorn{B}_0\vectorn{B}_0\cdot\tensor{F}\transp  - \tfrac{1}{2} \tensor{I}\tensor{C}:\vectorn{B}_0\vectorn{B}_0 \big) + J^{-2}\rho_0 \tensor{I}\tensor{F}:\vectorn{B}_0\vectorn{m} - J^{-2}\rho_0\tensor{F}\cdot\vectorn{B}_0\vectorn{m} , \label{eq:quasT} \\
  &\vectorn{B}_0 = -\mu_0 J \tensor{C}^{-1} \cdot \nabla_0\eta + \mu_0 \rho_0 \tensor{F}^{-1} \cdot \vectorn{m} , \label{eq:quasB}
\end{align}
The resulting PDEs must be solved with respect to unknown fields $\vectorn{u}$, $\vectorn{m}$, $\eta$. The PDEs are supplemented with the appropriate boundary conditions. Furthermore, the following interface conditions are imposed:
\begin{align}
  &\llbracket\vectorn{u}\rrbracket = \vectorn{0} , \quad\quad\quad 
  \llbracket\tensor{P}\rrbracket\cdot\vectorn{N}_* = - \vectorn{N}_* \cdot \llbracket J \tensor{F}^{-1} \cdot \tensor{\tau}_\mathrm{E}\rrbracket , 
  \quad\quad\quad\quad \vectorn{X}\in\varSigma , \label{eq:quasMechB} \\
  &\llbracket\vectorn{m}\rrbracket = \vectorn{0} , \quad\quad\quad 
  \llbracket\tensor{B}\rrbracket\cdot\vectorn{N}_* = \vectorn{0} , 
  \quad\quad\quad\quad \vectorn{X}\in\varSigma , \label{eq:quasMagB} \\
  &\llbracket\eta\rrbracket = 0 , \quad\quad\quad 
  \llbracket\vectorn{B}_0\rrbracket\cdot\vectorn{N}_* = 0 ,
  \quad\quad\quad\quad \vectorn{X}\in\varSigma . \label{eq:quasFieB}
\end{align}
Finally, a constitutive law for interface velocity $W_*$ must be imposed as a functional dependency on the driving force, such that it adheres to inequality \eqref{eq:intJumpMain}, for example,
\begin{equation}
  W_* = k_*\left(-\rho_0\llbracket\psi\rrbracket + \tfrac{1}{4}\llbracket J \rrbracket \llbracket J^{-1} \rrbracket \rho_0 \vectorn{m}\cdot\llbracket J^{-1}\tensor{F} \cdot\vectorn{B}_0\rrbracket + \llbracket\tensor{F}\rrbracket\transp : \langle\tensor{P}\rangle + \llbracket\nabla_0\vectorn{m}\rrbracket : \langle\tensor{B}\rangle \right) , 
\end{equation}
where $k_*$ is the kinetic parameter.

\subsection{General weak form of the quasistatic problem}
\label{sec:weak}

The weak formulation of the problem (for a non-dissipative system) can be obtained either from the strong form (i.e.\ via the principle of virtual work) or by a variation of an energy functional (i.e.\ via the principle of stationary potential energy) \cite{Holzapfel2000}. In the present section, a general quasistatic problem (with arbitrary number of unknowns) is considered, and a general energy functional is proposed, such that its variation results in the weak form, within which the interface conditions are enforced. This will be subsequently used to obtain the weak form of the quasistatic magneto-mechanical problem. 

As before, domain $\varOmega$ is arbitrary. The general energy functional can be represented by the following five terms:
\begin{equation}
  \varPi = \varPi_+ + \varPi_- + \varPi_* + \varPi_\varGamma + \varPi_\mathrm{I} ,
\end{equation}
where $\varPi_+$ and $\varPi_-$ are the bulk energies corresponding to domains $\varOmega_+$ and $\varOmega_-$, respectively, $\varPi_*$ is the interface energy, $\varPi_\varGamma$ is the energy term corresponding to the boundary, and $\varPi_\mathrm{I}$ is the numerical stabilisation energy term that is usually introduced in the cut-element-type methods and is outlined in section \ref{sec:stab}. The bulk energies are written as
\begin{equation}
  \varPi_\pm = \int_{\varOmega_\pm} w \dif \varOmega_\pm ,
\end{equation}
where $w$ is the energy density. The unknown that must be determined is vector field $\vectorn{z} = \vectorn{z}\big(\vectorn{X}\big)$, where $\vectorn{X}\in\varOmega$ is the spatial coordinate in the reference configuration. The unconventional element here is that vector $\vectorn{z}$ has $n$ components, i.e.\ $\vectorn{z}\in\mathbb{R}^n$. Introduction of this vector keeps all tensor notation intact, including tensor derivatives, but the number of the components of the tensors changes correspondingly. The energy density is then written as 
\begin{equation}
  w = w \left( \vectorn{z}, \nabla_0 \vectorn{z} \right) .
\end{equation}
Generalising the results from continuum mechanics \cite{Poluektov2019}, to enforce the interface conditions weakly, the following interface energy term is considered:
\begin{equation}
  \varPi_* = \int_\varSigma \left( \frac{\beta}{2} \llbracket\vectorn{z}\rrbracket - \vectorn{N}_* \cdot \left\langle\frac{\partial w}{\partial \nabla_0 \vectorn{z}} \right\rangle \right) \cdot \llbracket\vectorn{z}\rrbracket \dif \varSigma . \label{eq:piStz}
\end{equation}
The aim now is to find the corresponding weak and strong problems.

The variation of the total and bulk energies can be obtained in the straightforward way:
\begin{align}
  &\delta\varPi = \delta\varPi_+ + \delta\varPi_- + \delta\varPi_* + \delta\varPi_\varGamma + \delta\varPi_\mathrm{I} , \\
  &\delta\varPi_\pm = \int_{\varOmega_\pm} \left( \frac{\partial w}{\partial \vectorn{z}} \cdot \delta\vectorn{z} + \frac{\partial w}{\partial \nabla_0 \vectorn{z}} : \left(\nabla_0\delta\vectorn{z}\right)\transp \right) \dif \varOmega_\pm . \label{eq:piVol}
\end{align}
It is important to emphasise that in this section, the interface position is \emph{fixed} --- the purpose of the derived weak form is to solve the field equations for the \emph{given} configuration of the interface\footnote{This is followed by shifting the interface points by one increment (corresponding to one time step) to obtain the new position of the interface that is fixed again. The detailed algorithm is outlined in section \ref{sec:topo}.}; this is different to the variational approach presented in section \ref{sec:variat}. The variation of the interface energy results in
\begin{align}
  &\delta\varPi_* = \int_\varSigma \left( \beta \llbracket\vectorn{z}\rrbracket \cdot \llbracket\delta\vectorn{z}\rrbracket - \vectorn{N}_* \cdot \left\langle\frac{\partial w}{\partial \nabla_0 \vectorn{z}}\right\rangle\cdot \llbracket\delta\vectorn{z}\rrbracket - \vectorn{N}_* \cdot \left\langle \tensor{K}\left(\delta\vectorn{z}, \nabla_0 \delta\vectorn{z} \right) \right\rangle \cdot \llbracket\vectorn{z}\rrbracket \right) \dif \varSigma , \\
  &\tensor{K}\left(\delta\vectorn{z}, \nabla_0 \delta\vectorn{z} \right) = \frac{\partial^2 w}{\partial \nabla_0 \vectorn{z} \, \partial \vectorn{z}} \cdot \delta\vectorn{z} + \frac{\partial^2 w}{\partial \nabla_0 \vectorn{z} \, \partial \nabla_0 \vectorn{z}} : \left(\nabla_0\delta\vectorn{z}\right)\transp .
\end{align}
Thus, the resulting weak problem formulation is the following: find $\vectorn{z}_\pm$ such that $\delta\varPi = 0$ for $\forall\delta\vectorn{z}_\pm$. Here, boundary energy $\varPi_\varGamma$ is problem-specific and is written depending on the boundary conditions applied to domain $\varOmega$.

To obtain the strong form, the Gauss theorem is applied to equation \eqref{eq:piVol}, resulting in
\begin{equation}
  \delta\varPi_\pm = \int_{\varGamma_\pm} \vectorn{N}_\pm \cdot \frac{\partial w}{\partial \nabla_0 \vectorn{z}} \cdot \delta\vectorn{z} \dif \varGamma_\pm + \int_{\varOmega_\pm} \left( \frac{\partial w}{\partial \vectorn{z}} - \nabla_0\cdot\frac{\partial w}{\partial \nabla_0 \vectorn{z}} \right) \cdot \delta\vectorn{z} \dif \varOmega_\pm . \label{eq:varPiSt}
\end{equation}
The integrals over boundaries $\varGamma_\pm$ in equation \eqref{eq:varPiSt} can be split into the integrals over $\varSigma$ and the integrals over parts of the external boundary of the considered domain, resulting in
\begin{equation*}
  \delta\varPi_+ + \delta\varPi_- = \int_\varOmega \left(\ldots\right) \dif \varOmega + \int_{\varSigma} \vectorn{N}_* \cdot \left\llbracket\frac{\partial w}{\partial \nabla_0 \vectorn{z}} \cdot \delta\vectorn{z}\right\rrbracket \dif \varSigma + I_\varGamma ,
\end{equation*}
where $I_\varGamma$ is the integral over boundary $\varGamma$. It is assumed\footnote{When the weak form is obtained via the principle of virtual work, the closed form of $\varPi_\varGamma$ is not needed; one directly writes $\delta\varPi_\varGamma$ based on the boundary conditions of the specific problem.} that one can write such $\varPi_\varGamma$ that its variation is $\delta\varPi_\varGamma = -I_\varGamma$. Thus, the sum of all terms becomes (without $\delta\varPi_\mathrm{I}$)
\begin{align}
  &\delta\varPi - \delta\varPi_\mathrm{I} = \int_\varOmega \left( \frac{\partial w}{\partial \vectorn{z}} - \nabla_0\cdot\frac{\partial w}{\partial \nabla_0 \vectorn{z}} \right) \cdot \delta\vectorn{z} \dif \varOmega +   \int_{\varSigma} \vectorn{N}_* \cdot \left\llbracket\frac{\partial w}{\partial \nabla_0 \vectorn{z}} \right\rrbracket \cdot \left\langle \delta\vectorn{z} \right\rangle \dif \varSigma + 
  \vphantom{a} \nonumber \\ &\vphantom{a} \hspace{4cm} +
  \int_\varSigma \left( \beta \llbracket\delta\vectorn{z}\rrbracket - \vectorn{N}_* \cdot \left\langle \tensor{K}\left(\delta\vectorn{z}, \nabla_0 \delta\vectorn{z} \right) \right\rangle \right) \cdot \llbracket\vectorn{z}\rrbracket \dif \varSigma .
\end{align}
Since $\delta\vectorn{z}$ is arbitrary, to enforce $\delta\varPi = 0$, ignoring $\delta\varPi_\mathrm{I}$ for now, the terms in each integrand must be zero. This gives the following corresponding strong form:
\begin{equation}
   \frac{\partial w}{\partial \vectorn{z}} - \nabla_0\cdot\frac{\partial w}{\partial \nabla_0 \vectorn{z}} = \vectorn{0} , \quad\quad\quad\quad \vectorn{X}\in\varOmega_\pm , \label{eq:GenStrongPDE}
\end{equation}
with interface conditions
\begin{equation}
  \llbracket\vectorn{z}\rrbracket = \vectorn{0} , \quad\quad 
  \vectorn{N}_*\cdot\left\llbracket\frac{\partial w}{\partial \nabla_0 \vectorn{z}} \right\rrbracket = \vectorn{0} , \quad\quad \vectorn{X}\in\varSigma . \label{eq:GenStrongBC}
\end{equation}

\subsection{Consistency between the strong and the weak forms for magneto-mechanics}
\label{sec:consist}

\subsubsection{Enforcing the magnetisation length using the Lagrange multiplier}
\label{sec:larg}

The constant length of the magnetisation field can be enforced using various methods. Since the present paper aims at addressing the entire problem weakly, it is reasonable to enforce $\left|\vectorn{m}\right| = m_\mathrm{S}$ also weakly via the Lagrange multiplier, which is further denoted as $\lambda$. Thus, energy density $w$ is written as a function of the unknowns and their gradients:
\begin{equation*}
  w = w \left( \vectorn{u}, \vectorn{m}, \lambda, \nabla_0 \vectorn{u}, \nabla_0 \vectorn{m}, \nabla_0 \eta \right) .
\end{equation*}
In this case, unknown vector field $\vectorn{z}\in\mathbb{R}^8$ consists of components\footnote{Formally, if $\vectorn{e}_i$ are the basis vectors of $\mathbb{R}^8$, then $\vectorn{z} = u_1 \vectorn{e}_1 + u_2 \vectorn{e}_2 + u_3 \vectorn{e}_3 + m_1 \vectorn{e}_4 + m_2 \vectorn{e}_5 + m_3 \vectorn{e}_6 + \eta \vectorn{e}_7 + \lambda \vectorn{e}_8$.} of $\vectorn{u}$, $\vectorn{m}$, $\eta$, $\lambda$.

It is proposed to consider the following expression for the energy density:
\begin{equation}
  w = \rho_0\psi - \rho_0\vectorn{b}\cdot\vectorn{u} + w_\mathrm{M} + w_\mathrm{L} , \quad\quad\quad\quad  
  w_\mathrm{L} = \lambda \left( \vectorn{m}\cdot\vectorn{m} - m_\mathrm{S}^2 \right) , \label{eq:bigE}
\end{equation}
where $w_\mathrm{M}$ is the electromagnetic energy per unit of volume in the reference configuration (as if there was no material) with the following form:
\begin{align}
  &w_\mathrm{M} = -\tfrac{1}{2}\mu_0^{-1} J\vectorn{B}\cdot\vectorn{B} = -\tfrac{1}{2} \mu_0^{-1}J^{-1} \tensor{C} : \vectorn{B}_0 \vectorn{B}_0 = \nonumber \\
  &\vphantom{a} \hspace{3cm} = -\tfrac{1}{2} \mu_0 J \tensor{C}^{-1} : \nabla_0\eta \nabla_0\eta + \mu_0 \rho_0 \tensor{F}^{-1} : \vectorn{m} \nabla_0\eta - \tfrac{1}{2} \mu_0 \rho_0^2 m_\mathrm{S}^2 J^{-1} . \label{eq:magDens}
\end{align}
For simplicity, in the expression above, it is assumed that $\vectorn{b}$ is independent of the unknown variables; however, it is easy to construct the generalisation.

Now, it is necessary to show that the weak problem from section \ref{sec:weak} with the energy density given by equation \eqref{eq:bigE} results in the strong form from section \ref{sec:stat}.

First, the derivatives of the first term are found\footnote{A double-dot product with $\tensorFour{I}\transpRight$ transposes a second-order tensor, $\tensorFour{I}\transpRight : \tensor{A} = \tensor{A} : \tensorFour{I}\transpRight = \tensor{A}\transp$. The derivative of a second-order tensor by itself is $\partial\tensor{A}/\partial\tensor{A} = \tensorFour{I}\transpRight$.}:
\begin{align*}
  &\frac{\partial \left( \rho_0\psi \right)}{\partial \nabla_0 \vectorn{u}} = \rho_0 \frac{\partial \psi}{\partial \tensor{F}} : \tensorFour{I}\transpRight = \rho_0 \left( \frac{\partial \psi}{\partial \tensor{C}} : \frac{\partial \tensor{C}}{\partial \tensor{F}} + \frac{\partial \psi}{\partial \vectorn{m}_0} \cdot \frac{\partial \vectorn{m}_0}{\partial \tensor{F}} \right) : \tensorFour{I}\transpRight = \vphantom{a} \\
  &\vphantom{a} \hspace{4cm} = \rho_0 \left( 2\tensor{F}\cdot\frac{\partial \psi}{\partial \tensor{C}} + \vectorn{m} \frac{\partial \psi}{\partial \vectorn{m}_0} \right) : \tensorFour{I}\transpRight = \tensor{P}\transp , \\
  &\frac{\partial \left( \rho_0\psi \right)}{\partial \nabla_0 \vectorn{m}} = \rho_0 \frac{\partial \psi}{\partial \tensor{M}} : \frac{\partial \tensor{M}}{\partial \nabla_0 \vectorn{m}} = 2 \rho_0 \frac{\partial \psi}{\partial \tensor{M}} \cdot \nabla_0 \vectorn{m} = \tensor{B}\transp , \\
  &\frac{\partial \left( \rho_0\psi \right)}{\partial \vectorn{m}} = \rho_0 \frac{\partial \psi}{\partial \vectorn{m}_0} \cdot \frac{\partial \vectorn{m}_0}{\partial \vectorn{m}} = \rho_0 \frac{\partial \psi}{\partial \vectorn{m}_0} \cdot \tensor{F}\transp = -\rho_0 \vectorn{B}_\mathrm{L} .
\end{align*}
Other derivatives of $\rho_0\psi$ are obviously zeros. Next equation \eqref{eq:quasB} must be substituted into \eqref{eq:quasT} to obtain the Maxwell stress tensor in the reference configuration:
\begin{align*}
  &\tensor{\tau}_{\mathrm{E}0} = J \tensor{F}^{-1} \cdot \tensor{\tau}_\mathrm{E} = \mu_0 J \tensor{C}^{-1} \cdot \nabla_0\eta \nabla_0\eta \cdot \tensor{F}^{-1} - \mu_0 \rho_0 \tensor{F}^{-1} \cdot \vectorn{m} \nabla_0\eta \cdot \tensor{F}^{-1} - \vphantom{a} \\
  &\vphantom{a} \hspace{3cm} - \tfrac{1}{2} \mu_0 J \tensor{F}^{-1} \tensor{C}^{-1} : \nabla_0\eta \nabla_0\eta +  \tfrac{1}{2} \mu_0 \rho_0^2 m_\mathrm{S}^2 J^{-1} \tensor{F}^{-1} .
\end{align*}
It is not too difficult to check that\footnote{To do this it is useful to remember tensor derivatives $\partial \determ{\tensor{A}} / \partial\tensor{A} = \determ{\tensor{A}}\tensor{A}\transpm $ and \\$\partial \tensor{A}^{-1} / \partial\tensor{A} = -\tensor{A}^{-1} \cdot \tensorFour{I}\transpRight \cdot \tensor{A}\transpm$ for arbitrary second-order tensor $\tensor{A}$.}
\begin{equation*}
  \frac{\partial w_\mathrm{M}}{\partial \nabla_0 \vectorn{u}} = \tensor{\tau}_{\mathrm{E}0} , \quad\quad\quad\quad 
  \frac{\partial w_\mathrm{M}}{\partial \vectorn{m}} = -\rho_0 J^{-1}\tensor{F} \cdot\vectorn{B}_0 + \mu_0 \rho_0^2 J^{-1} \vectorn{m} , \quad\quad\quad\quad  
  \frac{\partial w_\mathrm{M}}{\partial \nabla_0\eta} = \vectorn{B}_0 .
\end{equation*}
Other derivatives of $w_\mathrm{M}$ are zeros. Finally, the non-zero derivatives of $w_\mathrm{L}$ are 
\begin{equation*}
  \frac{\partial w_\mathrm{L}}{\partial \vectorn{m}} = 2 \lambda \vectorn{m} , \quad\quad\quad\quad 
  \frac{\partial w_\mathrm{L}}{\partial \lambda} = \vectorn{m}\cdot\vectorn{m} - m_\mathrm{S}^2 ,
\end{equation*}
which finalises the set of required derivatives.

Assembling the derivatives of the terms of $w$ by $\nabla_0 \vectorn{u}$ and by $\vectorn{u}$ together and substituting them into equations \eqref{eq:GenStrongPDE} and \eqref{eq:GenStrongBC} directly results in equations \eqref{eq:quasMechP} and \eqref{eq:quasMechB} --- the first part of the strong problem. Equation \eqref{eq:quasMagP} has the form of $\vectorn{m} \times \left(\ldots\right) = 0$, which is equivalent to $\left(\ldots\right) = a\vectorn{m}$, where $a$ is some scalar. As above, substituting the derivatives of $w$ by $\nabla_0 \vectorn{m}$ and by $\vectorn{m}$ into equations \eqref{eq:GenStrongPDE} and \eqref{eq:GenStrongBC} results in such form of equation \eqref{eq:quasMagP} with interface conditions \eqref{eq:quasMagB}. Scalar $a$ is obtained via the magnetisation normalisation equation, which results from substituting the derivative of $w$ by $\lambda$ into equation \eqref{eq:GenStrongPDE}. This gives the second part of the strong problem. Finally, substituting the derivative of $w$ by $\nabla_0\eta$ into equations \eqref{eq:GenStrongPDE} and \eqref{eq:GenStrongBC} results in equations \eqref{eq:quasFieP} and \eqref{eq:quasFieB} --- the third part of the strong problem. This proves that the weak form with $w$ given by equation \eqref{eq:bigE} and the strong form are consistent.

\subsubsection{Representing magnetisation via spherical angles}

Sometimes, it may be convenient to represent the magnetisation via the spherical angles and thereby to enforce automatically the constant magnetisation length. In this case, energy density $w$ is written as a function of the following unknowns and their gradients:
\begin{equation*}
  w = w \left( \vectorn{u}, \vectorn{m}, \nabla_0 \vectorn{u}, \nabla_0 \vectorn{m}, \nabla_0 \eta \right) = w \left( \vectorn{u}, \alpha, \varphi, \nabla_0 \vectorn{u}, \nabla_0 \alpha, \nabla_0 \varphi, \nabla_0 \eta \right) ,
\end{equation*}
where angles $\alpha\in\left[-\pi/2;\pi/2\right]$ and $\varphi\in\left[0;2\pi\right]$ determine the magnetisation:
\begin{equation}
  \vectorn{m} = m_\mathrm{S} \cos\alpha \cos\varphi \,\vectorn{e}_1 + m_\mathrm{S} \cos\alpha \sin\varphi \,\vectorn{e}_2 + m_\mathrm{S} \sin\alpha \,\vectorn{e}_3 .
  \label{eq:magAng}
\end{equation}
Of course, $w$ cannot depend on $\alpha$, $\varphi$, and their gradients in an arbitrary way, and is written as a function of $\vectorn{m}$ and its gradient (more precisely, as a function of $\vectorn{m}_0$ and $\tensor{M}$ to adhere to the principle of objectivity, as outlined in section \ref{sec:constitut}), which are, in turn, functions of $\alpha$, $\varphi$, and their gradients.

Such approach is especially advantageous if it is \emph{a priori} known that the magnetisation is planar, reducing the number of corresponding variables from $4$ to $1$. However, the significant disadvantage of this approach is the potential problematic convergence of the Newton-Raphson method when the non-linear problem is solved --- for points with $\alpha=\pm\pi/2$ (or close), the Jacobian's rows corresponding to $\varphi$ will be zero (or close-to-zero), leading to ill-conditionality. This is easy to see from $\partial \vectorn{m} / \partial\varphi$ being zero at $\alpha=\pm\pi/2$. 

It is proposed to consider the following expression for the energy density:
\begin{equation}
  w = \rho_0\psi - \rho_0\vectorn{b}\cdot\vectorn{u} + w_\mathrm{M} ,
  \label{eq:bigEang}
\end{equation}
which is similar to expression \eqref{eq:bigE} but without the Lagrange multiplier.

Now, it is necessary to show that the weak problem from section \ref{sec:weak} with the energy density given by equation \eqref{eq:bigEang} and the magnetisation given by equation \eqref{eq:magAng} results in the strong form from section \ref{sec:stat}.

Obviously, derivatives of $w$ by $\vectorn{u}$, $\nabla_0 \vectorn{u}$, $\nabla_0 \eta$ do not change compared to section \ref{sec:larg}, resulting in the first and the third parts of the strong problem. What remains is to obtain the second part of the strong problem --- equation \eqref{eq:quasMagP} with interface conditions \eqref{eq:quasMagB}.

First, it is useful to write the magnetisation gradient:
\begin{align*}
  &\nabla_0\vectorn{m} = -m_\mathrm{S} \sin\alpha \cos\varphi \nabla_0\alpha \,\vectorn{e}_1 - m_\mathrm{S} \cos\alpha \sin\varphi \nabla_0\varphi \,\vectorn{e}_1 - m_\mathrm{S} \sin\alpha \sin\varphi \nabla_0\alpha \,\vectorn{e}_2 + \vphantom{a} \\
  &\hspace{1.5cm} \vphantom{a} + m_\mathrm{S} \cos\alpha \cos\varphi \nabla_0\varphi \,\vectorn{e}_2 + m_\mathrm{S} \cos\alpha \nabla_0\alpha\,\vectorn{e}_3 .
\end{align*}
The derivatives with respect to the angles can be easily written using the chain rule:
\begin{equation*}
  \frac{\partial w}{\partial \alpha} = \frac{\partial w}{\partial \vectorn{m}}\cdot \frac{\partial \vectorn{m}}{\partial \alpha} + \frac{\partial w}{\partial \nabla_0\vectorn{m}}:\frac{\partial \left(\nabla_0\vectorn{m}\right)\transp}{\partial \alpha} ,
  \quad\quad
  \frac{\partial w}{\partial \varphi} = \frac{\partial w}{\partial \vectorn{m}}\cdot \frac{\partial \vectorn{m}}{\partial \varphi} + \frac{\partial w}{\partial \nabla_0\vectorn{m}}:\frac{\partial \left(\nabla_0\vectorn{m}\right)\transp}{\partial \varphi} .
\end{equation*}
Similarly,
\begin{align}
  &\frac{\partial w}{\partial \nabla_0\alpha} = \frac{\partial w}{\partial \nabla_0\vectorn{m}}:\frac{\partial \left(\nabla_0\vectorn{m}\right)\transp}{\partial \nabla_0\alpha} = \frac{\partial w}{\partial \nabla_0\vectorn{m}}\cdot\frac{\partial\vectorn{m}}{\partial\alpha} , \label{eq:derWang1} \\
  &\frac{\partial w}{\partial \nabla_0\varphi} = \frac{\partial w}{\partial \nabla_0\vectorn{m}}:\frac{\partial \left(\nabla_0\vectorn{m}\right)\transp}{\partial \nabla_0\varphi} = \frac{\partial w}{\partial \nabla_0\vectorn{m}}\cdot\frac{\partial\vectorn{m}}{\partial\varphi} , \label{eq:derWang2}
\end{align}
where the second equality signs in the equations follow from the direct evaluation of $\nabla_0\vectorn{m}$ by the gradients of the angles. Substituting the derivatives into equation \eqref{eq:GenStrongPDE} results in
\begin{equation*}
  \left( \frac{\partial w}{\partial \vectorn{m}} - \nabla_0 \cdot \frac{\partial w}{\partial \nabla_0\vectorn{m}} \right) \cdot \frac{\partial \vectorn{m}}{\partial \alpha} = 0 ,
  \quad\quad
  \left( \frac{\partial w}{\partial \vectorn{m}} - \nabla_0 \cdot \frac{\partial w}{\partial \nabla_0\vectorn{m}} \right) \cdot \frac{\partial \vectorn{m}}{\partial \varphi} = 0 .
\end{equation*}
These are two scalar equations enforcing the components of the expression in brackets that are perpendicular to $\vectorn{m}$ to be zeros. They are equivalent to 
\begin{equation*}
  \vectorn{m}\times\left( \frac{\partial w}{\partial \vectorn{m}} - \nabla_0 \cdot \frac{\partial w}{\partial \nabla_0\vectorn{m}} \right) = \vectorn{0} .
\end{equation*}
Substituting the already-known derivatives of $w$ by $\vectorn{m}$ and $\nabla_0\vectorn{m}$, results in equation \eqref{eq:quasMagP}. Substituting derivatives \eqref{eq:derWang1} and \eqref{eq:derWang2} into interface conditions \eqref{eq:GenStrongBC} results in
\begin{equation*}
  \llbracket\alpha\rrbracket = 0 , \quad\quad
  \llbracket\varphi\rrbracket = 0 , \quad\quad
  \vectorn{N}_*\cdot\left\llbracket\frac{\partial w}{\partial \nabla_0\vectorn{m}} \right\rrbracket\cdot\frac{\partial\vectorn{m}}{\partial\alpha} = \vectorn{0} , \quad\quad
  \vectorn{N}_*\cdot\left\llbracket\frac{\partial w}{\partial \nabla_0\vectorn{m}} \right\rrbracket\cdot\frac{\partial\vectorn{m}}{\partial\varphi} = \vectorn{0} ,
\end{equation*}
where the continuity of derivatives of $\vectorn{m}$ by $\alpha$ and $\varphi$ due to the continuity of $\alpha$ and $\varphi$ is used. Since $\vectorn{m}\cdot\tensor{B} = \vectorn{0}$, it is sufficient to enforce the components of $\vectorn{N}_*\cdot\llbracket\tensor{B}\rrbracket\transp$ that are perpendicular to $\vectorn{m}$ to be zeros. This gives interface conditions \eqref{eq:quasMagB} and proves that the weak form with $w$ given by equation \eqref{eq:bigEang} and the strong form are consistent.

\subsection{Quasistatic problem and near-equilibrium kinetics from variational approach}
\label{sec:variat}

The approach outlined above (i.e.\ considering thermodynamic balance laws) results in the entropy production due to the phase boundary propagation, from which the thermodynamic driving force is obtained. The approach is general --- it covers the dynamics and arbitrary material constitutive behaviour.

There is also a complementary approach based on the variational principle and valid for non-dissipative quasistatic continua. It considers phase boundaries at the equilibrium. In this case, the variation of the total energy functional should vanish. Deriving such variation, while also varying the positions of the phase boundaries, gives the quantity that must be zero at the equilibrium. Taken with an appropriate sign, such quantity can be viewed as the driving force --- it drives the phase boundaries towards the equilibrium and is zero at the equilibrium. For example, work \cite{Grinfeld1980} was one of the first to write this for non-linear elasticity (see the introduction of \cite{Grabovsky2011} for a historical perspective).

It is important to emphasise that the approaches are complementary. In the case of non-linear elasticity, they result in the same expression (denoted as $\mathcal{H}$ for brevity). Then, from the thermodynamic approach, it can be seen that the expression for phase boundary velocity $W_*$ should be chosen as a function of $\mathcal{H}$, also ensuring that the entropy production is non-negative. While from the variational approach, it can be seen that $\mathcal{H}$ must be zero when $W_*$ is zero. Combined together, this motivates the choice of $W_*$ being e.g.\ a linear function of $\mathcal{H}$.

\subsubsection{General problem}

The minimisation of a functional of field $\vectorn{z}\in\mathbb{R}^n$ defined on a domain with an interface is a well-known problem, see e.g.\ \cite{Grabovsky2011}. The problem setting is similar to section \ref{sec:weak}:
\begin{equation}
  \varPi = \varPi_+ + \varPi_- , 
  \qquad\qquad \varPi_\pm = \int_{\varOmega_\pm} w \dif \varOmega_\pm ,
  \qquad\qquad w = w \left( \vectorn{z}, \nabla_0 \vectorn{z} \right) ,
\end{equation}
but now interface position $\vectorn{X}_* = \big\lbrace \vectorn{X} \, \big| \, \vectorn{X} \in \varSigma \big\rbrace$ is an unknown quantity, i.e.\ it will be varied. Field $\vectorn{z}$ is taken to be continuous across interface $\varSigma$ right from the start (as opposed to the weak problem formulation from section \ref{sec:weak}, where two solutions $\vectorn{z}_\pm$ are `glued' weakly, which is done for further numerical handling of the problem). The total energy functional consists of only the volumetric integrals of the energy densities of the phases and has the physical meaning of the free energy.

The same manipulations as in section \ref{sec:weak} lead to
\begin{align}
  &\delta\varPi = \int_\varOmega \left( \frac{\partial w}{\partial \vectorn{z}} - \nabla_0\cdot\frac{\partial w}{\partial \nabla_0 \vectorn{z}} \right) \cdot \delta\vectorn{z} \dif \varOmega + \int_{\varSigma} \vectorn{N}_* \cdot \left\llbracket\frac{\partial w}{\partial \nabla_0 \vectorn{z}} \right\rrbracket \cdot \left\langle \delta\vectorn{z} \right\rangle \dif \varSigma + 
  \vphantom{a} \nonumber \\ &\vphantom{a} \hspace{4cm} +
  \int_{\varSigma} \vectorn{N}_* \cdot \left\langle \frac{\partial w}{\partial \nabla_0 \vectorn{z}} \right\rangle \cdot \left\llbracket \delta\vectorn{z} \right\rrbracket \dif \varSigma + \int_{\varSigma} \left\llbracket w \right\rrbracket \vectorn{N}_* \cdot \delta\vectorn{X}_*  \dif \varSigma ,
  \label{eq:varPrincPi}
\end{align}
where the considered variations are such that $\delta\vectorn{z} = \vectorn{0}$ on external boundary $\varGamma$. The last term results from the variation of $\vectorn{X}_*$ in the same way as in the case of the time derivative of the energy, employing relation \eqref{eq:genBalR}. It can be easily obtained by writing the definition of the functional differential and treating the amplitude of $\delta\vectorn{z}$ as time. Analogously to relation \eqref{eq:genCompat}, 
\begin{equation}
  \llbracket \delta\vectorn{z} \rrbracket = -\delta\vectorn{X}_* \cdot \llbracket\nabla_0\vectorn{z}\rrbracket .
  \label{eq:varPrincJump}
\end{equation}
Equality $\delta\varPi = 0$ for all possible $\delta\vectorn{z}$ and $\delta\vectorn{X}_*$ then leads to strong problem \eqref{eq:GenStrongPDE} with interface conditions \eqref{eq:GenStrongBC}, which follows from the first and the second terms of equation \eqref{eq:varPrincPi}. Upon substitution of equation \eqref{eq:varPrincJump}, the third and the fourth terms, in turn, lead to the so-called second Weierstrass-Erdmann condition \cite{Grabovsky2011}:
\begin{equation}
  \left\llbracket w \right\rrbracket - \left\langle \frac{\partial w}{\partial \nabla_0 \vectorn{z}} \right\rangle : \llbracket\nabla_0\vectorn{z}\rrbracket\transp = 0 , \quad\quad \vectorn{X}\in\varSigma . \label{eq:intWeier}
\end{equation}

\subsubsection{Magneto-mechanics}

It transpires that if one uses expression \eqref{eq:bigE} for $w$, then the left-hand side of condition \eqref{eq:intWeier} directly leads to the expression in brackets from inequality \eqref{eq:intJumpMain} --- the thermodynamic driving force. Indeed, direct substitution of equation \eqref{eq:bigE} and its derivatives into \eqref{eq:intWeier}, taking $\vectorn{m}\cdot\vectorn{m} \equiv m_\mathrm{S}^2$, gives
\begin{equation}
  \rho_0 \llbracket \psi \rrbracket + \llbracket w_\mathrm{M} \rrbracket - \langle \tensor{P} \rangle\transp : \llbracket\tensor{F}\rrbracket - \langle \tensor{\tau}_{\mathrm{E}0} \rangle : \llbracket\tensor{F}\rrbracket - \langle \tensor{B} \rangle\transp : \llbracket\nabla_0\vectorn{m}\rrbracket\transp - \langle \vectorn{B}_0 \rangle \cdot \llbracket\nabla_0\eta\rrbracket = 0 .
\end{equation}
Through derivations outlined in appendix \ref{sec:addTran}, it is possible to show that 
\begin{align}
  &\llbracket w_\mathrm{M} \rrbracket - \langle \tensor{\tau}_{\mathrm{E}0} \rangle : \llbracket\tensor{F}\rrbracket - \langle \vectorn{B}_0 \rangle \cdot \llbracket\nabla_0\eta\rrbracket = \tfrac{1}{4} \mu_0^{-1} \llbracket J \rrbracket \llbracket\vectorn{B}\rrbracket \cdot \llbracket\vectorn{B}\rrbracket - \tfrac{1}{2} \llbracket J \rrbracket \llbracket J^{-1} \rrbracket \rho_0 \vectorn{m}\cdot \llbracket\vectorn{B}\rrbracket =
  \vphantom{a} \nonumber \\ &\vphantom{a} \hspace{8cm} =
  -\tfrac{1}{4} \llbracket J \rrbracket \llbracket J^{-1} \rrbracket \rho_0 \vectorn{m}\cdot \llbracket\vectorn{B}\rrbracket .
\end{align}
Substituting the latter into the former gives
\begin{equation}
  \rho_0 \llbracket \psi \rrbracket - \tfrac{1}{4} \llbracket J \rrbracket \llbracket J^{-1} \rrbracket \rho_0 \vectorn{m}\cdot \llbracket\vectorn{B}\rrbracket - \llbracket\tensor{F}\rrbracket\transp : \langle \tensor{P} \rangle - \llbracket\nabla_0\vectorn{m}\rrbracket : \langle \tensor{B} \rangle = 0 ,
\end{equation}
the left-hand side of which is the driving force from inequality \eqref{eq:intJumpMain}. Thus, the variational approach leads to (a) the strong form of the quasistatic problem and (b) the condition that the thermodynamic driving force must be zero when the phase boundaries are at the equilibrium. 

This also means that in the quasistatic non-dissipative case, the problem of finding the equilibrium configuration of the phase boundaries is the minimisation problem for the total free energy, consisting of the Helmholtz free energy of the spin-lattice continuum (that depends on the deformation gradient, the magnetisation, and the magnetisation gradient), the electromagnetic energy of the space occupied by the continuum (that depends on the deformation gradient, the magnetisation, and the gradient of the magnetic scalar potential), and the potential corresponding to the volumetric mechanical forces (e.g.\ gravity), with an optional Lagrange-multiplier-term to enforce the constant magnetisation length if needed. It is important to emphasise that the minimisation problem is written specifically with respect to variables $\vectorn{u}$, $\vectorn{m}$, and $\eta$.

\subsection{Constitutive law for magnetic shape-memory alloys}

To perform simulations of applied problems, the final element is needed --- a constitutive law for the bulk material behaviour in the form of the Helmholtz free energy. For the application to MSMAs outlined in the introduction, the following energy expression is proposed:
\begin{equation}
  \rho_0 \psi = \tfrac{1}{2} \left( K_\mathrm{e} - \tfrac{2}{3} G_\mathrm{e} \right) \left( \tensor{E}:\tensor{I} \right)^2 + G_\mathrm{e} \left( \tensor{E}:\tensor{E} \right) + \tfrac{1}{2} A_\mathrm{m} \left( \tensor{M}:\tensor{I} \right) - \tfrac{1}{2} K_\mathrm{m} \left( \vectorn{m}_0 \cdot \tensor{C}^{-1} \cdot \vectorn{p} \right)^2 ,
  \label{eq:constLaw}
\end{equation}
where $K_\mathrm{e}$ and $G_\mathrm{e}$ are the bulk and shear moduli, $A_\mathrm{m}$ is the ferromagnetic exchange parameter, $K_\mathrm{m}$ is magnetocrystalline anisotropy parameter, $\vectorn{p}$ is magnetocrystalline anisotropy axis, and $\tensor{E}$ is the Green-Lagrange strain: $\tensor{E} = \tfrac{1}{2} \left( \tensor{C} - \tensor{I} \right)$. The first two energy terms correspond to the standard elasticity (although non-linear to adhere to the objectivity principle). The third term is the standard ferromagnetic exchange. The final term provides magneto-mechanical coupling. The structure of the term is specifically selected such that when magnetisation $\vectorn{m}_0$ aligns with $\vectorn{p}$, the preferential state of the material is to contract along $\vectorn{p}$. This behaviour will be further demonstrated in the numerical examples.

\section{Computational approach based on cut elements}
\label{sec:comput}

The cut-finite-element method (CutFEM) is now widespread, which means that the description of how it is adapted to the magneto-mechanical coupled problem can be brief. The main ingredient of the approach is obtaining the weak form of the problem to be solved, where the interface/boundary conditions are enforced weakly, which was already done in sections \ref{sec:weak} and \ref{sec:consist}.

After the weak form is obtained, a finite-element mesh is introduced, as well as the standard nodal basis functions. Following the finite-element formulation, the system of non-linear algebraic equations with respect to the unknown nodal degrees of freedom is written. Although these two steps are rather straightforward and have been outlined many times in literature, e.g.\ sections 4.1.4 and 4.1.5 of \cite{Poluektov2019}, it is useful to repeat the summary of the finite-element formulation, aligning the notation with the present paper for the completeness.

\subsection{Finite-element formulation}

Up to now, arbitrary domain $\varOmega$ has been considered. For the finite-element formulation, the entire domain of the body must be considered; therefore, to avoid introducing new notation, $\varOmega$ will stand for the entire domain of the body in this section. The finite-element mesh covers the entire volume of the body $\varOmega = \varOmega_+ \cup \varOmega_-$ and is arbitrary with respect to interface $\varSigma$. Without loss of generality, the mesh is considered to be conforming to boundary $\varGamma$, which becomes the external boundary of the body in this section. 

The mesh contains $N$ nodes. The standard nodal basis function associated with node $i$ is denoted as $\nu_i$. These functions are continuous, piecewise-polynomial, are equal to $1$ at node $i$, and are equal to $0$ at all other nodes. The space of the standard nodal basis functions is denoted as
\begin{equation*}
  \mathcal{S}^h = \operatorname{span}\left\lbrace \nu_i \right\rbrace_{i=1}^N , \quad\quad
  \nu_i = \nu_i\big(\vectorn{X}\big) .
\end{equation*}
The following additional space is introduced:
\begin{equation*}
  \mathcal{Q}^h = \left\lbrace \vectorn{\chi} \, \big| \, \chi_j \in \mathcal{S}^h \right\rbrace , \quad
  j \in \left\lbrace 1,\ldots,n \right\rbrace ,
\end{equation*}
where $n$ is the number of degrees of freedom (as previously introduced in section \ref{sec:weak}) per node, and subscript $j$ denotes a component of a vector. Thus, space $\mathcal{Q}^h$ will contain finite-element solutions and finite-element test functions, each component of which belongs to space $\mathcal{S}^h$ of the standard nodal basis functions.

Globally-defined functions $\vectorn{z}_+^h, \vectorn{z}_-^h \in \mathcal{Q}^h$ are introduced. Restrictions of functions $\vectorn{z}_+^h$ and $\vectorn{z}_-^h$ to domains $\varOmega_+$ and $\varOmega_-$, respectively, approximate solutions $\vectorn{z}_+$ and $\vectorn{z}_-$, respectively.

The set of all elements is denoted as $\mathcal{T}$. Furthermore, the following sets can be defined:
\begin{equation*}
  \mathcal{T}_\pm = \left\lbrace E \, \big| \, E \in \mathcal{T}, \, E \cap \varOmega_\pm^h \neq \emptyset \right\rbrace ,
\end{equation*}
where $E$ denotes an element. Sets $\mathcal{T}_+$ and $\mathcal{T}_-$ overlap, i.e.\ they share the set of elements, which are intersected by $\varSigma^h$. The set of all elements intersected by the interface is denoted as 
\begin{equation*}
  \mathcal{T}_* = \mathcal{T}_+ \cap \mathcal{T}_- .
\end{equation*}
Here, superscript $h$ is added to $\varOmega_\pm$ and $\varSigma$ to indicate the discretisation of the boundaries of the domains.

Thus, the finite-element formulation of the problem is the following:
\begin{align*}
  &\text{find } \vectorn{z}_+^h , \vectorn{z}_-^h \in \mathcal{Q}^h \text{ such that} \quad \delta \varPi \left( \vectorn{z}_+^h, \vectorn{z}_-^h, \delta\vectorn{z}_+^h, \delta\vectorn{z}_-^h \right) = 0 , \quad
  \forall \delta\vectorn{z}_+^h , \delta\vectorn{z}_-^h \in \mathcal{Q}^h , \\
  &\vectorn{z}_\pm^h = \vectorn{0} \text{ at nodes that do not belong to elements } \mathcal{T}_\pm ,
\end{align*}
where functional $\delta \varPi$ was obtained in section \ref{sec:weak}, and $\delta\vectorn{z}_\pm^h$ fulfil the role of the finite-element test functions (with superscript $h$ added to indicate that they are restricted to space $\mathcal{Q}^h$). Numerical parameter $\beta$ in equation \eqref{eq:piStz}, which is introduced in the Nitsche-like methods, is usually taken to be inversely proportional to spatial step $h$,
\begin{equation}
  \beta = \frac{\kappa_\mathrm{p}}{h} ,
\end{equation}
where $\kappa_\mathrm{p}$ is the so-called penalty parameter. The above finite-element problem represents a system of non-linear algebraic equations with respect to nodal values of $\vectorn{z}_+^h$ and $\vectorn{z}_-^h$, which can be solved using the standard Newton-Raphson method.

\subsection{Stabilisation term}
\label{sec:stab}

Since the interface can partition some elements into highly unequal area fractions, additional numerical stabilisation terms are introduced in CutFEM to avid ill-conditionality of the system of equations \cite{Burman2012}. The stabilisation term outlined below is valid only for linear elements; for higher-order elements, the stabilisation term should contain the normal derivatives of the higher order \cite{Hansbo2017}, up to the order of the elements. It should be noted that the code supplementing the present paper is based on linear elements.

The set of all element boundaries is denoted as $\mathcal{F}$. To construct the stabilisation term, it is necessary to introduce the set of boundaries of all elements that are intersected by the interface:
\begin{equation*}
  \mathcal{F}_* = \left\lbrace F \, \big| \, F \in \mathcal{F}, \, E \in \mathcal{T}_*, \, F \cap E \neq \emptyset \right\rbrace .
\end{equation*}
Next, it is necessary to define the sets of element boundaries that have two adjacent elements from the same element set ($\mathcal{T}_+$ or $\mathcal{T}_-$):
\begin{equation*}
  \mathcal{F}_\pm^* = \left\lbrace F \, \big| \, F \in \mathcal{F_*}, \, \exists E_1 \in \mathcal{T}_\pm, \, \exists E_2 \in \mathcal{T}_\pm, \, F \cap E_1 \neq \emptyset, \, F \cap E_2 \neq \emptyset \right\rbrace .
\end{equation*}
The stabilisation term is added as an additional term $\varPi_\mathrm{I}$ to the energy:
\begin{equation}
  \varPi_\mathrm{I} = \sum_{\varGamma_\mathrm{f}\in\mathcal{F}_+^*} \frac{\kappa_\mathrm{s} h}{2} \int_{\varGamma_\mathrm{f}} \llbracket \vectorn{N}_\mathrm{f} \cdot \nabla_0 \vectorn{z}_+ \rrbracket_\mathrm{e}^2 \dif \varGamma_\mathrm{f} + 
  \sum_{\varGamma_\mathrm{f}\in\mathcal{F}_-^*} \frac{\kappa_\mathrm{s} h}{2} \int_{\varGamma_\mathrm{f}} \llbracket \vectorn{N}_\mathrm{f} \cdot \nabla_0 \vectorn{z}_- \rrbracket_\mathrm{e}^2 \dif \varGamma_\mathrm{f} ,
  \label{eq:numStabTermEnergy}
\end{equation}
where (with the slight abuse of the notation) the square of a vector implies a scalar product of the vector with itself, vector $\vectorn{N}_\mathrm{f}$ is the normal to boundary $\varGamma_\mathrm{f}$, scalar $\kappa_\mathrm{s}$ is the stabilisation parameter, and $\llbracket \cdot \rrbracket_\mathrm{e}$ denotes the jump of the quantity across the element boundary. The orientation of normal $\vectorn{N}_\mathrm{f}$ is not important, as the jump is squared. The variation of the stabilisation term leads to
\begin{align}
  &\delta\varPi_\mathrm{I} =
  \sum_{\varGamma_\mathrm{f}\in\mathcal{F}_+^*} \kappa_\mathrm{s} h \int_{\varGamma_\mathrm{f}} \llbracket \vectorn{N}_\mathrm{f} \cdot \nabla_0 \vectorn{z}_+ \rrbracket_\mathrm{e} \cdot 
  \llbracket \vectorn{N}_\mathrm{f} \cdot \nabla_0 \delta\vectorn{z}_+ \rrbracket_\mathrm{e} \dif \varGamma_\mathrm{f} +
  \vphantom{a}\nonumber\\&\vphantom{a} \hspace{1.5cm} + 
  \sum_{\varGamma_\mathrm{f}\in\mathcal{F}_-^*} \kappa_\mathrm{s} h \int_{\varGamma_\mathrm{f}} \llbracket \vectorn{N}_\mathrm{f} \cdot \nabla_0 \vectorn{z}_- \rrbracket_\mathrm{e} \cdot 
  \llbracket \vectorn{N}_\mathrm{f} \cdot \nabla_0 \delta\vectorn{z}_- \rrbracket_\mathrm{e} \dif \varGamma_\mathrm{f} .
  \label{eq:numStabTerm}
\end{align}

\subsection{Interface evolution and capturing arbitrary topology change}
\label{sec:topo}

The original interface evolution procedure was described in section 4.4 of \cite{Poluektov2019}. The original implementation\footnote{The first version of the 2D CutFEM MATLAB code aimed at non-linear chemo-mechanical coupled problems, was developed in autumn 2017 by the author of the present paper, and was demonstrated at ECCM 2018.} did include handling of the topology change, but the algorithm was long, hard-to-read, and relied on numerous conditional statements. Since then, the implementation has undergone various modifications, and a completely new algorithm for the interface evolution has been implemented, resulting in a short and robust code\footnote{Although developed in summer 2022 and shared via GitHub, up to now, it has been unpublished.}, as described below.

The implementation is limited to problems with 2D geometries. Hence, interface $\varSigma$ is either a single oriented curve or a disjoint set of oriented curves that cut through the finite-element mesh (the `oriented' property is important). If interface $\varSigma$ is given, then the first task is to construct a discretised version of the interface, denoted as $\varSigma^h$, also identifying which elements belong to which set (e.g.\ $\mathcal{T}_+$ or $\mathcal{T}_-$). This is done as follows.

First, the case when $\varSigma$ is either one or several closed curves is considered. The key idea is to check whether each node of the finite-element mesh falls inside or outside of $\varSigma$. It is safe to assume that $\varSigma$ is piecewise-linear (the lengths of linear segments can be arbitrarily small). This is a well-known problem in computational geometry and is called the `point-in-polygon' problem. An elegant solution is calculating the node's (point's) winding number with respect to the interface (polygon). For each linear segment of the interface, the angle between vectors $\vectorn{p}_1$ and $\vectorn{p}_2$ is calculated\footnote{The angle can be positive or negative --- it is calculated by how much $\vectorn{p}_1$ should be rotated to be aligned with $\vectorn{p}_2$. The counter-clockwise and the clockwise rotations correspond to the positive and the negative angles, respectively.}, where $\vectorn{p}_1$ is the vector connecting the node and the first point of the segment and $\vectorn{p}_2$ is the vector connecting the node and the second point of the segment. Angles corresponding to all linear segments are summed, resulting in the node's winding number. If it is $2\pi$ or greater\footnote{This statement assumes counter-clockwise orientation of the interface. For the clockwise orientation, the inside node will have the winding number of $-2\pi$ or smaller.}, then the node lies inside $\varSigma$; otherwise, it is outside.

If $\varSigma$ is not closed, e.g.\ a flat interface, the calculation results in the winding numbers with the absolute values smaller than $2\pi$. From the geometrical meaning of the winding number, it is then easy to choose a threshold that will clearly split the nodes into groups located at the different sides of the interface.

Once the inside/outside location of the nodes is known, it is also known which elements belong to $\mathcal{T}_+ \setminus \mathcal{T}_*$, which elements belong to $\mathcal{T}_- \setminus \mathcal{T}_*$, and which elements belong to $\mathcal{T}_*$. For each intersected element, interface $\varSigma$ inside is approximated by strictly one linear segment (keeping orientation consistent). A set of the constructed connected linear segments is the sought discretised interface $\varSigma^h$.

For the given discretised interface $\varSigma^h$, the PDEs are solved, resulting in the finite-element solution $\vectorn{z}^h$ at the nodes. Energy density $w^h$ and its derivatives by $\vectorn{z}^h$ and $\nabla_0\vectorn{z}^h$ are defined at the integration points of the elements. Standard inter-element averaging is used to obtain their values at the nodes. Using linear interpolation, the values of $w^h$, $\vectorn{z}^h$, $\partial w^h/\partial \vectorn{z}^h$, $\partial w^h/\partial \nabla_0\vectorn{z}^h$ are obtained at the intersection points of $\varSigma^h$ and the mesh. The latter are finally used to calculate the interface velocity at these intersection points. The interface is then evolved by moving the intersection points according to their velocities, resulting in new interface $\hat{\varSigma}^h$. The same procedure of `discretising' it again with the use of nodes' winding numbers is applied, giving the discretised interface at the new time step. This approach automatically handles self-intersections, interface splitting, interface merging, and other topological changes.

\section{Numerical examples}
\label{sec:num}

The purpose of this section is to show four examples demonstrating how the proposed approach has been implemented in the code:
\begin{itemize}
\item the first example focuses purely on micromagnetics (only magnetisation distribution) and demonstrates the mesh convergence;
\item the second example covers coupled micromagnetics-magnetostatics and illustrates the domain wall emergence due to the demagnetising field;
\item the third example focuses on purely mechanical behaviour and illustrates the automatic handling of the interface topology change by the code;
\item the final example provides the simulations of magnetically-driven and mechanically-driven structural phase transitions in MSMAs.
\end{itemize}
Since the focus of the present paper is on the theory, units are omitted for all the quantities and parameters are varied within large ranges, which is common in applied mathematics.

The multi-physics CutFEM implementation supplements the present paper and is publicly available via GitHub\footnote{https://github.com/mpoluektov/cutfem-multiphys-2d}. The previous versions of the code have been specifically tailored to non-linear mechanics and chemo-mechanics. For the present paper, the core part of the code has been re-written to make it an intrinsically multi-physics code. It now solves the general PDE in the form of equation \eqref{eq:GenStrongPDE} with interface conditions \eqref{eq:GenStrongBC}, where energy density $w$ is arbitrary and can be defined by the user in the \texttt{constLaw} section of the \texttt{calcRes} file.

The code relies on the structured finite-element mesh, where linear finite elements are used. All elements of the mesh are isosceles right triangles with side lengths of $h$. The modelled geometry is rectangular, with the user-defined dimensions. In this version of the code, \emph{four} Gauss integration points are used per element to avoid errors due to inaccurate integration. The latter is especially important for the case when the constant magnetisation length is enforced using the Lagrange multiplier (it is not difficult to check that one Gauss integration point leads to an ill-conditioned Jacobian in this case).

For the numerical examples below, the absolute tolerance of $10^{-11}$ is set in the Newton-Raphson method both for the $l^\infty$-norm of the solution update and the $l^\infty$-norm of the residual. The Nitsche penalty parameter $\kappa_\mathrm{p} = 10^4$ is used. The inter-element stabilisation parameter $\kappa_\mathrm{s} = 0.1$ is used for the first and third examples and $\kappa_\mathrm{s} = 1$ for the fourth example.

\subsection{Magnetic domain wall, mesh convergence}

The first example is a purely micromagnetic simulation to estimate the convergence rate numerically. There are four degrees of freedom --- three of the magnetisation vector and the Lagrange multiplier. Thus, the following strain energy density is used (same as in equations \eqref{eq:bigE} and \eqref{eq:constLaw}, but excluding $\vectorn{u}$ and $\eta$):
\begin{equation}
  w = \tfrac{1}{2} A_\mathrm{m} \left( \tensor{M}:\tensor{I} \right) - \tfrac{1}{2} K_\mathrm{m} \left( \vectorn{m} \cdot \vectorn{p} \right)^2 + \lambda \left( \vectorn{m}\cdot\vectorn{m} - m_\mathrm{S}^2 \right) , \label{eq:constMagOnly}
\end{equation}
where the following parameter values are taken: $\vectorn{p} = \vectorn{e}_2$ corresponding to the vertical anisotropy axis, $m_\mathrm{S} = 1$, $A_\mathrm{m} = 1$, and $K_\mathrm{m} = 2 \pi^2$. Such energy density has a local minimum in the form of the domain wall (with the parameter values substituted):
\begin{equation}
  \vectorn{m} = \cos\vartheta \, \vectorn{e}_1 - \sin\vartheta \, \vectorn{e}_2 , \quad\quad
  \vartheta = \arcsin \big( \tanh \big( \pi \sqrt{2} \, ( X_1 - X_\mathrm{c} ) \big) \big) ,
  \label{eq:DWan}
\end{equation}
where $X_1$ is the coordinate along the horizontal axis and $X_\mathrm{c}$ is the position of the centre of the domain wall.

The test is reminiscent of the patch test in finite elements; however, in this case, a constant gradient of the solution cannot be created ($\vectorn{m}$ must have unit length). The unit square geometry is selected. At the centre of the geometry, a circular interface with the radius of $2/7$ is created, separating the domain into two phases (inside and outside), both having the same properties. The numerical solution then tests how the enforcement of the interface conditions weakly affects the accuracy. The boundary conditions are enforced on the \emph{left} and the \emph{right} boundaries from analytical solution \eqref{eq:DWan} with $X_\mathrm{c} = 1/2$ (domain wall located in the centre); the \emph{bottom} and the \emph{top} boundaries are unconstrained.

\begin{figure}
  \begin{center}
    \includegraphics{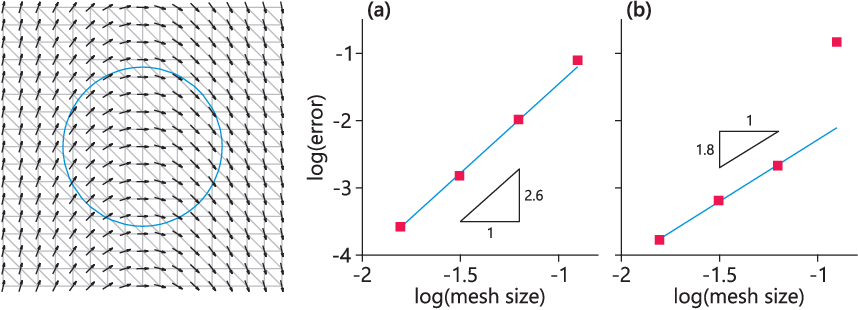}
  \end{center}
  \caption{The numerical solution of the magnetic domain wall (left); dependence of the solution error (a) and the magnetisation length error (b) on the mesh size in the first example.}
  \label{fig:DW}
\end{figure}

The solution (the magnetisation vectors at the finite-element nodes) is shown in figure \ref{fig:DW}, also showing the interface. In figure \ref{fig:DW}a, the $L^2$-norm of the error of the numerical solution (the difference between numerically-obtained magnetisation $\vectorn{m}^h$ and the analytical solution) is plotted vs.\ the mesh size. The convergence rate of $2.6$ is observed. In figure \ref{fig:DW}b, the $L^2$-norm of the error of the magnetisation length (deviation from $m_\mathrm{S} = 1$) is plotted vs.\ the mesh size. The convergence rate of $1.8$ is observed. 

\subsection{Effect of demagnetising field}

The second example is a coupled micromagnetics-magnetostatics simulation to illustrate the presence and the effects of the demagnetising field --- the magnetic field generated by the magnetisation inside the material. This field originates from long-range dipole-dipole interactions between spin magnetic moments in the material and is crucial for modelling relatively large scales. The outlined theory of course includes it as a constituent part\footnote{It can be seen from equations \eqref{eq:quasMagP}, \eqref{eq:quasFieP}, and \eqref{eq:quasB}. With constitutive law \eqref{eq:constLaw} and the absence of material deformation, $\tensor{F} = \tensor{I}$, the governing equations become $\vectorn{m} \times \big( A_\mathrm{m} \Delta \vectorn{m} + K_\mathrm{m} \vectorn{p} \vectorn{p} \cdot \vectorn{m} - \mu_0 \rho_0 \nabla \eta \big) = \vectorn{0}$ and $\Delta \eta = \rho_0 \nabla \cdot \vectorn{m}$ together with enforcement of $\left|\vectorn{m}\right| = m_\mathrm{S}$. One can recognise standard static micromagnetics with the gradient of the magnetic scalar potential providing the demagnetising field, which can be found in classical textbooks, e.g.\ \cite{Aharoni1996}.}.

In this example, there are five degrees of freedom --- three of the magnetisation vector, the Lagrange multiplier, and field $\eta$. Thus, the following strain energy density is used (same as in equations \eqref{eq:bigE} and \eqref{eq:constLaw}, but excluding $\vectorn{u}$ and $K_\mathrm{m}$):
\begin{equation}
  w = \tfrac{1}{2} A_\mathrm{m} \left( \tensor{M}:\tensor{I} \right) - \tfrac{1}{2} \mu_0 \nabla\eta \cdot \nabla\eta + \mu_0 \rho_0 \vectorn{m} \cdot \nabla\eta + \lambda \left( \vectorn{m}\cdot\vectorn{m} - m_\mathrm{S}^2 \right) , \label{eq:constDemag}
\end{equation}
where the following parameter values are taken: $A_\mathrm{m} = 1/100$, $\rho_0 = 1$, $m_\mathrm{S} = 1$, and the value of $\mu_0$ varied. Because the deformation is excluded, symbols $\nabla_0$ and $\nabla$ can be used interchangeably.

It is well known that the demagnetising field acts as anisotropy in a magnet. With the magnetocrystalline anisotropy (the term with $K_\mathrm{m}$) explicitly excluded from the energy density, the demagnetising field becomes the only source of anisotropy. Parameter $\mu_0$ is responsible for the strength of the field\footnote{Because the entire energy density can be rescaled, from the mathematical point of view, varying $\mu_0$ is equivalent to varying $A_\mathrm{m}$.}. Under weak fields, the exchange energy should dominate and the magnetisation should tend to be uniform, while under strong fields, the demagnetising field should create a strong anisotropy leading to the emergence of domain walls (a range of computational examples demonstrating this can be found in e.g.\ \cite{Arjmand2020}). This effect is illustrated by the present example.

The unit square geometry is selected. For the magnetisation, the Dirichlet boundary conditions are enforced on the \emph{left} boundary, $\vectorn{m} = \sin(\pi X_2) \, \vectorn{e}_2 + \cos(\pi X_2) \, \vectorn{e}_3$, where $X_2$ is the coordinate along the vertical axis, and the Neumann boundary conditions are enforced on \emph{all other} boundaries. For field $\eta$, the Neumann boundary conditions are enforced on all boundaries (and $\eta=0$ is enforced at the bottom-left corner). The mesh size $h=1/64$ is used.

\begin{figure}
  \begin{center}
    \includegraphics{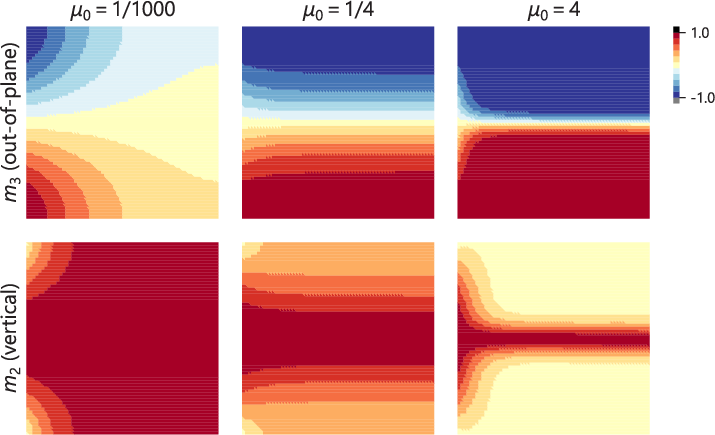}
  \end{center}
  \caption{The contour plots of the components of the magnetisation under different strengths of the demagnetising field in the second example.}
  \label{fig:emag}
\end{figure}

The results for different $\mu_0$ are shown in figure \ref{fig:emag}. The field plots show the spatial distributions of $m_2$ and $m_3$ components of the magnetisation. On the left boundary, the magnetisation changes orientation from out-of-plane towards the viewer to out-of-plane away from the viewer, exactly as imposed by the boundary conditions. Under a weak demagnetising field ($\mu_0 = 1/1000$), the magnetisation tends to orient in the vertical direction (the average of the imposed by the Dirichlet boundary conditions), creating one domain. Under a strong demagnetising field ($\mu_0 = 4$), the domain wall emerges in the middle of the computational region, separating out-of-plane orientations towards and away from the viewer.

\subsection{Stress-induced phase transition, topology change}

The third example is a purely mechanical simulation to demonstrate the automatic handling of the topology change of the evolving phase boundary. There are two degrees of freedom --- the in-plane displacements (the out-of-plane displacement is zero, corresponding to the plane strain formulation). Thus, the following strain energy density is used (similar to equations \eqref{eq:bigE} and \eqref{eq:constLaw}, but excluding $\vectorn{m}$ and $\eta$, while adding a transformation strain):
\begin{align}
  &w_\pm = \rho_0 \psi_\pm = w_\pm^0 + \tfrac{1}{2} \left( K_\pm - \tfrac{2}{3} G_\pm \right) \big( \tilde{\tensor{E}}_\pm : \tensor{I} \big)^2 + G_\pm \big( \tilde{\tensor{E}}_\pm : \tilde{\tensor{E}}_\pm \big) , \\
  &\tilde{\tensor{E}}_\pm = \tfrac{1}{2} \big( \tilde{\tensor{C}}_\pm - \tensor{I} \big) ,
  \quad\quad\quad
  \tilde{\tensor{C}}_+ = \tensor{G}\transpm \cdot \tensor{C}_+ \cdot \tensor{G}^{-1} ,
  \quad\quad\quad
  \tilde{\tensor{C}}_- = \tensor{C}_- ,
\end{align}
where different bulk and shear moduli are used for different phases ($K_\pm$ and $G_\pm$), and where phase `$+$' undergoes a volumetric transformation characterised by tensor $\tensor{G}$:
\begin{equation}
  \tensor{G} = g \left( \vectorn{e}_1 \vectorn{e}_1 + \vectorn{e}_2 \vectorn{e}_2 \right) + \vectorn{e}_3 \vectorn{e}_3 .
\end{equation}
It corresponds to the in-plane expansion of phase `$+$' compared to phase `$-$'. Constants $w_+^0$ and $w_-^0$ are the free energy volume densities of the phases in the stress-free states (jump term $\rho_0 \llbracket \psi \rrbracket$ enters the expression for the phase boundary velocity). The following parameter values are taken: $G_- = 24$, $G_+ = 10$, $K_- = 75$, $K_+ = 32$, $g = 1.05$, $\llbracket w^0 \rrbracket = 0.15$, $k_* = 0.043$.

The geometry is the unit square. Initially, three circular interfaces, having the radius of $0.11$, located at $(0.65;0.65)$, $(0.31;0.56)$, and $(0.56;0.31)$ are created, with phase `$-$' inside (inclusions) and phase `$+$' outside (matrix). Biaxial stretching boundary conditions are applied, with the \emph{left} and the \emph{bottom} boundaries fixed, and the \emph{right} and the \emph{top} boundaries displaced by $0.038$ in the normal direction. The mesh size $h=1/64$ and the time step $\Delta t = 0.25$ are used.

\begin{figure}
  \begin{center}
    \includegraphics{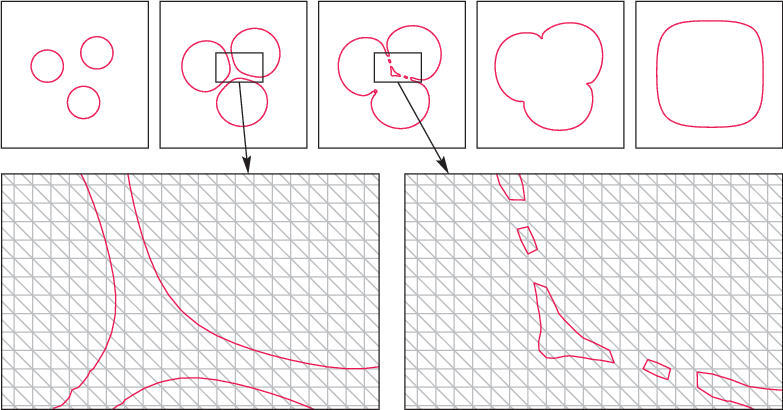}
  \end{center}
  \caption{The evolution of the phase boundaries in the stress-induced phase transition case, the third example.}
  \label{fig:topo}
\end{figure}

The numerical simulation of the phase transition process is shown in figure \ref{fig:topo}, where the snapshots at $t=0,10,15,25,200$ are shown. It is clearly seen how evolution and merging of the phase boundaries take place, eventually creating one single rounded-square-shaped inclusion of phase `$-$'. This is the stable equilibrium configuration of the phase boundary. Stability of phase boundaries was studied across many publications, e.g.\ \cite{Eremeyev2007,Morozov2023,Morozov2024}. In \cite{Morozov2024}, almost identical problem was solved (but in the linear-elastic setting, for chemo-mechanics, and without the topology change) using the CutFEM and the remeshing approaches; visible artefacts distorting the final equilibrium shape were present (and it was speculated that inaccurate stresses and strains lead to these artefacts). In the present simulations, the artefacts are absent. Throughout the development of the code after the publication of \cite{Morozov2024}, it transpired that there has been a bug in the post-processing calculation of stresses and strains at the interface (knowing stresses and strains in the elements). Correction of the bug eliminated the artefacts and started producing `smooth' interfaces at the equilibrium configuration.

\subsection{Twin boundary propagation in MSMAs}

The final example is a fully-coupled magneto-mechanical simulation of a structural phase tradition process. In this example, it is \emph{a priori} known that the magnetisation is planar; hence, it is convenient to represent it via angles, utilising energy density in form \eqref{eq:bigEang}. There are four degrees of freedom --- two in-plane displacements, angle $\varphi$, and field $\eta$ (the plane strain formulation and planar magnetisation, corresponding to $\alpha = 0$). The following parameter values are taken: $K_\mathrm{e} = 100$, $G_\mathrm{e} = 2$, $A_\mathrm{m} = 1/6000$, $K_\mathrm{m} = \pi^2/30$, $\mu_0 = 1/100$, $\rho_0 = 1/10$, $m_\mathrm{S} = 1$, $k_* = 0.05$. Such parameters correspond to nearly incompressible material; therefore, the last (third) term of $w_\mathrm{M}$ and the second term of $W_*$ (containing $\llbracket J \rrbracket$) are neglected.

With these parameter values, if a homogeneous single-phase unit-square domain is created with $\vectorn{p} = \vectorn{e}_1$ anisotropy axis (no rigid body motion boundary conditions for displacements and $\eta = 0$ on the whole external boundary), then the energy minimisation results in the orientation of the magnetisation vectors along $\vectorn{p}$, zero magnetic field ($\eta = 0$) in the bulk, and homogeneous deformation gradient --- the domain shrinks by $0.05376$ in the horizontal direction and expands by $0.04907$ in the vertical direction.

\begin{figure}
  \begin{center}
    \includegraphics{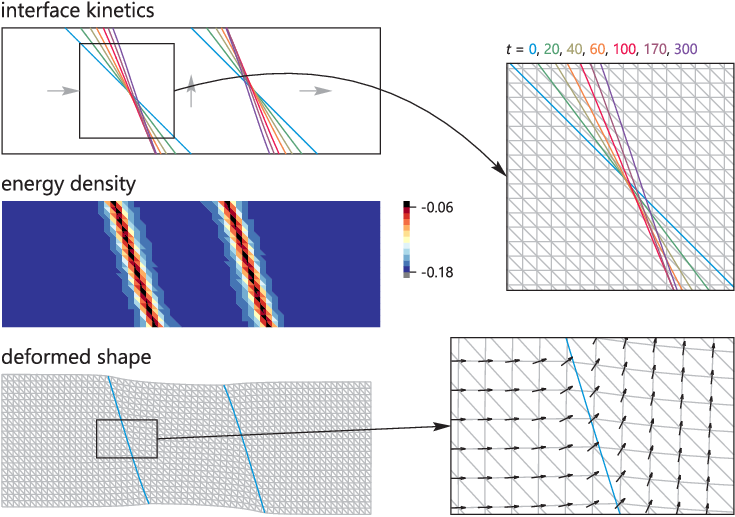}
  \end{center}
  \caption{The phase boundary kinetics, the contour plot of the energy density, and the deformed shape of the free-standing sample consisting of domains with different orientation of the magnetocrystalline anisotropy (indicated by grey arrows).}
  \label{fig:mainInit}
\end{figure}

For this example, the rectangular geometry with the height equal to $1/3$ and the length equal to $1$ is selected. The mesh size $h=1/63$ and the time step $\Delta t = 1$ are used. Initially, two flat interfaces, inclined by $45^\circ$ and crossing the bottom boundary at $X_1 = 1/2$ and $X_1 = 5/6$, are created, as illustrated in figure \ref{fig:mainInit} (top) in the light blue colour ($t=0$). The interfaces surround the phase with $\vectorn{p} = \vectorn{e}_2$ in the middle, while the phases with $\vectorn{p} = \vectorn{e}_1$ are located at the left and the right sides, which is indicated by grey arrows.

For the first simulation, no rigid body motion boundary conditions are enforced for displacements (two degrees of freedom are constrained at the bottom-left corner, one degree of freedom is constrained at the top-left corner) and $\eta = 0$ is enforced on the whole external boundary. Other degrees of freedom are unconstrained on the external boundary --- the Neumann boundary conditions are used. The phase boundaries are evolved until the stable equilibrium position is reached, with the kinetics shown in figure \ref{fig:mainInit} (top and top-right). The figure also shows the contour plot of the energy density at the final state ($t=300$, middle) as well as the deformed shape of the geometry with the magnetisation vectors in the vicinity of the phase boundary (bottom). At the equilibrium state of the phase boundaries, the sample shrinks by approximately $0.024$ compared to its reference length.

\begin{figure}
  \begin{center}
    \includegraphics[scale=0.95]{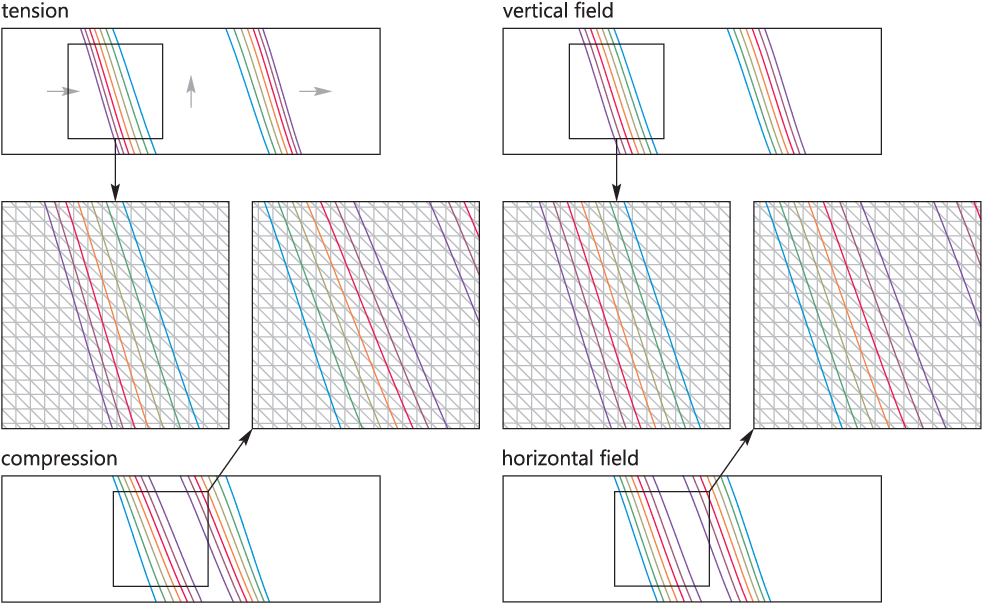}
  \end{center}
  \caption{The phase boundary kinetics under application of tension, compression, vertical and horizontal magnetic fields in the fourth example.}
  \label{fig:mainCases}
\end{figure}

For the second simulation, the starting point is the final equilibrium state of the first simulation. Four different cases are recreated: tension ($u_1 = 0$ on the left boundary, $u_2 = 0$ at the bottom-left corner, $u_1 = 0.02$ on the right boundary, $\eta = 0$ on all boundaries), compression (same but $u_1 = -0.07$ on the right boundary), horizontally-oriented magnetic field ($\eta = 4$ on the left boundary, $\eta = 0$ on the right boundary, no rigid body motion), vertically-oriented magnetic field ($\eta = 4/3$ on the bottom boundary, $\eta = 0$ on the top boundary, no rigid body motion). In figure \ref{fig:mainCases}, it can be seen that the phase boundaries behave as qualitatively expected for a magnetic shape-memory alloy: upon tension or under vertical magnetic field, the middle domain grows; upon compression or under horizontal magnetic field, the middle domain shrinks. The middle domain has the magnetisation vectors oriented vertically; hence, it is elongated in the horizontal direction and shrunk in the vertical direction compared to the reference configuration. The application of the tension makes such domain to be energetically preferential, leading to its growth. The converse is true for the application of the compression. The application of the vertical magnetic field also makes the magnetisation orientation of the middle domain to be energetically preferential, while the horizontal field makes the magnetisation state of the left and the right domains (horizontal orientation) to be energetically preferential. In figure \ref{fig:mainCases}, the coloured lines correspond to the simulation times of $t=0,15,30,45,60,75,90$ for tension, compression, and vertical field, and to the simulation times of $t=0,12,24,36,48,60,72$ for horizontal field.

\section{Conclusions}

In the present paper, a general derivation of the entropy production due to propagation of transformation fronts in deformable ferromagnets has been demonstrated. The quasistatic part of the resulting driving force coincides with the previous derivations obtained under more restrictive assumptions, e.g.\ relying on non-linear elastic constitutive behaviour. 

The case of non-dissipative solids and quasistatics has then been considered, and it has been shown that solving such problem is equivalent to minimisation of the total energy functional, which consists of the free energy of the solid and the electromagnetic energy of the space occupied by the solid. The variational approach leads to the same expression for the driving force as the thermodynamic approach and, thus, complements it for the purposes of writing the kinetic constitutive law.

Based on the previous results for non-linear solid mechanics, a general weak form of a coupled quasistatic problem has been derived via the variation of an energy functional; it has then been shown to be consistent with the strong form. Within the weak formulation, all interface conditions are enforced weakly. This weak form is then used to formulate the cut-finite-element method (CutFEM) for magneto-mechanics with propagating phase boundaries.

The present paper is supplemented by the 2D implementation of CutFEM in MATLAB, which has been heavily modified compared to the previous versions (supplementing previous publications by the author). It is now an intrinsically multi-physics code, solving a general PDE that results from the energy minimisation (with the continuity of the unknown field and the derivative of the energy density by the gradient of the unknown field across the interface), and the user can specify arbitrary energy density and arbitrary interface kinetics law. Furthermore, a novel interface topology change algorithm has been outlined and demonstrated using a numerical example. The code is available via GitHub (see link above).

The developments have been applied to the case study of the magnetically-driven and mechanically-driven structural phase transitions in magnetic shape-memory alloys (MSMAs). The simplest continuum constitutive law for the bulk magneto-mechanical behaviour of MSMAs has been proposed. Numerical simulations have been used to model qualitatively the behaviour of the twin boundaries in MSMAs.

\section*{Acknowledgements}

The author would like to express a sincere gratitude to Prof.\ Gunilla Kreiss (Uppsala University) who introduced the author to cut-element methods and who encouraged the author to research the modelling of magneto-mechanical coupling in solids. 

The author is immensely grateful to Prof.\ Alexander B.\ Freidin (Institute for Problems in Mechanical Engineering of Russian Academy of Sciences and St.\ Petersburg Polytechnic University) for invaluable discussions on mechanics of configuration forces and on phase transitions. 

The original MATLAB implementation of CutFEM was developed during the author's research fellowship at the University of Warwick. 

The author acknowledges funding from Research England's ``Expanding Excellence in England'' (E3) fund via the ``Multi-scale Multi-disciplinary Modelling for Impact'' programme (M$^3$4Impact).

\appendix
\titleformat{\section}[hang]{\Large\bfseries\raggedright\sffamily}{Appendix \thesection}{1em}{}

\section{Derivation of some interface terms}
\label{sec:addDeriv}

The first step is to push forward interface conditions \eqref{eq:MaxwBCref1} and \eqref{eq:MaxwBCref2} for the electromagnetic quantities to the current configuration. The following jump relations are used throughout the derivation:
\begin{equation*}
  \vectorn{N}_* \cdot \llbracket J \tensor{F}^{-1} \rrbracket = \vectorn{0} , \quad\quad
  \vectorn{T}_* \cdot \llbracket \tensor{F}\transp \rrbracket = \vectorn{0} ,
\end{equation*}
where the first relation follows from equation \eqref{eq:Nanson}, while the second relation is the compatibility condition on a coherent phase boundary. Applying the former directly results in 
\begin{equation}
  \llbracket\vectorn{B}\rrbracket\cdot\vectorn{n}_* = 0 , \quad\quad
  \llbracket\vectorn{D}\rrbracket\cdot\vectorn{n}_* = 0 ,
  \label{eq:MaxwBCur}
\end{equation}
where $\vectorn{n}_*$ is the interface normal in the current configuration. The interface conditions for $\vectorn{H}_0$ and $\vectorn{E}_0$ are similar; therefore, it is sufficient to give details only for one of them. The interface condition for $\vectorn{H}_0$ can also be written as (see the footnote at equations \eqref{eq:MaxwBCref1} and \eqref{eq:MaxwBCref2})
\begin{equation*}
  \vectorn{T}_*\cdot\big(\llbracket\vectorn{H}_0\rrbracket + \llbracket\vectorn{D}_0\rrbracket\times\vectorn{W}\big) = 0 .
\end{equation*}
Given that for arbitrary vectors $\vectorn{a}$, $\vectorn{b}$, $\vectorn{c}$, 
\begin{equation*}
  \vectorn{a}\cdot\big(\vectorn{b}\times\vectorn{c}\big) = J^{-1}\big(\tensor{F}\cdot\vectorn{a}\big)\cdot\big(\big(\tensor{F}\cdot\vectorn{b}\big)\times\big(\tensor{F}\cdot\vectorn{c}\big)\big) ,
\end{equation*}
the interface condition transforms to
\begin{equation*}
  \vectorn{T}_*\cdot\tensor{F}_\pm\transp\cdot\llbracket\tensor{F}\transpm\cdot\vectorn{H}_0\rrbracket + \vectorn{T}_*\cdot\tensor{F}_\pm\transp\cdot\llbracket\big(J^{-1}\tensor{F}\cdot\vectorn{D}_0\big)\times\big(\tensor{F}\cdot\vectorn{W}\big)\rrbracket = 0 ,
\end{equation*}
which transforms to
\begin{equation*}
  \vectorn{t}_*\cdot\llbracket\vectorn{H} - \vectorn{v}\times\vectorn{D} + \vectorn{D}\times\big(\vectorn{w} - \vectorn{v}\big)\rrbracket = 0 ,
\end{equation*}
where $\vectorn{t}_*$ is the tangent to the interface in the current configuration and $\vectorn{w}$ is the interface velocity in the current configuration. In the above, the standard relation between the interface velocities in the reference and the current configurations is used:
\begin{equation*}
  \vectorn{w} = \tensor{F}_\pm \cdot \vectorn{W} + \vectorn{v}_\pm = \langle\tensor{F}\rangle \cdot \vectorn{W} + \langle\vectorn{v}\rangle ,
\end{equation*}
from which it is also evident that $\llbracket\vectorn{v}\rrbracket = -\llbracket\tensor{F}\rrbracket\cdot \vectorn{W}$. The resulting interface conditions for $\vectorn{H}$ and $\vectorn{E}$ are
\begin{equation}
  \vectorn{n}_*\times\llbracket\vectorn{H}\rrbracket = -\vectorn{n}_*\times\big(\llbracket\vectorn{D}\rrbracket\times\vectorn{w}\big) , \quad\quad
  \vectorn{n}_*\times\llbracket\vectorn{E}\rrbracket = \vectorn{n}_*\times\big(\llbracket\vectorn{B}\rrbracket\times\vectorn{w}\big) .
  \label{eq:MaxwHCur}
\end{equation}
These conditions can also be rewritten as 
\begin{equation}
  \llbracket\vectorn{H}\rrbracket = a_\mathrm{h} \vectorn{n}_* - \llbracket\vectorn{D}\rrbracket\times\vectorn{w} , \quad\quad
  \llbracket\vectorn{E}\rrbracket = a_\mathrm{e} \vectorn{n}_* + \llbracket\vectorn{B}\rrbracket\times\vectorn{w} ,
\end{equation}
where $a_\mathrm{h}$ and $a_\mathrm{e}$ are some scalars. 

Now, it is necessary to transform some terms that are used later. First, one encounters the normal component of the jump of $\vectorn{H}\times\vectorn{E}$, which transforms as
\begin{align*}
  &\vectorn{n}_*\cdot\llbracket\vectorn{H}\times\vectorn{E}\rrbracket =
  \vectorn{n}_*\cdot\big(\langle\vectorn{H}\rangle\times\llbracket\vectorn{E}\rrbracket + \llbracket\vectorn{H}\rrbracket\times\langle\vectorn{E}\rangle\big) = \\
  &\hspace{1cm}=\vectorn{n}_*\cdot\big(\langle\vectorn{H}\rangle\times\big(\llbracket\vectorn{B}\rrbracket\times\vectorn{w}\big) - \big( \llbracket\vectorn{D}\rrbracket\times\vectorn{w}\big)\times\langle\vectorn{E}\rangle\big) = \\
  &\hspace{1cm}=\vectorn{n}_*\cdot\big(\llbracket\vectorn{B}\rrbracket\langle\vectorn{H}\rangle\cdot\vectorn{w} - \vectorn{w}\langle\vectorn{H}\rangle\cdot\llbracket\vectorn{B}\rrbracket + \llbracket\vectorn{D}\rrbracket\langle\vectorn{E}\rangle\cdot\vectorn{w} - \vectorn{w}\langle\vectorn{E}\rangle\cdot\llbracket\vectorn{D}\rrbracket\big) = \\
  &\hspace{1cm}=-\vectorn{n}_*\cdot\vectorn{w}\big(\langle\vectorn{H}\rangle\cdot\llbracket\vectorn{B}\rrbracket + \langle\vectorn{E}\rangle\cdot\llbracket\vectorn{D}\rrbracket\big) .
\end{align*}
Second, the following term with the Maxwell's stress tensor is needed:
\begin{equation*}
  \tensor{\tau}_\mathrm{E} - \zeta_\mathrm{E} \tensor{I} = \vectorn{B}\vectorn{H} -\vectorn{B}\cdot\vectorn{H} \tensor{I} + \vectorn{D}\vectorn{E} - \vectorn{D}\cdot\vectorn{E} \tensor{I} + \vectorn{v}\vectorn{D}\times\vectorn{B} .
\end{equation*}
When its jump is multiplied by the normal and $\langle\vectorn{v}\rangle$, it becomes
\begin{align*}
  &\vectorn{n}_*\cdot\llbracket\tensor{\tau}_\mathrm{E} - \zeta_\mathrm{E} \tensor{I}\rrbracket\cdot\langle\vectorn{v}\rangle = \vectorn{n}_*\cdot\langle\vectorn{B}\rangle\llbracket\vectorn{H}\rrbracket\cdot\langle\vectorn{v}\rangle - \vectorn{n}_*\cdot\langle\vectorn{v}\rangle\llbracket\vectorn{B}\cdot\vectorn{H}\rrbracket +
  \vphantom{a} \\ &\hspace{3cm}\vphantom{a} +
  \vectorn{n}_*\cdot\langle\vectorn{D}\rangle\llbracket\vectorn{E}\rrbracket\cdot\langle\vectorn{v}\rangle - \vectorn{n}_*\cdot\langle\vectorn{v}\rangle\llbracket\vectorn{D}\cdot\vectorn{E}\rrbracket + \vectorn{n}_*\cdot\llbracket\vectorn{v}\vectorn{D}\times\vectorn{B}\rrbracket\cdot\langle\vectorn{v}\rangle .
\end{align*}
Finally, the following large term must be transformed, using the already-obtained terms:
\begin{align*}
  &\vectorn{n}_*\cdot\llbracket\tensor{\tau}_\mathrm{E} - \zeta_\mathrm{E} \tensor{I}\rrbracket\cdot\langle\vectorn{v}\rangle - \vectorn{n}_*\cdot\llbracket\vectorn{H}\times\vectorn{E}\rrbracket = \\
  &\hspace{0.2cm} = \vectorn{n}_*\cdot\langle\vectorn{B}\rangle\llbracket\vectorn{H}\rrbracket\cdot\langle\vectorn{v}\rangle - \vectorn{n}_*\cdot\langle\vectorn{v}\rangle\langle\vectorn{B}\rangle\cdot\llbracket\vectorn{H}\rrbracket +
  \vectorn{n}_*\cdot\langle\vectorn{D}\rangle\llbracket\vectorn{E}\rrbracket\cdot\langle\vectorn{v}\rangle - \vectorn{n}_*\cdot\langle\vectorn{v}\rangle\langle\vectorn{D}\rangle\cdot\llbracket\vectorn{E}\rrbracket +
  \vphantom{a} \\ &\hspace{0.6cm}\vphantom{a} +
  \vectorn{n}_*\cdot\langle\vectorn{v}\rangle\llbracket\vectorn{D}\times\vectorn{B}\rrbracket\cdot\langle\vectorn{v}\rangle + \vectorn{n}_*\cdot\llbracket\vectorn{v}\rrbracket\langle\vectorn{D}\times\vectorn{B}\rangle\cdot\langle\vectorn{v}\rangle +
  \vectorn{n}_*\cdot\langle\tensor{F}\rangle\cdot\vectorn{W}\big(\langle\vectorn{H}\rangle\cdot\llbracket\vectorn{B}\rrbracket + \langle\vectorn{E}\rangle\cdot\llbracket\vectorn{D}\rrbracket\big) = \\
  &\hspace{0.2cm} = -\vectorn{n}_*\cdot\langle\vectorn{B}\rangle\big(\llbracket\vectorn{D}\rrbracket\times\vectorn{w}\big)\cdot\langle\vectorn{v}\rangle + 
  \vectorn{n}_*\cdot\langle\vectorn{v}\rangle\langle\vectorn{B}\rangle\cdot\big(\llbracket\vectorn{D}\rrbracket\times\vectorn{w}\big) +
  \vectorn{n}_*\cdot\langle\vectorn{D}\rangle\big(\llbracket\vectorn{B}\rrbracket\times\vectorn{w}\big)\cdot\langle\vectorn{v}\rangle - 
  \vphantom{a} \\ &\hspace{0.6cm}\vphantom{a} -
  \vectorn{n}_*\cdot\langle\vectorn{v}\rangle\langle\vectorn{D}\rangle\cdot\big(\llbracket\vectorn{B}\rrbracket\times\vectorn{w}\big) + 
  \vectorn{n}_*\cdot\langle\vectorn{v}\rangle\big(\langle\vectorn{D}\rangle\times\llbracket\vectorn{B}\rrbracket\big)\cdot\langle\vectorn{v}\rangle + \vectorn{n}_*\cdot\langle\vectorn{v}\rangle\big(\llbracket\vectorn{D}\rrbracket\times\langle\vectorn{B}\rangle\big)\cdot\langle\vectorn{v}\rangle - 
  \vphantom{a} \\ &\hspace{0.6cm}\vphantom{a} -
  \vectorn{n}_*\cdot\llbracket\tensor{F}\rrbracket\cdot\vectorn{W}\langle\vectorn{D}\times\vectorn{B}\rangle\cdot\langle\vectorn{v}\rangle +
  \vectorn{n}_*\cdot\langle\tensor{F}\rangle\cdot\vectorn{W}\big(\langle\vectorn{H}\rangle\cdot\llbracket\vectorn{B}\rrbracket + \langle\vectorn{E}\rangle\cdot\llbracket\vectorn{D}\rrbracket\big) = \\
  &\hspace{0.2cm} = \vectorn{n}_*\cdot\big(\langle\vectorn{v}\rangle\langle\vectorn{B}\rangle - \langle\vectorn{B}\rangle\langle\vectorn{v}\rangle\big)\cdot\big(\llbracket\vectorn{D}\rrbracket\times\langle\tensor{F}\rangle\cdot\vectorn{W}\big) + 
  \vectorn{n}_*\cdot\big(\langle\vectorn{D}\rangle\langle\vectorn{v}\rangle - \langle\vectorn{v}\rangle\langle\vectorn{D}\rangle\big)\cdot\big(\llbracket\vectorn{B}\rrbracket\times\langle\tensor{F}\rangle\cdot\vectorn{W}\big) - 
  \vphantom{a} \\ &\hspace{0.6cm}\vphantom{a} -
  \vectorn{n}_*\cdot\llbracket\tensor{F}\rrbracket\cdot\vectorn{W}\langle\vectorn{D}\times\vectorn{B}\rangle\cdot\langle\vectorn{v}\rangle +
  \vectorn{n}_*\cdot\langle\tensor{F}\rangle\cdot\vectorn{W}\big(\langle\vectorn{H}\rangle\cdot\llbracket\vectorn{B}\rrbracket + \langle\vectorn{E}\rangle\cdot\llbracket\vectorn{D}\rrbracket\big) .
\end{align*}
This expression is multiplied by the ratio between the infinitesimal areas in the current and the reference configurations, $\dif\sigma/\dif\varSigma$. Normal $\vectorn{n}_*$ then transforms into $\vectorn{N}_*\cdot \langle J \tensor{F}^{-1}\rangle$. After this transformation, it is easy to see that the the left-hand side of the obtained expression is the first two terms of the left-hand side of equation \eqref{eq:bigTermsS}. To obtain the right-hand side of equation \eqref{eq:bigTermsS}, the following relations are needed, the validity of which is straightforward to check:
\begin{align*}
  &\vectorn{N}_*\cdot \langle J \tensor{F}^{-1}\rangle \cdot\langle\tensor{F}\rangle\cdot\vectorn{N}_* = \langle J \rangle , \\
  &\vectorn{N}_*\cdot \langle J \tensor{F}^{-1}\rangle \cdot\llbracket\tensor{F}\rrbracket\cdot\vectorn{N}_* = \llbracket J \rrbracket .
\end{align*}

\section{Transformation of some magnetic terms}
\label{sec:addTran}

The term considered in section \ref{sec:variat} is the following:
\begin{align*}
  &\mathcal{H}_\mathrm{M} = \llbracket w_\mathrm{M} \rrbracket - \langle \tensor{\tau}_{\mathrm{E}0} \rangle : \llbracket\tensor{F}\rrbracket - \langle \vectorn{B}_0 \rangle \cdot \llbracket\nabla_0\eta\rrbracket = -\tfrac{1}{2}\mu_0^{-1} \llbracket J\vectorn{B}\cdot\vectorn{B} \rrbracket -
  \vphantom{a} \\ &\hspace{0.6cm}\vphantom{a} -  
  \langle J \tensor{F}^{-1} \cdot (\mu_0^{-1} \vectorn{B}\vectorn{B} + J^{-1}\rho_0\vectorn{m}\cdot\vectorn{B} \tensor{I} - J^{-1}\rho_0\vectorn{B}\vectorn{m} - \tfrac{1}{2}\mu_0^{-1}\vectorn{B}\cdot\vectorn{B} \tensor{I}) \rangle : \llbracket\tensor{F}\rrbracket - 
  \vphantom{a} \\ &\hspace{0.6cm}\vphantom{a} -
  \langle J \tensor{F}^{-1} \cdot \vectorn{B} \rangle \cdot \llbracket -\tensor{F}\transp\cdot(\mu_0^{-1}\vectorn{B}-J^{-1}\rho_0\vectorn{m}) \rrbracket ,
\end{align*}
where expressions \eqref{eq:BHlink}, \eqref{eq:volForceTau}, \eqref{eq:Bref}, \eqref{eq:magDens} have been substituted together with the relations between $\tensor{\tau}_{\mathrm{E}0}$ and $\tensor{\tau}_\mathrm{E}$, between $\rho_0$ and $\rho$, between $\nabla_0\eta$ and $\vectorn{H}_0$, while taking $\vectorn{E} = \vectorn{0}$.

First, for arbitrary scalar $a$, the validity of the following relation is straightforward to check:
\begin{equation*}
  \langle J \tensor{F}^{-1} a \rangle : \llbracket\tensor{F}\rrbracket = \llbracket J \rrbracket\langle a \rangle .
\end{equation*}
Next, it is necessary to transform the following terms:
\begin{align*}
  &\llbracket\tensor{F}\rrbracket \cdot \langle \tensor{F}^{-1}\cdot\vectorn{B} \rangle - \llbracket J^{-1}\tensor{F}\rrbracket \cdot \langle J\tensor{F}^{-1}\cdot\vectorn{B} \rangle = \\
  &\hspace{0.2cm} = \tfrac{1}{2} \big((1-J_+^{-1}J_-)\tensor{F}_+\cdot\tensor{F}_-^{-1}\cdot\vectorn{B}_- - (1-J_-^{-1}J_+)\tensor{F}_-\cdot\tensor{F}_+^{-1}\cdot\vectorn{B}_+\big) = \\
  &\hspace{0.2cm} = \tfrac{1}{2} \big((1-J_+^{-1}J_-)\vectorn{B}_- - (1-J_-^{-1}J_+)\vectorn{B}_+ + (1-J_+^{-1}J_-)\llbracket\tensor{F}\rrbracket\cdot\tensor{F}_-^{-1}\cdot\vectorn{B}_- + 
  \vphantom{a} \\ &\hspace{0.6cm}\vphantom{a} + 
  (1-J_-^{-1}J_+)\llbracket\tensor{F}\rrbracket\cdot\tensor{F}_+^{-1}\cdot\vectorn{B}_+\big) = 
  -\llbracket J^{-1} \rrbracket\langle J\vectorn{B} \rangle ,
\end{align*}
where the brackets are directly opened in the first equality, tensors $\tensor{F}_\pm$ are expressed via $\llbracket\tensor{F}\rrbracket$ and $\tensor{F}_\mp$ in the second equality, and interface condition $\vectorn{N}_* \cdot J_\pm \tensor{F}_\pm^{-1} \cdot \llbracket\vectorn{B}\rrbracket = 0$ is used in the third equality (thus, the terms with $\tensor{F}$ are eliminated), which, in turn, follows from condition \eqref{eq:MaxwBCur} and the relation between normals $\vectorn{n}_*$ and $\vectorn{N}_*$. The interface condition for $\llbracket\vectorn{B}\rrbracket$ can be used due to the structure of $\llbracket\tensor{F}\rrbracket$ according to relation \eqref{eq:genGradForm}. This allows collecting all terms with $\vectorn{m}$ in $\mathcal{H}_\mathrm{M}$ resulting in
\begin{equation*}
  \mathcal{H}_{\mathrm{M}1} = -\llbracket J \rrbracket\langle J^{-1}\vectorn{B} \rangle\cdot \rho_0\vectorn{m} - \llbracket J^{-1} \rrbracket\langle J\vectorn{B} \rangle \cdot \rho_0\vectorn{m} = -\tfrac{1}{2} \llbracket J \rrbracket \llbracket J^{-1} \rrbracket \llbracket\vectorn{B}\rrbracket \cdot \rho_0\vectorn{m} ,
\end{equation*}
which is done by directly opening the brackets.

To collect all terms with $B^2$, the following terms are transformed:
\begin{align*}
  &-\llbracket\tensor{F}\rrbracket : \langle J \tensor{F}^{-1}\cdot\vectorn{B}\vectorn{B} \rangle + \llbracket \vectorn{B}\cdot\tensor{F}\rrbracket \cdot \langle J\tensor{F}^{-1}\cdot\vectorn{B} \rangle = \\
  &\hspace{0.2cm} = \tfrac{1}{2} \big( J_- \llbracket\vectorn{B}\rrbracket \cdot\tensor{F}_+\cdot\tensor{F}_-^{-1}\cdot\vectorn{B}_- + J_+ \llbracket\vectorn{B}\rrbracket \cdot\tensor{F}_-\cdot\tensor{F}_+^{-1}\cdot\vectorn{B}_+ \big) = \llbracket\vectorn{B}\rrbracket \cdot \langle J\vectorn{B} \rangle ,
\end{align*}
where the brackets are directly opened in the first equality. For the second equality, exactly as above, tensors $\tensor{F}_\pm$ are expressed via $\llbracket\tensor{F}\rrbracket$ and $\tensor{F}_\mp$, which is followed by the application of the interface condition for $\llbracket\vectorn{B}\rrbracket$, eliminating the terms with $\tensor{F}$. This allows collecting all terms with $B^2$ in $\mathcal{H}_\mathrm{M}$ resulting in
\begin{equation*}
  \mathcal{H}_{\mathrm{M}2} = -\tfrac{1}{2}\mu_0^{-1} \llbracket J\vectorn{B}\cdot\vectorn{B} \rrbracket + \mu_0^{-1} \llbracket\vectorn{B}\rrbracket \cdot \langle J\vectorn{B} \rangle + \tfrac{1}{2}\mu_0^{-1}  \llbracket J \rrbracket \langle \vectorn{B}\cdot\vectorn{B} \rangle =
  \tfrac{1}{4}\mu_0^{-1} \llbracket J \rrbracket \llbracket\vectorn{B}\rrbracket \cdot \llbracket\vectorn{B}\rrbracket ,
\end{equation*}
which is done by directly opening the brackets.

From interface conditions \eqref{eq:MaxwBCur} and \eqref{eq:MaxwHCur} in absence of $\vectorn{E}$, it is evident that $\llbracket\vectorn{B}\rrbracket \perp \vectorn{n}_*$ and $\llbracket\vectorn{H}\rrbracket \parallel \vectorn{n}_*$. Next, from equation \eqref{eq:BHlink}, is follows that
\begin{equation*}
  \llbracket\vectorn{H}\rrbracket = \mu_0^{-1} \llbracket\vectorn{B}\rrbracket - \llbracket J^{-1} \rrbracket \rho_0\vectorn{m} .
\end{equation*}
Combining these together leads to the following:
\begin{equation*}
  \mu_0^{-1}\llbracket\vectorn{B}\rrbracket = \llbracket J^{-1} \rrbracket \rho_0\vectorn{m} \cdot \big( \tensor{I} - \vectorn{n}_*\vectorn{n}_* \big) .
\end{equation*}
Substituting into $\mathcal{H}_{\mathrm{M}2}$ results in
\begin{equation*}
  \mathcal{H}_\mathrm{M} = \mathcal{H}_{\mathrm{M}1} + \mathcal{H}_{\mathrm{M}2} = -\tfrac{1}{4} \llbracket J \rrbracket \llbracket J^{-1} \rrbracket \rho_0\vectorn{m} \cdot \llbracket\vectorn{B}\rrbracket .
\end{equation*}



\begin{flushleft}

\bibliographystyle{unsrt}
\bibliography{refs}

@Book{Abeyaratne2006,
  Title                    = {Evolution of phase transitions: {A} continuum theory},
  Author                   = {Abeyaratne, R. and Knowles, J. K.},
  Publisher                = {Cambridge University Press},
  Year                     = {2006}
}

@Book{Aharoni1996,
  author    = {Aharoni, A.},
  publisher = {Oxford University Press},
  title     = {Introduction to the {T}heory of {F}erromagnetism},
  year      = {1996},
}

@Article{Arjmand2020,
  author  = {Arjmand, D. and Poluektov, M. and Kreiss, G.},
  journal = {Advances in Computational Mathematics},
  title   = {Modelling long-range interactions in multiscale simulations of ferromagnetic materials},
  year    = {2020},
  number  = {1},
  pages   = {2},
  volume  = {46},
  doi     = {10.1007/s10444-020-09747-5},
  url     = {https://doi.org/10.1007/s10444-020-09747-5},
}

@Book{Brown1966,
  author    = {Brown, W. F.},
  editor    = {Truesdell, C.},
  publisher = {Springer-Verlag Berlin Heidelberg},
  title     = {Magnetoelastic interactions},
  year      = {1966},
  series    = {Springer Tracts in Natural Philosophy},
  volume    = {9},
}

@Article{Burman2012,
  Title                    = {Fictitious domain finite element methods using cut elements: {II}. {A} stabilized {N}itsche method},
  Author                   = {Burman, E. and Hansbo, P.},
  Journal                  = {Applied Numerical Mathematics},
  Year                     = {2012},
  Number                   = {4},
  Pages                    = {328--341},
  Volume                   = {62},

  Af                       = {Burman, ErikEOLEOLHansbo, Peter},
  C1                       = {[Hansbo, Peter] Univ Gothenburg, SE-41296 Gothenburg, Sweden.EOLEOL[Burman, Erik] Univ Sussex, Dept Math, Brighton BN1 9RF, E Sussex, England.},
  Da                       = {2018-05-29},
  De                       = {Interior penalty; Fictitious domain; Finite element},
  Doi                      = {10.1016/j.apnum.2011.01.008},
  Em                       = {peter.hansbo@chalmers.se},
  Ga                       = {913UF},
  J9                       = {APPL NUMER MATH},
  Ji                       = {Appl. Numer. Math.},
  La                       = {English},
  Nr                       = {7},
  Oi                       = {Burman, Erik/0000-0003-4287-7241},
  Pa                       = {PO BOX 211, 1000 AE AMSTERDAM, NETHERLANDS},
  Pg                       = {14},
  Pi                       = {AMSTERDAM},
  Publisher                = {Elsevier Science Bv},
  Ri                       = {Burman, Erik/G-2414-2011},
  Rp                       = {Hansbo, P (reprint author), Chalmers, Dept Math Sci, SE-41296 Gothenburg, Sweden.},
  Sc                       = {Mathematics},
  Si                       = {SI},
  Sn                       = {0168-9274},
  Tc                       = {88},
  U1                       = {0},
  U2                       = {21},
  Ut                       = {WOS:000301902200008},
  Wc                       = {Mathematics, Applied},
  Z9                       = {88}
}

@Article{Burman2015,
  author    = {Burman, E. and Claus, S. and Hansbo, P. and Larson, M. G. and Massing, A.},
  journal   = {International Journal for Numerical Methods in Engineering},
  title     = {{CutFEM}: {D}iscretizing geometry and partial differential equations},
  year      = {2015},
  number    = {7},
  pages     = {472--501},
  volume    = {104},
  af        = {Burman, ErikEOLEOLClaus, SusanneEOLEOLHansbo, PeterEOLEOLLarson, Mats G.EOLEOLMassing, Andre},
  c1        = {[Burman, Erik; Claus, Susanne] UCL, Dept Math, London WC1E 6BT, England.EOLEOL[Hansbo, Peter] Jonkoping Univ, Dept Mech Engn, SE-55111 Jonkoping, Sweden.EOLEOL[Larson, Mats G.] Umea Univ, Dept Math & Math Stat, SE-90187 Umea, Sweden.EOLEOL[Massing, Andre] Simula Res Lab, Ctr Biomed Comp, NO-1325 Lysaker, Norway.},
  da        = {2021-03-16},
  de        = {extended finite element method; unfitted methods; finite elementEOLEOLmethods; meshfree methods; Galerkin; level sets; stability},
  doi       = {10.1002/nme.4823},
  ei        = {1097-0207},
  em        = {e.burman@ucl.ac.uk},
  fu        = {EPSRCUK Research & Innovation (UKRI)Engineering & Physical SciencesEOLEOLResearch Council (EPSRC) [EP/J002312/2]; Swedish Foundation forEOLEOLStrategic ResearchSwedish Foundation for Strategic Research [AM13-0029];EOLEOLSwedish Research CouncilSwedish Research CouncilEuropean CommissionEOLEOL[2011-4992, 2013-4708]; Center of Excellence grant from the ResearchEOLEOLCouncil of Norway; Engineering and Physical Sciences Research CouncilUKEOLEOLResearch & Innovation (UKRI)Engineering & Physical Sciences ResearchEOLEOLCouncil (EPSRC) [EP/J002313/2, EP/J002313/1] Funding Source:EOLEOLresearchfish},
  fx        = {This research was supported in part by EPSRC (first and second authors,EOLEOLGrant No. EP/J002312/2), in part by the Swedish Foundation for StrategicEOLEOLResearch (second and third author, Grant No. AM13-0029) and the SwedishEOLEOLResearch Council (second and third author, Grants No. 2011-4992 (PH) andEOLEOLNo. 2013-4708 (MGL)). The work for this article was also supported by aEOLEOLCenter of Excellence grant from the Research Council of Norway to theEOLEOLCenter for Biomedical Computing at Simula Research Laboratory.},
  ga        = {CT1IR},
  j9        = {INT J NUMER METH ENG},
  ji        = {Int. J. Numer. Methods Eng.},
  la        = {English},
  nr        = {49},
  oa        = {Other Gold, Green Published},
  oi        = {Burman, Erik/0000-0003-4287-7241},
  pa        = {111 RIVER ST, HOBOKEN 07030-5774, NJ USA},
  pg        = {30},
  pi        = {HOBOKEN},
  publisher = {Wiley},
  ri        = {Burman, Erik/G-2414-2011},
  rp        = {Burman, E (corresponding author), UCL, Dept Math, Mortimer St, London WC1E 6BT, England.},
  sc        = {Engineering; Mathematics},
  si        = {SI},
  sn        = {0029-5981},
  tc        = {202},
  u1        = {3},
  u2        = {19},
  ut        = {WOS:000362552500002},
  wc        = {Engineering, Multidisciplinary; Mathematics, InterdisciplinaryEOLEOLApplications},
  z9        = {202},
}

@Article{Burman2018,
  author    = {Burman, E. and Elfverson, D. and Hansbo, P. and Larson, M. G. and Larsson, K.},
  journal   = {Computer Methods in Applied Mechanics and Engineering},
  title     = {Shape optimization using the cut finite element method},
  year      = {2018},
  pages     = {242--261},
  volume    = {328},
  af        = {Burman, ErikEOLEOLElfverson, DanielEOLEOLHansbo, PeterEOLEOLLarson, Mats G.EOLEOLLarsson, Karl},
  c1        = {[Burman, Erik; Larson, Mats G.; Larsson, Karl] UCL, Dept Math, Gower St, London WC1E 6BT, England.EOLEOL[Elfverson, Daniel] Umea Univ, Dept Math & Math Stat, SE-90187 Umea, Sweden.EOLEOL[Hansbo, Peter] Jonkoping Univ, Dept Mech Engn, SE-55111 Jonkoping, Sweden.},
  da        = {2020-03-16},
  de        = {CutFEM; Shape optimization; Level-set; Fictitious domain method; LinearEOLEOLelasticity},
  doi       = {10.1016/j.cma.2017.09.005},
  ei        = {1879-2138},
  em        = {e.burman@ucl.ac.uk; daniel.elfverson@umu.se; peter.hansbo@ju.se;EOLEOLmats.larson@umu.se; karl.larsson@umu.se},
  fu        = {Swedish Foundation [AM13-0029]; Swedish Research CouncilSwedish ResearchEOLEOLCouncil [2011-4992, 2013-4708]; Swedish Research Programme Essence;EOLEOLEPSRC, UKEngineering & Physical Sciences Research Council (EPSRC)EOLEOL[EP/P01576X/1]},
  fx        = {This research was supported in part by the Swedish Foundation forEOLEOLStrategic Research Grant No. AM13-0029, the Swedish Research CouncilEOLEOLGrant Nos. 2011-4992, 2013-4708, the Swedish Research Programme Essence,EOLEOLand EPSRC, UK, Grant No. EP/P01576X/1.},
  ga        = {FN7SJ},
  j9        = {COMPUT METHOD APPL M},
  ji        = {Comput. Meth. Appl. Mech. Eng.},
  la        = {English},
  nr        = {21},
  oi        = {Larsson, Karl/0000-0001-7838-1307},
  pa        = {PO BOX 564, 1001 LAUSANNE, SWITZERLAND},
  pg        = {20},
  pi        = {LAUSANNE},
  publisher = {Elsevier Science Sa},
  rp        = {Elfverson, D (reprint author), Umea Univ, Dept Math & Math Stat, SE-90187 Umea, Sweden.},
  sc        = {Engineering; Mathematics; Mechanics},
  sn        = {0045-7825},
  tc        = {24},
  u1        = {2},
  u2        = {13},
  ut        = {WOS:000416218500011},
  wc        = {Engineering, Multidisciplinary; Mathematics, InterdisciplinaryEOLEOLApplications; Mechanics},
  z9        = {24},
}

@Article{Burman2019,
  author    = {Burman, E. and Elfverson, D. and Hansbo, P. and Larson, M. G. and Larsson, K.},
  journal   = {Computer Methods in Applied Mechanics and Engineering},
  title     = {Cut topology optimization for linear elasticity with coupling to parametric nondesign domain regions},
  year      = {2019},
  pages     = {462--479},
  volume    = {350},
  af        = {Burman, ErikEOLEOLElfverson, DanielEOLEOLHansbo, PeterEOLEOLLarson, Mats G.EOLEOLLarsson, Karl},
  c1        = {[Burman, Erik] UCL, Dept Math, Gower St, London WC1E 6BT, England.EOLEOL[Elfverson, Daniel; Larson, Mats G.; Larsson, Karl] Umea Univ, Dept Math & Math Stat, SE-90187 Umea, Sweden.EOLEOL[Hansbo, Peter] Jonkoping Univ, Dept Mech Engn, SE-55111 Jonkoping, Sweden.},
  da        = {2020-03-16},
  de        = {Material distribution topology optimization; Design and nondesign domainEOLEOLregions; Cut finite element method},
  doi       = {10.1016/j.cma.2019.03.016},
  ei        = {1879-2138},
  em        = {e.burman@ucl.ac.uk; daniel.elfverson@umu.se; peter.hansbo@ju.se;EOLEOLmats.larson@umu.se; karl.larsson@umu.se},
  fu        = {Swedish Foundation for Strategic ResearchSwedish Foundation forEOLEOLStrategic Research [AM13-0029]; Swedish Research CouncilSwedish ResearchEOLEOLCouncil [2013-4708, 2017-03911, 2018-05262]; Swedish Research ProgrammeEOLEOLEssence; EPSRCEngineering & Physical Sciences Research Council (EPSRC)EOLEOL[EP/P01576X/1, EP/P012434/1]},
  fx        = {This research was supported in part by the Swedish Foundation forEOLEOLStrategic Research Grant No. AM13-0029, the Swedish Research CouncilEOLEOLGrants Nos. 2013-4708, 2017-03911, 2018-05262, and the Swedish ResearchEOLEOLProgramme Essence. EB was supported by EPSRC research grantsEOLEOLEP/P01576X/1 and EP/P012434/1.},
  ga        = {HY5JG},
  j9        = {COMPUT METHOD APPL M},
  ji        = {Comput. Meth. Appl. Mech. Eng.},
  la        = {English},
  nr        = {41},
  oi        = {Burman, Erik/0000-0003-4287-7241; Larsson, Karl/0000-0001-7838-1307;EOLEOLHansbo, Peter/0000-0001-7352-1550},
  pa        = {PO BOX 564, 1001 LAUSANNE, SWITZERLAND},
  pg        = {18},
  pi        = {LAUSANNE},
  publisher = {Elsevier Science Sa},
  ri        = {Burman, Erik/G-2414-2011},
  rp        = {Larsson, K (reprint author), Umea Univ, Dept Math & Math Stat, SE-90187 Umea, Sweden.},
  sc        = {Engineering; Mathematics; Mechanics},
  sn        = {0045-7825},
  tc        = {1},
  u1        = {2},
  u2        = {3},
  ut        = {WOS:000468163500019},
  wc        = {Engineering, Multidisciplinary; Mathematics, InterdisciplinaryEOLEOLApplications; Mechanics},
  z9        = {1},
}

@Article{Burman2025,
  author  = {Burman, E. and Hansbo, P. and Larson, M. G. and Zahedi, S.},
  journal = {Acta Numerica},
  title   = {Cut finite element methods},
  year    = {2025},
  pages   = {1–121},
  volume  = {34},
  doi     = {10.1017/S0962492925000017},
}

@Article{Coleman1963,
  author  = {Coleman, B. D. and Noll, W.},
  journal = {Archive for Rational Mechanics and Analysis},
  title   = {The thermodynamics of elastic materials with heat conduction and viscosity},
  year    = {1963},
  number  = {3},
  pages   = {167--178},
  volume  = {13},
  af      = {COLEMAN, BDEOLEOLNOLL, W},
  doi     = {10.1007/BF01262690},
  sn      = {0003-9527},
  ut      = {WOS:A1963XL62900001},
}

@Article{Eremenko2001,
  author  = {Eremenko, V. and Gnatchenko, S. and Makedonska, N. and Shabakayeva, Yu. and Shvedun, M. and Sirenko, V. and Fink-Finowicki, J. and Kamenev, K. V. and Balakrishnan, G. and Mck Paul, D.},
  journal = {Low Temperature Physics},
  title   = {X-ray study of {N}d0.5{S}r0.5{M}n{O}3 manganite structure above and below the ferromagnetic metal-antiferromagnetic insulator spontaneous phase transition},
  year    = {2001},
  number  = {11},
  pages   = {930--934},
  volume  = {27},
  doi     = {10.1063/1.1421458},
  eprint  = {https://pubs.aip.org/aip/ltp/article-pdf/27/11/930/8231858/930_1_online.pdf},
  url     = {https://doi.org/10.1063/1.1421458},
}

@Article{Eremeyev2007,
  author    = {Yeremeyev, V. A. and Freidin, A. B. and Sharipova, L. L.},
  journal   = {Journal of Applied Mathematics and Mechanics},
  title     = {The stability of the equilibrium of two-phase elastic solids},
  year      = {2007},
  number    = {1},
  pages     = {61--84},
  volume    = {71},
  af        = {Yeremeyev, V. A.EOLEOLFreidin, A. B.EOLEOLSharipova, L. L.},
  da        = {2020-03-16},
  doi       = {10.1016/j.jappmathmech.2007.03.007},
  ei        = {1873-4855},
  em        = {eremeyev@math.rsu.ru; freidin@mechanies.ipme.ru; leah@mechanics.ipme.ru},
  ga        = {179MJ},
  j9        = {PMM-J APPL MATH MEC+},
  ji        = {Pmm-J. Appl. Math. Mech.},
  la        = {English},
  nr        = {49},
  oi        = {Eremeyev, Victor A./0000-0002-8128-3262; Freidin,EOLEOLAlexander/0000-0003-2916-3375},
  pa        = {THE BOULEVARD, LANGFORD LANE, KIDLINGTON, OXFORD OX5 1GB, ENGLAND},
  pg        = {24},
  pi        = {OXFORD},
  publisher = {Pergamon-elsevier Science Ltd},
  ri        = {Eremeyev, Victor A./B-1478-2010; Freidin, Alexander/N-2089-2015},
  sc        = {Mathematics; Mechanics},
  sn        = {0021-8928},
  tc        = {43},
  u1        = {0},
  u2        = {1},
  ut        = {WOS:000247293600007},
  wc        = {Mathematics, Applied; Mechanics},
  z9        = {44},
}

@Book{Eringen1990,
  author    = {Eringen, A. C. and Maugin, G. A.},
  publisher = {Springer-Verlag New York},
  title     = {Electrodynamics of {C}ontinua},
  year      = {1990},
  volume    = {I and II},
}

@Article{Faran2011,
  author  = {Faran, E. and Shilo, D.},
  journal = {Journal of the Mechanics and Physics of Solids},
  title   = {The kinetic relation for twin wall motion in {N}i{M}n{G}a},
  year    = {2011},
  number  = {5},
  pages   = {975--987},
  volume  = {59},
  doi     = {https://doi.org/10.1016/j.jmps.2011.02.009},
  url     = {https://www.sciencedirect.com/science/article/pii/S0022509611000469},
}

@Article{Faran2013,
  author  = {Faran, E. and Shilo, D.},
  journal = {Journal of the Mechanics and Physics of Solids},
  title   = {The kinetic relation for twin wall motion in {N}i{M}n{G}a---part 2},
  year    = {2013},
  number  = {3},
  pages   = {726--741},
  volume  = {61},
  doi     = {https://doi.org/10.1016/j.jmps.2012.11.004},
  url     = {https://www.sciencedirect.com/science/article/pii/S0022509612002426},
}

@Article{Fomethe1996,
  author  = {Fomethe, A. and Maugin, G. A.},
  journal = {Continuum Mechanics and Thermodynamics},
  title   = {Material forces in thermoelastic ferromagnets},
  year    = {1996},
  number  = {5},
  pages   = {275--292},
  volume  = {8},
  doi     = {10.1007/s001610050044},
  url     = {https://doi.org/10.1007/s001610050044},
}

@Article{Fomethe1997,
  author  = {Fomethe, A. and Maugin, G. A.},
  journal = {International Journal of Applied Electromagnetics and Mechanics},
  title   = {Propagation of phase-transition fronts and domain walls in thermoelastic ferromagnets},
  year    = {1997},
  number  = {2},
  pages   = {143--165},
  volume  = {8},
  doi     = {10.1177/138354169700800203},
  eprint  = {https://doi.org/10.1177/138354169700800203},
}

@Article{Freidin2014,
  author    = {Freidin, A. B. and Vilchevskaya, E. N. and Korolev, I. K.},
  journal   = {International Journal of Engineering Science},
  title     = {Stress-assist chemical reactions front propagation in deformable solids},
  year      = {2014},
  pages     = {57--75},
  volume    = {83},
  af        = {Freidin, A. B.EOLEOLVilchevskaya, E. N.EOLEOLKorolev, I. K.},
  c1        = {[Freidin, A. B.; Vilchevskaya, E. N.; Korolev, I. K.] Russian Acad Sci, Inst Problems Mech Engn, St Petersburg 199178, Russia.EOLEOL[Freidin, A. B.] St Petersburg State Univ, St Petersburg 198504, Russia.EOLEOL[Vilchevskaya, E. N.] St Petersburg State Polytech Univ, St Petersburg 195251, Russia.},
  de        = {Chemical affinity tensor; Mechanochemistry; Chemical transformationEOLEOLstrain; Reaction front kinetics; Eshelby stress tensor; Locking effect},
  doi       = {10.1016/j.ijengsci.2014.03.008},
  ei        = {1879-2197},
  em        = {alexander.freidin@gmail.com; vilchevska@gmail.com; i.korolev82@gmail.com},
  fu        = {Russian Foundation for Basic Research [13-01-00687, 14-01-31433]; SandiaEOLEOLNational Laboratories (USA); Scientific Programs of Russian Academy ofEOLEOLSciences},
  fx        = {This work was supported by Russian Foundation for Basic Research (GrantsEOLEOLNos. 13-01-00687, 14-01-31433), Sandia National Laboratories (USA) andEOLEOLScientific Programs of Russian Academy of Sciences.},
  ga        = {AP7IX},
  j9        = {INT J ENG SCI},
  ji        = {Int. J. Eng. Sci.},
  la        = {English},
  nr        = {43},
  oi        = {Freidin, Alexander/0000-0003-2916-3375; Vilchevskaya,EOLEOLElena/0000-0002-5173-3218},
  pa        = {THE BOULEVARD, LANGFORD LANE, KIDLINGTON, OXFORD OX5 1GB, ENGLAND},
  pg        = {19},
  pi        = {OXFORD},
  publisher = {Pergamon-elsevier Science Ltd},
  ri        = {Freidin, Alexander/N-2089-2015; Vilchevskaya, Elena/P-3007-2016},
  rp        = {Freidin, AB (reprint author), Russian Acad Sci, Inst Problems Mech Engn, Bolshoy Pr,61,VO, St Petersburg 199178, Russia.},
  sc        = {Engineering},
  si        = {SI},
  sn        = {0020-7225},
  tc        = {6},
  u1        = {3},
  u2        = {12},
  ut        = {WOS:000342252100004},
  wc        = {Engineering, Multidisciplinary},
  z9        = {6},
}

@Article{Gebbia2018,
  author  = {Gebbia, J. F. and Cast\'{a}n, T. and Lloveras, P. and Porta, M. and Saxena, A. and Planes, A.},
  journal = {Physica Status Solidi B},
  title   = {Multiferroic and related hysteretic behavior in ferromagnetic shape memory alloys},
  year    = {2018},
  number  = {2},
  pages   = {1700327},
  volume  = {255},
  doi     = {https://doi.org/10.1002/pssb.201700327},
  eprint  = {https://onlinelibrary.wiley.com/doi/pdf/10.1002/pssb.201700327},
  url     = {https://onlinelibrary.wiley.com/doi/abs/10.1002/pssb.201700327},
}

@Book{Glansdorff1971,
  author    = {Glansdorff, P. and Prigogine, I.},
  publisher = {John Wiley \& Sons Ltd.},
  title     = {Thermodynamic theory of structure, stability and fluctuations},
  year      = {1971},
}

@Article{Grabovsky2011,
  author    = {Grabovsky, Y. and Truskinovsky, L.},
  journal   = {Archive for Rational Mechanics and Analysis},
  title     = {Roughening instability of broken extremals},
  year      = {2011},
  number    = {1},
  pages     = {183--202},
  volume    = {200},
  af        = {Grabovsky, YuryEOLEOLTruskinovsky, Lev},
  c1        = {[Grabovsky, Yury] Temple Univ, Dept Math, Philadelphia, PA 19122 USA.EOLEOL[Truskinovsky, Lev] Ecole Polytech, Mecan Solides Lab, F-91128 Palaiseau, France.},
  da        = {2020-03-16},
  doi       = {10.1007/s00205-010-0377-8},
  em        = {yury@temple.edu; trusk@lms.polytechnique.fr},
  fu        = {National Science FoundationNational Science Foundation (NSF) [1008092];EOLEOLEUEuropean Union (EU) [MRTN-CT-2004-505226]},
  fx        = {The authors are grateful to G. Allaire, A. Cherkaev, A. Freidin, R.EOLEOLFosdick, M. Grinfeld, R. V. Kohn, K. Lurie, S. Muller, P.EOLEOLPodio-Guidugli, M. Silhavy and an anonymous reviewer for helpful andEOLEOLinsightful remarks. This material is based upon work supported by theEOLEOLNational Science Foundation under Grant No. 1008092. LT was alsoEOLEOLsupported by the EU contract MRTN-CT-2004-505226.},
  ga        = {736QE},
  j9        = {ARCH RATION MECH AN},
  ji        = {Arch. Ration. Mech. Anal.},
  la        = {English},
  nr        = {78},
  pa        = {233 SPRING ST, NEW YORK, NY 10013 USA},
  pg        = {20},
  pi        = {NEW YORK},
  publisher = {Springer},
  rp        = {Grabovsky, Y (reprint author), Temple Univ, Dept Math, Philadelphia, PA 19122 USA.},
  sc        = {Mathematics; Mechanics},
  sn        = {0003-9527},
  tc        = {9},
  u1        = {0},
  u2        = {5},
  ut        = {WOS:000288508400006},
  wc        = {Mathematics, Applied; Mechanics},
  z9        = {10},
}

@Article{Grinfeld1980,
  author  = {Grinfeld, M. A.},
  journal = {Doklady Akademii Nauk SSSR},
  title   = {On the conditions of thermodynamic equilibrium of nonlinearly elastic material phases},
  year    = {1980},
  number  = {4},
  pages   = {824--828},
  volume  = {251},
}

@Article{Hansbo2002,
  author    = {Hansbo, A. and Hansbo, P.},
  journal   = {Computer Methods in Applied Mechanics and Engineering},
  title     = {An unfitted finite element method, based on {N}itsche's method, for elliptic interface problems},
  year      = {2002},
  pages     = {5537--5552},
  volume    = {191},
  af        = {Hansbo, AEOLEOLHansbo, P},
  c1        = {Univ Trollhattan Uddevalla, Dept Math & Informat, S-46139 Trollhattan, Sweden.EOLEOLChalmers Univ Technol, Dept Appl Mech, S-41296 Gothenburg, Sweden.},
  da        = {2021-03-16},
  doi       = {10.1016/S0045-7825(02)00524-8},
  ga        = {621NN},
  j9        = {COMPUT METHOD APPL M},
  ji        = {Comput. Meth. Appl. Mech. Eng.},
  la        = {English},
  nr        = {10},
  pa        = {PO BOX 564, 1001 LAUSANNE, SWITZERLAND},
  pg        = {16},
  pi        = {LAUSANNE},
  publisher = {Elsevier Science Sa},
  rp        = {Hansbo, P (corresponding author), Chalmers Univ Technol, Dept Solid Mech, S-41296 Gothenburg, Sweden.},
  sc        = {Engineering; Mathematics; Mechanics},
  sn        = {0045-7825},
  tc        = {465},
  u1        = {1},
  u2        = {30},
  ut        = {WOS:000179596200008},
  wc        = {Engineering, Multidisciplinary; Mathematics, InterdisciplinaryEOLEOLApplications; Mechanics},
  z9        = {468},
}

@InProceedings{Hansbo2017,
  Title                    = {Cut finite element methods for linear elasticity problems},
  Author                   = {Hansbo, P. and Larson, M. G. and Larsson, K.},
  Booktitle                = {Geometrically Unfitted Finite Element Methods and Applications},
  Year                     = {2017},
  Editor                   = {Bordas, S. P. A. and Burman, E. and Larson, M. G. and Olshanskii, M. A.},
  Pages                    = {25--63},
  Publisher                = {Springer International Publishing},

  ISBN                     = {978-3-319-71431-8}
}

@Book{Holzapfel2000,
  author    = {Holzapfel, G. A.},
  publisher = {John Wiley \& Sons Ltd.},
  title     = {Nonlinear solid mechanics: {A} continuum approach for engineering},
  year      = {2000},
}

@Article{James2002,
  author  = {James, R. D.},
  journal = {Continuum Mechanics and Thermodynamics},
  title   = {Configurational forces in magnetism with application to the dynamics of a small-scale ferromagnetic shape memory cantilever},
  year    = {2002},
  number  = {1},
  pages   = {55--86},
  volume  = {14},
  doi     = {10.1007/s001610100072},
  url     = {https://doi.org/10.1007/s001610100072},
}

@Article{Kuwahara1995,
  author  = {Kuwahara, H. and Tomioka, Y. and Asamitsu, A. and Moritomo, Y. and Tokura, Y.},
  journal = {Science},
  title   = {A first-order phase transition induced by a magnetic field},
  year    = {1995},
  number  = {5238},
  pages   = {961--963},
  volume  = {270},
  doi     = {10.1126/science.270.5238.961},
  eprint  = {https://www.science.org/doi/pdf/10.1126/science.270.5238.961},
  url     = {https://www.science.org/doi/abs/10.1126/science.270.5238.961},
}

@Book{Maugin1988,
  author    = {Maugin, G. A.},
  publisher = {Elsevier Science Publishers},
  title     = {Continuum mechanics of electromagnetic solids},
  year      = {1988},
  series    = {North-Holland Series in Applied Mathematics and Mechanics, Volume 33},
}

@Book{Maugin1991,
  author    = {Maugin, G. A.},
  editor    = {Dunayev, I. M. and Parton, V. Z.},
  publisher = {Moscow Mir},
  title     = {Mehanika elektromagnitnyh sploshnyh sred},
  year      = {1991},
}

@Article{Maugin1997,
  author  = {Maugin, G. A. and Fomethe, A.},
  journal = {Meccanica},
  title   = {Phase-transition fronts in deformable ferromagnets},
  year    = {1997},
  number  = {4},
  pages   = {347--362},
  volume  = {32},
  doi     = {10.1023/A:1004227307949},
  url     = {https://doi.org/10.1023/A:1004227307949},
}

@InCollection{Maugin2011,
  author    = {Maugin, G. A.},
  booktitle = {Mechanics and Electrodynamics of Magneto- and Electro-Elastic Materials},
  publisher = {Springer Wien New York},
  title     = {Electromagnetics in deformable solids},
  year      = {2011},
  editor    = {Ogden, R. W. and Steigmann, D. J.},
  pages     = {1--55},
  series    = {CISM International Centre for Mechanical Sciences Courses and Lectures},
  volume    = {527},
  doi       = {10.1007/978-3-7091-0701-0_1},
  url       = {https://doi.org/10.1007/978-3-7091-0701-0_1},
}

@Article{Mennerich2011,
  author  = {Mennerich, C. and Wendler, F. and Jainta, M. and Nestler, B.},
  journal = {Archives of Mechanics},
  title   = {A phase-field model for the magnetic shape memory effect},
  year    = {2011},
  number  = {5-6},
  pages   = {549--571},
  volume  = {63},
}

@Article{Morozov2023,
  author  = {Morozov, A. and Freidin, A. B. and M\"{u}ller, W. H.},
  journal = {International Journal of Engineering Science},
  title   = {On stress-affected propagation and stability of chemical reaction fronts in solids},
  year    = {2023},
  pages   = {103876},
  volume  = {189},
  doi     = {10.1016/j.ijengsci.2023.103876},
}

@Article{Morozov2024,
  author  = {Morozov, A. and Poluektov, M. and Freidin, A. B. and Figiel, \L{}. and M\"{u}ller, W. H.},
  journal = {European Journal of Mechanics --- A/Solids},
  title   = {Stability of chemical reaction fronts in solids: {A}nalytical and numerical approaches},
  year    = {2024},
  pages   = {105211},
  volume  = {104},
  doi     = {https://doi.org/10.1016/j.euromechsol.2023.105211},
  url     = {https://www.sciencedirect.com/science/article/pii/S0997753823003030},
}

@Article{Nitsche1971,
  author  = {Nitsche, J.},
  journal = {Abhandlungen aus dem Mathematischen Seminar der Universit\"{a}t Hamburg},
  title   = {\"{U}ber ein variationsprinzip zur l\"{o}sung von {D}irichlet-problemen bei verwendung von teilr\"{a}umen, die keinen randbedingungen unterworfen sind},
  year    = {1971},
  number  = {1},
  pages   = {9--15},
  volume  = {36},
  url     = {https://doi.org/10.1007/BF02995904},
}

@Article{Planes2009,
  author  = {Planes, A. and Ma\~{n}osa, L. and Acet, M.},
  journal = {Journal of Physics: Condensed Matter},
  title   = {Magnetocaloric effect and its relation to shape-memory properties in ferromagnetic {H}eusler alloys},
  year    = {2009},
  number  = {23},
  pages   = {233201},
  volume  = {21},
  doi     = {10.1088/0953-8984/21/23/233201},
  url     = {https://doi.org/10.1088/0953-8984/21/23/233201},
}

@Article{Poluektov2019,
  author  = {Poluektov, M. and Figiel, \L{}.},
  journal = {Computational Mechanics},
  title   = {A numerical method for finite-strain mechanochemistry with localised chemical reactions treated using a {N}itsche approach},
  year    = {2019},
  number  = {5},
  pages   = {885--911},
  volume  = {63},
  url     = {https://doi.org/10.1007/s00466-018-1628-z},
}

@Article{Poluektov2022,
  author  = {Poluektov, M. and Figiel, \L{}.},
  journal = {Computer Methods in Applied Mechanics and Engineering},
  title   = {A cut finite-element method for fracture and contact problems in large-deformation solid mechanics},
  year    = {2022},
  pages   = {114234},
  volume  = {388},
  af      = {Poluektov, MichaelEOLEOLFigiel, Lukasz},
  ea      = {NOV 2021},
  ei      = {1879-2138},
  oi      = {Figiel, Lukasz/0000-0002-5826-8320},
  sn      = {0045-7825},
  ut      = {WOS:000720450400006},
  doi     = {10.1016/j.cma.2021.114234},
}

@Article{RenukaBalakrishna2024,
  author  = {A. Renuka Balakrishna},
  journal = {MRS Communications},
  title   = {Rethinking hysteresis in magnetic materials},
  year    = {2024},
  number  = {5},
  pages   = {835--845},
  volume  = {14},
  doi     = {10.1557/s43579-024-00624-6},
  url     = {https://doi.org/10.1557/s43579-024-00624-6},
}

@InCollection{Truesdell1960,
  author    = {Truesdell, C. and Toupin, R.},
  booktitle = {Principles of Classical Mechanics and Field Theory / Prinzipien der Klassischen Mechanik und Feldtheorie},
  publisher = {Springer-Verlag Berlin G\"{o}ttingen Heidelberg},
  title     = {The classical field theories},
  year      = {1960},
  editor    = {Fl{\"u}gge, S.},
  pages     = {226--858},
  series    = {Handbuch der Physik},
  volume    = {III/1},
  doi       = {10.1007/978-3-642-45943-6_2},
  url       = {https://doi.org/10.1007/978-3-642-45943-6_2},
}

@Article{Wang2014,
  author  = {Wang, J. and Steinmann, P.},
  journal = {Continuum Mechanics and Thermodynamics},
  title   = {On the modeling of equilibrium twin interfaces in a single-crystalline magnetic shape memory alloy sample. {I}: {T}heoretical formulation},
  year    = {2014},
  number  = {4},
  pages   = {563--592},
  volume  = {26},
  doi     = {10.1007/s00161-013-0319-4},
  url     = {https://doi.org/10.1007/s00161-013-0319-4},
}

@Article{Wang2016,
  author  = {Wang, J. and Steinmann, P.},
  journal = {Continuum Mechanics and Thermodynamics},
  title   = {On the modeling of equilibrium twin interfaces in a single-crystalline magnetic shape memory alloy sample. {II}: {N}umerical algorithm},
  year    = {2016},
  number  = {3},
  pages   = {669--698},
  volume  = {28},
  doi     = {10.1007/s00161-014-0403-4},
  url     = {https://doi.org/10.1007/s00161-014-0403-4},
}

@Article{Wang2016a,
  author  = {Wang, J. and Steinmann, P.},
  journal = {Continuum Mechanics and Thermodynamics},
  title   = {On the modeling of equilibrium twin interfaces in a single-crystalline magnetic shape-memory alloy sample. {III}: {M}agneto-mechanical behaviors},
  year    = {2016},
  number  = {3},
  pages   = {885--913},
  volume  = {28},
  doi     = {10.1007/s00161-015-0452-3},
  url     = {https://doi.org/10.1007/s00161-015-0452-3},
}

\end{flushleft}

\end{document}